\documentclass[twocolumn,aps,prd,amsmath,nofootinbib,superscriptaddress]{revtex4-1}
\usepackage{cancel}
\usepackage{graphicx}
\usepackage{color}
\usepackage{setspace}
\usepackage{fancyhdr}
\usepackage{amssymb}
\usepackage{amsmath}
\usepackage{float}
\usepackage{todonotes}
\usepackage{booktabs}
\usepackage{multirow}
\usepackage[utf8]{inputenc}

\usepackage{hyperref}
\usepackage{epstopdf}
\usepackage{bm}
\usepackage{times}
\hypersetup{
  colorlinks=true,        
  linkcolor=blue,         
  citecolor=magenta,      
}
\newcommand{\cf}{{cf.,}~}
\newcommand{\ie}{{i.e.,}~}
\newcommand{\eg}{{e.g.,}~}

\begin{document}

\title{Constraining twin stars with GW170817}

\date{\today}

\author{Gl\`oria~Monta\~na}
\affiliation{Departament de F\'{\i}sica
  Qu\`antica i Astrof\'{\i}sica and Institut de Ci\`encies del Cosmos
  (ICCUB), Universitat de Barcelona, Mart\'{\i} i Franqu\`es 1, 08028
  Barcelona, Spain}

\author{Laura~Tol\'os}
\affiliation{Institut f{\"u}r Theoretische Physik,
  Max-von-Laue-Stra{\ss}e 1, 60438 Frankfurt, Germany}
\affiliation{Frankfurt Institute for Advanced Studies,
  Ruth-Moufang-Stra{\ss}e 1, 60438 Frankfurt, Germany}
\affiliation{Institute of Space Sciences (ICE, CSIC), Campus UAB, Carrer de Can Magrans, 08193, Barcelona, Spain}
\affiliation{Institut d'Estudis Espacials de Catalunya (IEEC), 08034 Barcelona, Spain}
\author{Matthias~Hanauske}
\affiliation{Institut f{\"u}r Theoretische Physik,
  Max-von-Laue-Stra{\ss}e 1, 60438 Frankfurt, Germany}
\affiliation{Frankfurt Institute for Advanced Studies,
  Ruth-Moufang-Stra{\ss}e 1, 60438 Frankfurt, Germany}

\author{Luciano~Rezzolla} \affiliation{Institut f{\"u}r Theoretische
  Physik, Max-von-Laue-Stra{\ss}e 1, 60438 Frankfurt, Germany}


\begin{abstract} 
If a phase transition is allowed to take place in the core of a compact
star, a new stable branch of equilibrium configurations can appear,
providing solutions with the same mass as the purely hadronic branch and
hence giving rise to ``twin-star'' configurations. We perform an
extensive analysis of the features of the phase transition leading to
twin-star configurations and, at the same time, fulfilling the
constraints coming from the maximum mass of $2~M_{\odot}$ and the
information following gravitational-wave event GW170817. In particular,
we use a general equation of state for the neutron-star matter that
parametrizes the hadron--quark phase transition between the model
describing the hadronic phase and a constant speed of sound for the quark
phase. We find that the largest number of twin-star solutions has masses
in the neutron-star branch that are in the range $1$--$2\,M_{\odot}$ and
maximum masses $\gtrsim$ $2\,M_{\odot}$ in the twin-star branch. The
analysis of the masses, radii and tidal deformabilities also reveals that
when twin stars appear, the tidal deformability shows two distinct
branches with the same mass, thus differing considerably from the
behaviour expected for normal neutron stars. In addition, we find that
the data from GW170817 is compatible with the existence of hybrid stars
and could also be interpreted as produced by the merger of a binary
system of hybrid stars or of a hybrid star with a neutron star. Indeed,
with the use of a well-established hadronic EOS the presence of a hybrid
star in the inspiral phase could be revealed if future gravitational-wave
detections measure chirp masses $\mathcal{M} \lesssim 1.2\,M_\odot$ and
tidal deformabilities of $\Lambda_{1.4} \lesssim 400$ for
$1.4\,M_{\odot}$ stars. Finally, combining all observational information
available so far, we set constraints on the parameters that characterise
the phase transition, the maximum masses, and the radii of
$1.4\,M_{\odot}$ stars described by equations of state leading to
twin-star configurations.
\end{abstract}

\maketitle

\section{Introduction}

Compact stars have been the subject of great attention over the years as
natural laboratories for testing the different phases of matter under
extreme conditions. Depending on the type of matter in their interior,
several possibilities for their nature have been postulated: strange
quark stars, (pure) neutron stars or hybrid stars. Whereas strange quark
stars are made of deconfined quark matter \cite{Itoh:1970uw, Cheng1996PRL,
  Bombaci2000ApJ, Hanauske2001PRD, Weber:2004kj, Zacchi2015PRD}, pure
neutron stars are composed of hadrons
\cite{Pal1999PRC,Hanauske2000ApJ,Pal2000NuPhA,Lattimer:2006xb}. Hybrid
stars are compact stars with a core consisting of quark matter and outer
layers of hadronic matter \cite{Heiselberg1993PRL, Oestgaard:1994gy,
  Schertler1999PRC, Steiner2000PLB, Hanauske2001GRG, Mishustin2003PhLB,
  Banik2004PRD, Zacchi2016PRD, WeberBook2017, Roark:2018boj}. Present and future
observations of neutron-star features, such as masses, radii and tidal
deformabilities, will help to constrain the equation of state (EOS) in
the high-density regime in the upcoming years.

High-precision measurements, obtained using post-Keplerian parameters,
have shown that the EOS of neutron stars must be able to support masses
of $2\,M_\odot$ \cite{Demorest:2010bx, Antoniadis:2013pzd,
  Arzoumanian:2017puf}. The radii, on the other hand, are more difficult
to be determined observationally. The uncertainties in the modeling of
the X-ray emission result in different radii determinations, which still
lay in a rather wide range. Several astrophysical analyses for the
extraction of the radii \cite{Verbiest:2008gy, Ozel:2010fw,
  Suleimanov:2010th, Lattimer:2012xj, Steiner:2012xt, Bogdanov:2012md,
  Guver:2013xa, Guillot:2013wu, Lattimer:2013hma, Poutanen:2014xqa,
  Heinke:2014xaa, Guillot:2014lla, Ozel:2015fia, Ozel:2015gia,
  Lattimer:2015nhk, Ozel:2016oaf, Steiner:2017vmg} are favouring small
values, mostly in the range of $9$--$13\,{\rm km}$. High-precision X-ray
space missions such as the ongoing Neutron star Interior Composition
ExploreR (NICER) \cite{2014SPIE.9144E..20A} or the future enhanced X-ray
Timing and Polarimetry Mission (eXTP) \cite{Watts:2018iom} are expected
to offer precise and simultaneous measurements of masses and radii. Also
promising constraints on the mass--radius relation are expected to be
obtained from gravitational waves and multi-messenger astronomy
\cite{Margalit:2017dij, Bauswein:2017vtn, Shibata:2017xdx,
  Rezzolla:2017aly, Ruiz:2017due, Annala:2017llu}.
 
The recent detection by the Advanced LIGO and Virgo collaborations
\cite{TheLIGOScientific:2017qsa,Abbott:2018wiz} of gravitational waves
from merging compact stars, GW170817, has provided important new insights
on the maximum mass and on the radius of neutron stars by means of the
measurement of tidal deformabilities in a binary system
\cite{Margalit:2017dij, Shibata:2017xdx, Rezzolla:2017aly, Ruiz:2017due,
  Annala:2017llu, Kumar:2017wqp, Fattoyev:2017jql, Most:2018hfd,
  Lim:2018bkq, Raithel:2018ncd, Burgio:2018yix, Tews:2018chv, De:2018uhw,
  Abbott:2018exr, Malik:2018zcf}. We recall that the tidal deformability
measures the induced quadrupole moment of a star in response to the tidal
field of its companion, \ie it determines how easily a star is deformed
in a binary system \cite{Hinderer:2009ca}. This quantity is therefore
strongly correlated with the properties of the phases of matter in the
compact-star interior and that are described by the EOS.

In fact, a number of works have explored the possibility of using the
detection of GW170817 to probe the occurrence of a hadron--quark phase
transition (HQPT), finding that GW170817 is consistent with
the coalescence of neutron stars and hybrid stars
\cite{Paschalidis:2017qmb, Nandi:2017rhy, Most:2018hfd, Burgio:2018yix,
  Alvarez-Castillo:2018pve, Gomes:2018eiv, Sieniawska:2018zzj,
  Li:2018ayl, Christian:2018jyd, Han:2018mtj}. We also recall that
depending on the features of the phase transition between the inner quark
core and outer hadronic parts of the hybrid star, twin-star solutions
might appear as the mass--radius relation could exhibit two stable
branches with similar masses \cite{Gerlach:1968zz, Kampfer:1981yr, Glendenning:1998ag}. Indeed,
information from gravitational waves can also be exploited to better
understand the twin-star scenario \cite{Paschalidis:2017qmb,
  Alvarez-Castillo:2018pve, Burgio:2018yix, Sieniawska:2018zzj,
  Christian:2018jyd}.

We here present a systematic and detailed study of the features of the
HQPT in order to obtain twin-star configurations and, at the same time,
fulfil the $2\,M_{\odot}$ observations and the information on
multi-messenger observation of the GW170817 event. Our results show that
the GW170817 event is compatible with either the merger of a binary
hybrid-star system or the merger of a hybrid star with a neutron
star. We place constraints on the parametrization of the HQPT so as to be
consistent with the GW170817 information and obtain the resulting allowed
ranges for the maximum mass $M_{_{\rm TOV}}^{^{\uparrow}}$ and radius of
a $1.4\,M_\odot$ star. We note that in a very recent work, Han and
Steiner \cite{Han:2018mtj} investigate the sensitivity of the tidal
deformability to the properties of a sharp HQPT, not necessarily
producing twin stars, and report that a smoothing of the transition has
appreciable effects only for central densities close to the onset of the
quark phase. In our work we evaluate in more detail the similarities and
differences between a model with a sharp phase transition and a model
allowing for a mixed phase of hadrons and quarks in the context of twin
stars and we particularly show that with a non-sharp HQPT twin-star
solutions are harder to be found and the parameters characterizing the
transition are better constrained.

The article is organized as follows. In Section~\ref{sec:models} we
describe the details of the general two EOS models used through the work
that implement HQPTs using a Maxwell (sharp) or a Gibbs (smooth)
construction, while in Section~\ref{sec:results} we present our
constraints for the HQPT parameter space, as well as the mass, radius and
tidal deformability of binary neutron stars. Our conclusions and outlook
are summarized in Section~\ref{sec:conclusions}.

\section{Models of the Equation of State}
\label{sec:models}

In this exploratory work we systematically construct two classes of
physically plausible EOSs of the neutron-star matter. In particular, for
the ``low-density'' region of the inner core we make use of EOSs that
share the same properties of a hadronic EOS recently discussed in
Refs. \cite{Tolos:2016hhl,Tolos:2017lgv}, whereas for the inner and outer
crust we employ the EOS of Ref.~\citep{Sharma:2015bna}. For the
``high-density'' region, on the other hand, we consider two distinct
models that provide different parametrizations of the HQPT, assuming
either a ``Maxwell construction'' or a ``Gibbs construction'' (referred
to as \texttt{Model-1} and \texttt{Model-2} EOSs below). Finally, the
quark phase is modelled using a parametrization with a constant speed of
sound (CSS). The two models are described below.

\subsection{Hadronic EOS}
\label{subsec:hadronicEOS}

\begin{figure}
 \centering
 \includegraphics[width=\columnwidth]{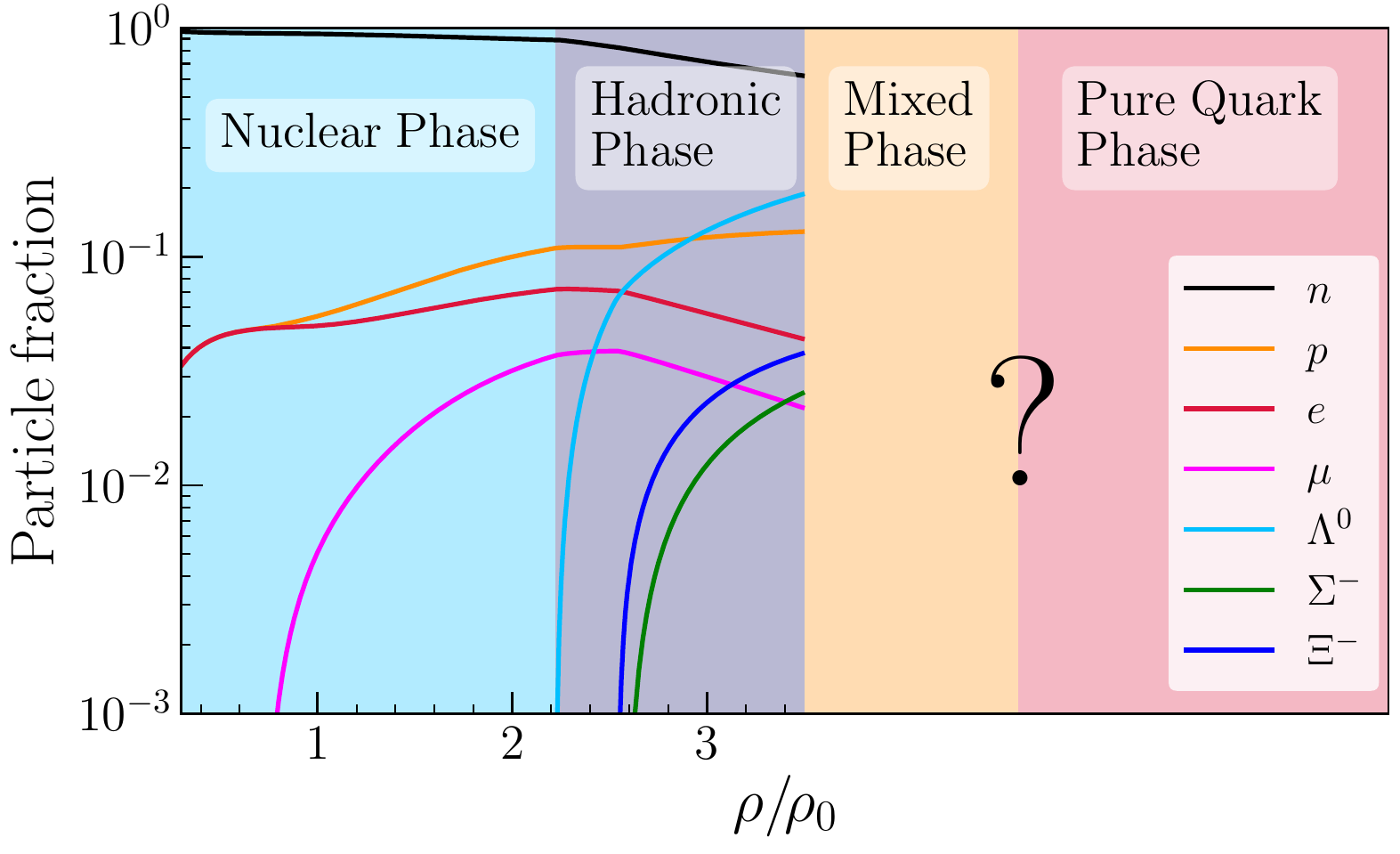}
 \caption{Particle fractions as functions of the baryonic density for the
   FSU2H model \cite{Tolos:2016hhl,Tolos:2017lgv} up to the point where
   the HQPT is implemented, giving rise to a phase of deconfined quark
   matter which can be separated from the nuclear (or hadronic) phase by
   a mixed phase of hadrons and quarks. We note that the actual fractions
   of nucleons/hyperons and quarks $u,d,s$ in the mixed and quark phases
   cannot be determined with the parametrizations used in this
   work. }
\label{fig:particlefractions}
\end{figure}

For the hadronic phase we use the FSU2H EOS of
Refs.~\cite{Tolos:2016hhl,Tolos:2017lgv}, which is a recent
relativistic-mean field (RMF) model based on the nucleonic FSU2 model of
\citep{Chen:2014sca} that considers not only nucleons but also hyperons
in the inner core of neutron stars by reproducing the available
hypernuclear structure data \cite{Dover:1982ng, Millener:1988hp,
  Fukuda:1998bi, Khaustov:1999bz, Noumi:2001tx, Harada:2006yj,
  Kohno:2006iq}. This scheme reconciles the $2\,M_{\odot}$ mass
observations with the recent analyses of radii below 13 km for neutron
stars \cite{Verbiest:2008gy, Ozel:2010fw, Suleimanov:2010th,
  Lattimer:2012xj, Steiner:2012xt, Bogdanov:2012md, Guver:2013xa,
  Guillot:2013wu, Lattimer:2013hma, Poutanen:2014xqa, Heinke:2014xaa,
  Guillot:2014lla, Ozel:2015fia, Ozel:2015gia, Lattimer:2015nhk,
  Ozel:2016oaf, Steiner:2017vmg}, while fulfilling the saturation
properties of nuclear matter and finite nuclei
\citep{Tsang:2012se,Chen:2014sca}, as well as the constraints extracted
from nuclear collective flow \citep{Danielewicz:2002pu} and kaon
production \citep{Fuchs:2000kp,Lynch:2009vc} in heavy-ion
collisions. Moreover, cooling simulations for isolated neutron stars
using the FSU2H model are in very good agreement with observational data
\cite{Negreiros:2018cho}.

The particle fractions as functions of the baryonic density for the FSU2H
model are shown in Fig.~\ref{fig:particlefractions} up to a density
$\rho_{\rm tr}$, where the HQPT is implemented. As already seen in
Refs. \cite{Tolos:2016hhl, Tolos:2017lgv, Negreiros:2018cho}, the first
hyperon to appear is the $\Lambda$ particle, followed by $\Xi^-$ and
$\Sigma^-$, as beta-equilibrium and charge conservation are fulfilled
taking into account the most plausible hyperon potentials extracted from
hypernuclear data.

\subsection{High-density EOS}
\label{subsec:highdensityEOS}

As the density is increased, hadrons might undergo deconfinement,
liberating quarks and enabling the existence of a quark-matter
core. Although the low temperature and large chemical potential regime
occurring in neutron-stars interiors is still far from being well
understood, two frameworks have been mainly used in the literature to
describe quark matter in compact objects: the MIT bag model 
and the Nambu-Jona-Lasinio (NJL) model. A simpler description assuming a
density-independent speed of sound mimicking these sophisticated models\footnote{Calculations
of the speed of sound for these models can be found in recent works \cite{Zacchi:2015oma,Ranea-Sandoval:2015ldr}. }
was first investigated in \cite{Chamel:2012ea, Zdunik:2012dj,
  Alford:2013aca}.

Due to its simplicity, the phenomenological CSS parametrization is
well-suited for the systematic investigation of EOSs with twin stars
developed in this paper. The value $c_s^2:= \partial p/\partial e ={\rm
  const}=1$, where $p$ and $e$ are respectively the pressure and internal
energy density \cite{Rezzolla_book:2013}, has been previously used in
Refs. \cite{Alford:2013aca, AlfordSedrakian2017, Christian:2017jni, Paschalidis:2017qmb}. We
have checked that the lower value $c_s^2=1/3$ provided by perturbative
QCD calculations \cite{Kurkela:2009gj} does not give rise to EOSs with
twin stars that satisfy the $>2\,M_\odot$ maximum-mass constraint. Yet,
simply setting $c^2_s=1$ allows us to carry out the extended analysis
that will be presented in the following sections.

\subsection{Phase Transition}
\label{subsec:phasetransition}

The first-order transition between the hadronic and the quark phases is
attained by either a Gibbs or a Maxwell construction. In the former case,
the transition is modelled with a polytrope $p(\rho)=K_{\rm m}\rho^{
  \Gamma_{\rm m}}$ to account for a mixed soft phase of hadrons and
quarks \cite{Glendenning:1992vb,Macher:2004vw}, and has been investigated
in view of the recent gravitational-wave observations
\cite{Nandi:2017rhy}.  The latter, on the other hand, is equivalent to a
$\Gamma_{\rm m}=0$ polytrope, generates a sharp transition between the
low- and high-density phases, and has been widely used in recent works,
\eg Refs. \cite{Alford:2013aca, Christian:2017jni, Most:2018hfd,
  Paschalidis:2017qmb, Alvarez-Castillo:2018pve, Most:2018hfd,
  Han:2018mtj}. However, if the surface tension of the deconfined quark
phase has moderate values, a mixed phase between the pure hadronic and
pure quark phases is expected to be present. The construction of such a
continuous HQPT, where charge is only globally conserved, depends on the
properties of the pure hadronic and quark models and, additionally, on
possible effects of pasta structures within the mixed phase. As a result,
the amount to which the EOS is softened at the beginning of the mixed
phase is quite uncertain (see \eg \cite{Ayriyan2018}) and we have used a
value of $\Gamma_{\rm m}=1.03$ to mimic this effect.  
This choice is restricted by the fact that in our approach it is essential
that $\Gamma_{\rm m}$ has a low value around 1 in order to get a strong 
enough softening of the EOS and to be considerably different from the 
Maxwell construction. We have checked that increasing/decreasing
$\Gamma_{\rm m}$ by $\sim 2\%$ can shift the energy-density jump $\Delta
e$ up to a $\sim 5\%$ higher/lower values at a given transition pressure
$p_{\rm tr}$ of the parameter space of \texttt{Model-2} discussed below.
Thus the polytropic approach is a reasonable first approximation of a HQPT
using the Gibbs construction and the specific choice of the value
$\Gamma_{\rm m}=1.03$ does not affect significantly the discussion in the
following sections.

\begin{figure}
 \centering
     \includegraphics[width=\columnwidth]{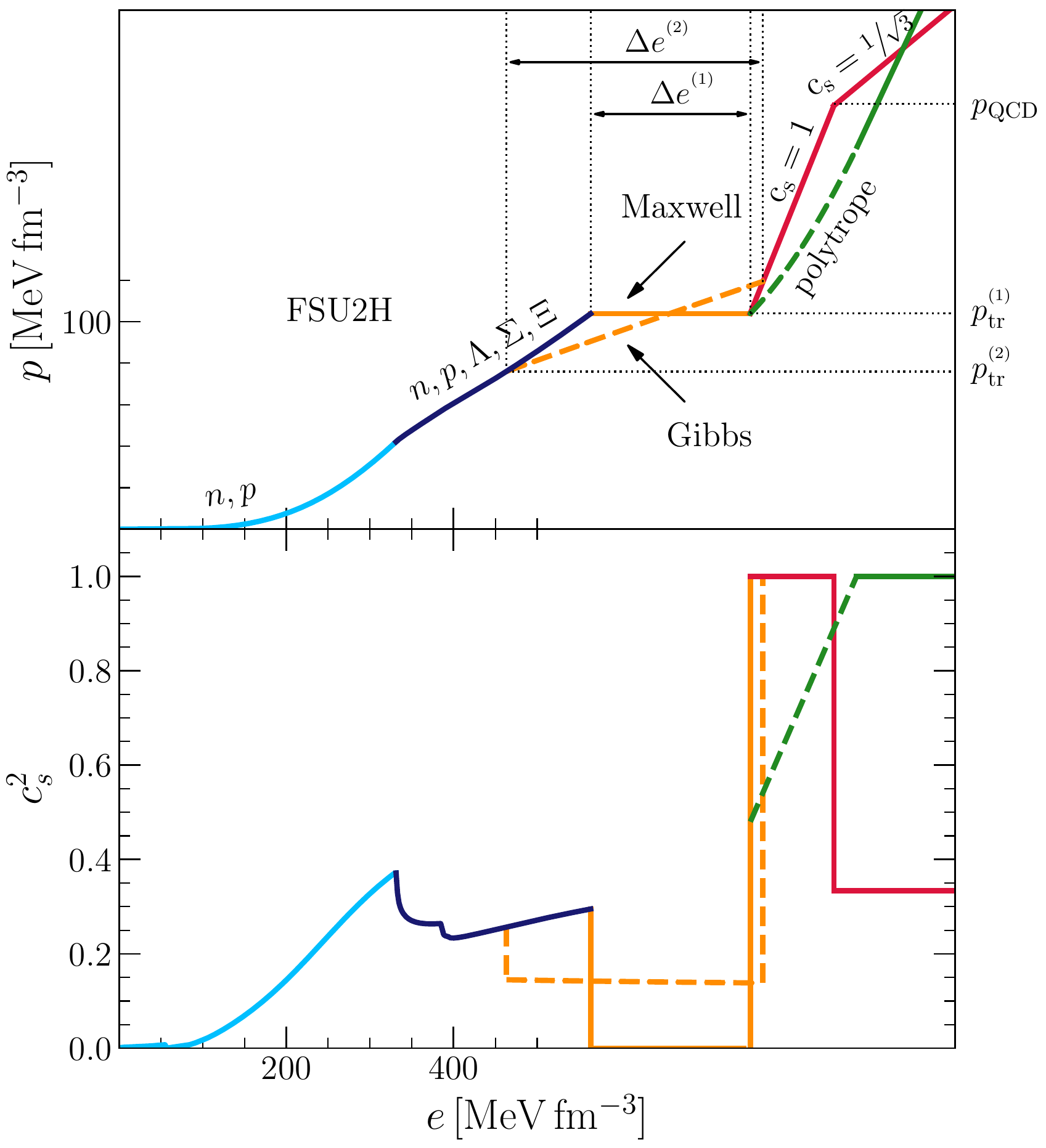}
 \caption{Upper panel: energy density as a function of the pressure
   corresponding to an EOS that implements a HQPT. Lower panel: The
   corresponding sound-speed squared in units of the speed of light. The
   colours are related to the composition of matter at increasing
   densities: $\beta$-equilibrated nucleonic matter (light blue),
   $\beta$-equilibrated nucleonic and hyperonic matter \mbox{-- hadronic
     matter --} (dark blue), mixed phase of hadrons and deconfined quarks
   (orange, only Gibbs), and pure quark matter (red). Solid and dashed
   lines are different ways of modelling the phase transition and the
   quark phase (see details in the text). }
\label{fig:eos}
\end{figure}

\subsection{Summary of the EOS models}

In view of the considerations above, the adoption of the two types of
phase transition will give rise to the following two models where the
relation between the specific internal energy and the pressure,
$e(p)$, is given by:
\begin{itemize}
 \item \texttt{Model-1}: FSU2H + \emph{Maxwell} + CSS
 \begin{equation}\label{eq:model1}
  e=\begin{cases} 
      e_{\rm FSU2H}(p) & p\leq p_{\rm tr} \\
      e_{\rm FSU2H}(p_{\rm tr}) + \Delta e+c_s^{-2}(p-p_{\rm tr}) & p \geq p_{\rm tr}
   \end{cases}
 \end{equation}
 with $c_s^{2}=1$.
 \item \texttt{Model-2}: FSU2H + \emph{Gibbs} + CSS
 \begin{equation}\label{eq:model2}
  e=\begin{cases} 
      e_{\rm FSU2H}(p) & p\leq p_{\rm tr} \\
      (1+a_{\rm m})\left({p}/{K_{\rm m}}\right)^{1/\Gamma_{\rm m}}+
      {p}/({\Gamma_{\rm m}-1}) & p_{\rm tr} \leq p \leq p_{_{\rm CSS}} \\
      e(p_{_{\rm CSS}}) + c_s^{-2}(p-p_{_{\rm CSS}}) & p \geq p_{_{\rm CSS}}
   \end{cases}
 \end{equation}
 with $c_s^{2}=1$ and $\Gamma_{\rm m}=1.03$.
\end{itemize}

The values of the polytropic constant $K_{\rm m}$ and the coefficient
$a_{\rm m}$ are obtained by ensuring that $p$ and $e$ are continuous at
the transition points. We note that in the Gibbs construction an
energy-density jump $\Delta e$ is not explicitly defined. In this case,
we assign to its value the increase in $e(p)$ during the mixed phase (see
Fig.~\ref{fig:eos}).

The possible models for the EOSs are schematically shown in the upper
panel of Fig.~\ref{fig:eos}, where one can clearly see the comparison
between \texttt{Model-1$^*$} and \texttt{Model-2$^*$} EOSs, in which the
speed of sound is set to reach the perturbative QCD
limit $c_s^2=1/3$ for quark matter above a certain energy density,
$e(p_{\rm QCD})$, and \texttt{Model-1$^\dagger$}, in
which there is a softer EOS for the quark matter right after the phase
transition. Following Refs. \cite{Paschalidis:2017qmb, Most:2018hfd},
this is modelled with a polytrope similar to that of the mixed phase
[middle piece of Eq.~(\ref{eq:model2}) with $K_{\rm m}\rightarrow K_{\rm
    q}$ and $a_{\rm m}\rightarrow a_{\rm q}$, guaranteeing the continuity
  of $p$ and $e$ after the phase transition], which is then combined with
a CSS parametrization with $c_s^2=1$ in the high-density quark phase. 
We note that \texttt{Model-1} and
\texttt{Model-2} constitute a particular case of \texttt{Model-1$^*$} and
\texttt{Model-2$^*$}, respectively, for which the perturbative QCD limit
is reached at densities higher than those in the interior of neutron stars.
In the lower panel of
Fig.~\ref{fig:eos} we display the square of the speed of sound (in units
of the speed of light), noting that in all cases it fulfils the causal
condition of $c_s^{2} \leq 1$. 

\begin{figure*}[tbp!]
 \centering
 \includegraphics[width=0.77\textwidth]{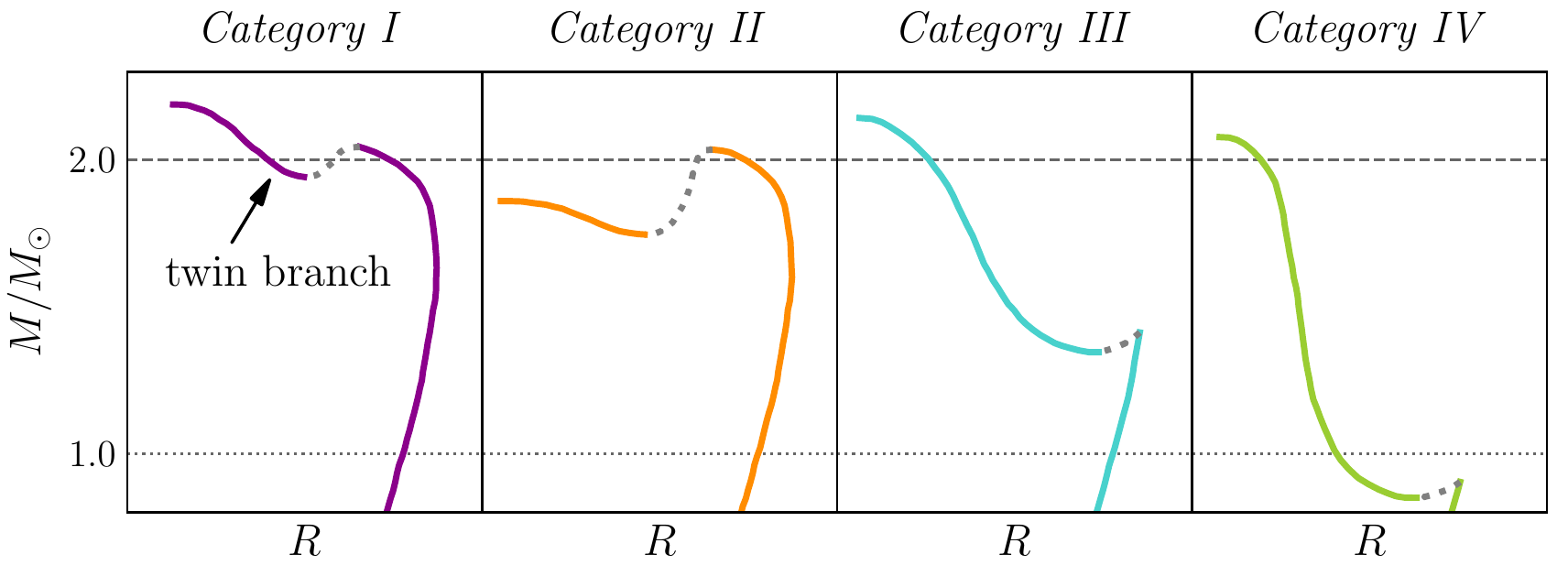}
 \caption{Schematic behaviour of the mass--radius relation for the
   twin-star categories \textit{I--IV} defined in the text. Note the
   appearance of a ``twin'' branch with a mixed or pure-quark phase; the
   twin branch has systematically smaller radii than the branch with a
   nuclear or hadronic phase. The colours used for these categories will
   be employed also in the subsequent figures.}
\label{fig:cartoon_MvsR}
\end{figure*}

Before discussing in the next section the similarities and differences of
the various models discussed above, it is useful to remark that our
construction of a HQPT is not based on the ``strange matter hypothesis''
\cite{Bodmer:1971we, Witten:1984rs} for which the strange-quark phase is
the true ground state of elementary matter. Under such an assumption, the
underlying EOS would separate in two different branches describing
neutron star and pure quark matter and as a consequence, a neutron star
would transform into a pure-quark star after exceeding a certain
deconfinement barrier \cite{Drago2014, Bombaci2016, Lugones2016,
  Drago2018}. This scenario is normally referred to as the
``two-families'' scenario and is different from the twin-star scenario, where
the two branches of compact stars are described by a single EOS.

\section{Results}
\label{sec:results}

\subsection{Parameter Space}
\label{subsec:parameterspace}

In order to analyse the implications of the EOS models discussed in the
previous section on the masses, radii and tidal deformabilities of
twin-star solutions, we vary the two free parameters of the models: the
density at which the phase transition to the mixed phase takes place,
$\rho_{\rm tr}$, and the density discontinuity (\texttt{Model-1}) or
density extension of the mixed phase (\texttt{Model-2}) up to the pure
quark phase, $\Delta\rho$. This is also equivalent to setting the
transition pressure, $p_{\rm tr}$, and the energy-density jump, $\Delta
e$. We note that the mass density $\rho$ in the quark phase is obtained
from Eqs.~(\ref{eq:model1})--(\ref{eq:model2}) together with the
thermodynamic relation at zero temperature \cite{Rezzolla_book:2013}
\begin{equation}
  p=\rho\frac{\partial e}{\partial\rho}-e\,,
\end{equation}
so as to use the values of $\rho_{\rm QCD}$ of the order of those
displayed in Fig.~7 of Ref.~\cite{Tews:2018kmu} when discussing the density
at which the perturbative QCD limit is reached. 

To allow for a wide range of EOSs, the parameter space analysed is
$\rho_{\rm tr}\in [1.4-6.5]\,\rho_0$ and $\Delta\rho\in [0.2-3.0]\,
\rho_0$ with variations of $0.1\,\rho_0$, where $\rho_0$ is the nuclear
saturation density.

We recall that by using the maximum masses in the two branches, twin-star
solutions can be classified in the four distinct categories shown schematically
in Fig. \ref{fig:cartoon_MvsR}:
\begin{itemize}
 \item  \textit{Category I}: \newline $M_{_{\rm TOV}}\geq 2.0\,M_\odot$ and $M_{_{\rm TOV,\,T}}\geq 2.\,M_\odot$
 \item \textit{Category II}: \newline $M_{_{\rm TOV}}\geq 2.0\,M_\odot$ and $M_{_{\rm TOV,\,T}} < 2.0\,M_\odot$
 \item \textit{Category III}: \newline $1.0\,M_\odot \leq M_{_{\rm TOV}} < 2.0\,M_\odot$ and $M_{_{\rm TOV,\,T}}\geq 2.0\,M_\odot$
 \item \textit{Category IV}: \newline $M_{_{\rm TOV}} < 1.0\,M_\odot$ and $M_{_{\rm TOV,\,T}}\geq 2.0\,M_\odot$,
\end{itemize}
where $M_{_{{\rm TOV}}}$ and $M_{_{\rm TOV,\,T}}$ are the maximum masses
of the branches with large (``normal-neutron-star'' branch) and small
radii (``twin'' branch), respectively, of a nonrotating neutron star
obtained by solving the Tolman-Oppenheimer-Volkoff (TOV) equations. Since
our aim is to focus on those configurations that allow for maximum masses
larger than $2\,M_\odot$ while having a twin-star solution, we will not
consider those cases where the EOSs lead to twin-star solutions that
violate the $M_{_{\rm TOV}}^{^{\uparrow}}:=\max\{ M_{_{\rm TOV}},M_{_{\rm
    TOV,\,T}}\}\ge 2.0\,M_\odot$ constraint.  Similarly, EOSs that do not
produce twin stars will also be rejected from our analysis.

\begin{figure*}[htb!]
 \centering
 \includegraphics[width=0.95\textwidth]{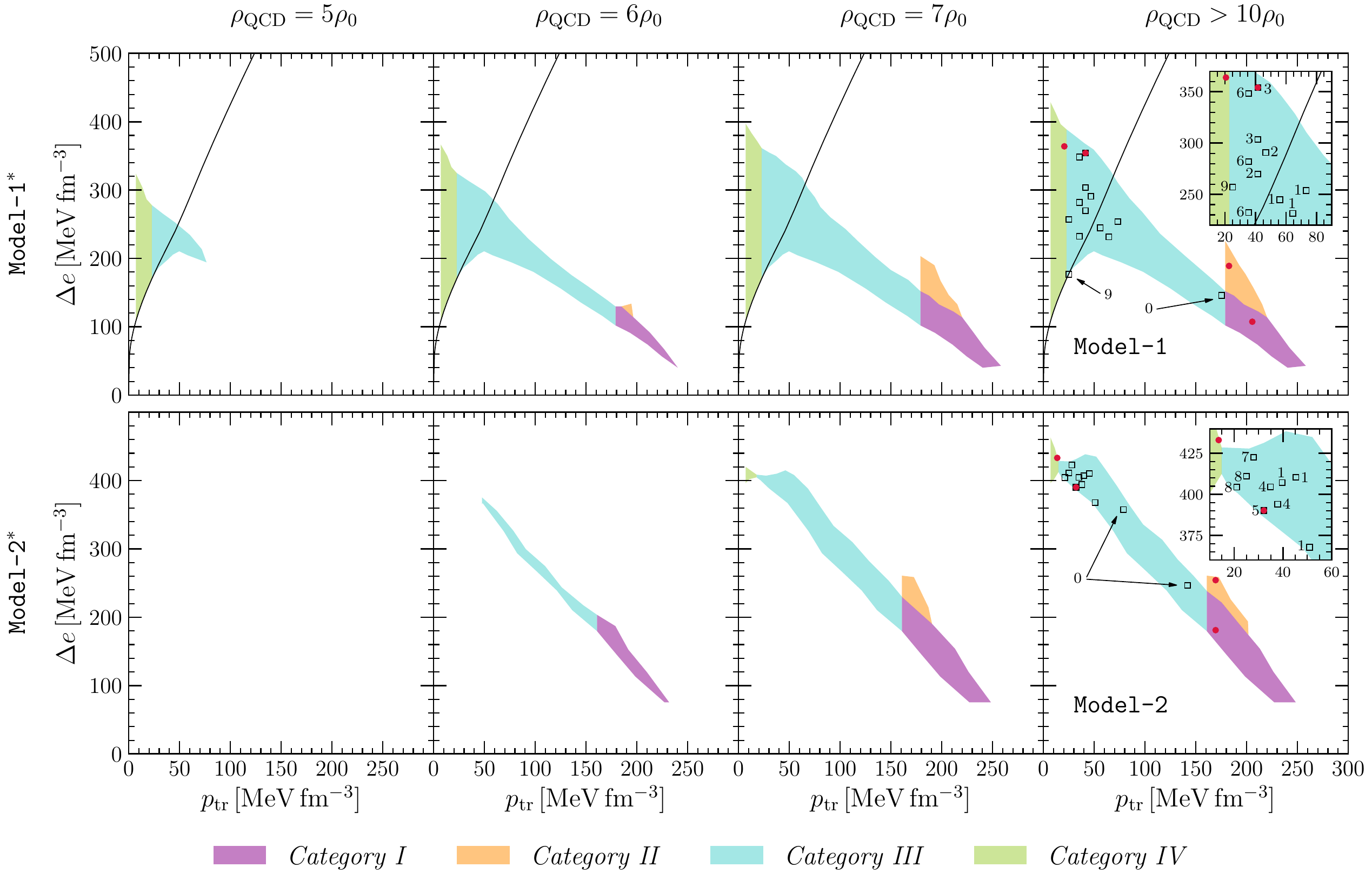}
 \caption{Areas containing the various categories of twin stars in the
   $\Delta e$--$p_{\rm tr}$ parameter space for EOS models with Maxwell
   construction (upper panels) and models with Gibbs construction (lower
   panels) for the phase transition, and increasing values of the density
   $\rho_{\rm QCD}$ corresponding to the onset of the perturbative QCD
   limit, $c_s=1/\sqrt{3}$, labelled \texttt{Model-1$^*$} and
   \texttt{Model-2$^*$} in the text. The \texttt{Model-1} and
   \texttt{Model-2} EOSs are depicted in the rightmost upper and lower
   panels, respectively. The solid line in the upper panels is the Seidov
   limit of Eq.~(\ref{eq:Seidov}). Circles, squares and numbers identify
   the specific cases studied in the figures in the rest of the paper and
   are also shown in the magnified insets.}
\label{fig:parameters_rhoqcd}
\end{figure*}

\begin{figure*}[htb!]
 \centering
 \includegraphics[width=0.95\textwidth]{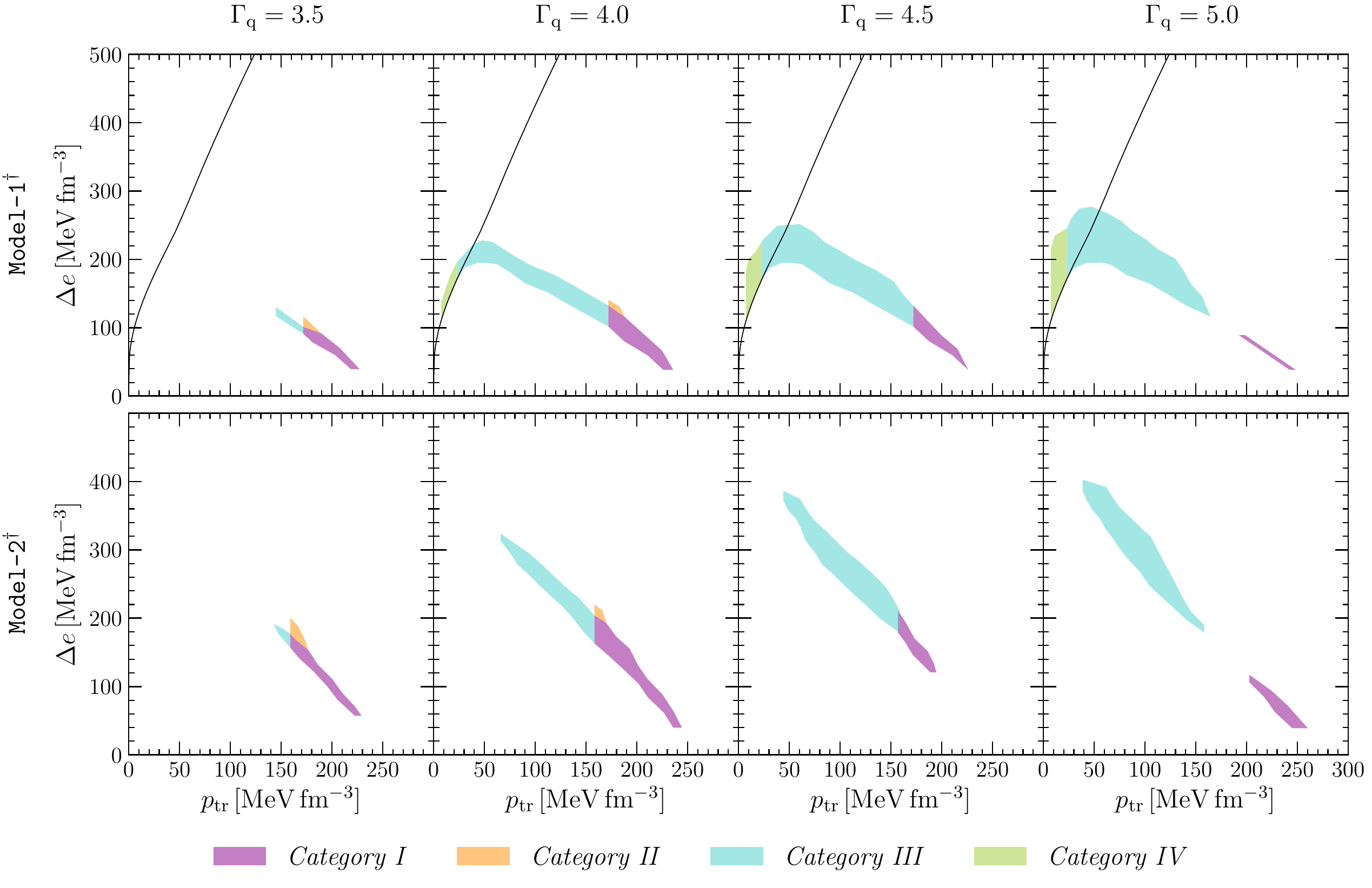}
\caption{Areas containing the various categories of twin stars in the
  $\Delta e$--$p_{\rm tr}$ parameter space for \texttt{Model-1$^\dagger$}
   with a Maxwell (upper panels) and \texttt{Model-2$^\dagger$} with a 
   Gibbs (lower panels) phase transition, that use for
  the quark phase a polytrope $p(\rho)=K_{\rm q}\rho^{\Gamma_{\rm q}}$
  combined with a constant speed of sound parametrization when the speed
  of sound $c_s^2=\partial p/\partial e$ reaches the value of one. The
  value of the polytropic index $\Gamma_{\rm q}$ is varied between 3.5 to
  5.0, so as to obtain twin-star solutions and masses $M\ge
  2.0\,M_\odot$. The solid line in the upper panels is the Seidov limit
  of Eq.~(\ref{eq:Seidov}).}
\label{fig:parameters_gamma}
\end{figure*}

Figure~\ref{fig:parameters_rhoqcd} shows the parameter space in the
$\Delta e$--$p_{\rm tr}$ plane for EOS models having either a Maxwell
construction (\texttt{Model-1$^*$}, upper panels) or a Gibbs construction
(\texttt{Model-2$^*$}, lower panels) for the phase transition. From the
left to the right, we vary the density
$\rho_{\rm QCD}$ at which the asymptotic perturbative $c_s^{2}=1/3$ limit
in the quark-matter phase is reached.

Only the combinations of parameters within the shaded regions correspond
to EOSs that allow for a twin-star configuration together with $M_{_{\rm
    TOV}}^{^{\uparrow}}\ge 2.0\,M_\odot$, with different colours referring
to \textit{Categories~I-IV}. We note that the allowed parameter space is
determined by the intersection of the two regions satisfying each of
these conditions, the boundaries of which are similar to those shown in
Fig. 2 of Ref.~\cite{Alford:2015dpa} with a Maxwell construction. The
solid line in the top panels of Fig.~\ref{fig:parameters_rhoqcd}
corresponds to the limiting condition for hybrid stars appearing in the
normal-neutron-star branch, which can be derived solely in the presence
of a sharp discontinuity in the energy density by performing an expansion
in powers of the size of the quark-matter core \cite{1971SvA....15..347S,
  1983A&A...126..121S, Lindblom:1998dp, Alford:2013aca}, written in the
following form:
\begin{equation}
  \label{eq:Seidov}
 \Delta e=\frac{1}{2}e_{\rm tr}+\frac{3}{2} p_{\rm tr} \,.
\end{equation}
For combinations of parameters above the Seidov line \eqref{eq:Seidov},
the sequence of stars will become unstable immediately after the central
pressure reaches the value $p_{\rm c}=p_{\rm tr}$, \ie the stars in the
normal-neutron-star branch will be purely hadronic, while the combinations
below the line correspond to solutions for which the normal-neutron-star
branch can support hybrid stars (with quark core) before turning unstable.
Note also that the circles and squares
in the figure identify the specific cases studied more in detail below.

When inspecting Fig. \ref{fig:parameters_rhoqcd} it is evident that
\texttt{Model-1} and \texttt{Model-2} are the most effective EOSs in
fulfilling both requirements, as shown by the corresponding largest
coverage of the $\Delta e$--$p_{\rm tr}$ parameter space. Moreover, since
it is not yet clear at what value of $\rho_{\rm QCD}/\rho_0$ the
asymptotic perturbative QCD limit takes place, our analysis hereafter
will be focused on \texttt{Model-1} and \texttt{Model-2} EOSs only. For
these two cases, \textit{Categories I} (purple areas) and \textit{III}
(blue) are easily produced. Twin-star solutions of \textit{Category II}
(orange) also appear in both models, although the area filled in the
parameter space is rather small, while twin stars of \textit{Category IV}
(green) are abundant for the \texttt{Model-1} EOSs, but become more
difficult to be found for \texttt{Model-2} having the Gibbs construction.

By analysing \texttt{Model-1} [\texttt{Model-2}] EOSs we observe that, in
order to reach $2\,M_\odot$ in the normal-neutron-star branch for 
\textit{Categories I-II}, we need $p_{\rm tr} > 180~\rm MeV\,fm^{-3}\;
 [160~\rm MeV\,fm^{-3}]$. Twin-star solutions of \textit{Category II} are 
located in the same range of $p_{\rm tr}$ occupied by those of 
\textit{Category I} (see Fig.~\ref{fig:parameters_rhoqcd}), but at slightly
higher values of $\Delta e$. We recall that the two classes of solutions 
differ in whether the twin branch is above (\textit{Category I}) or below
(\textit{Category II}) the $2\,M_{\odot}$ value (see
Fig. \ref{fig:cartoon_MvsR}). In fact, using our two EOS models, these
two categories are difficult to be differentiated since the values of the
maximum masses for the two branches lie within a rather small range, \ie
$1.95\,M_\odot\lesssim M_{_{\rm TOV,\,T}}\lesssim
2.05\,M_\odot$\footnote{When using our hadronic EOS we obtain twin-star
  solutions in the twin branch that in
  \textit{Category~I}/\textit{Category~II} do not have maximum masses
  much larger/smaller than the observational constraint, \ie $M_{_{\rm
      TOV,\,T}} \gtrsim 2~M_\odot$ for \textit{Category~I} and $M_{_{\rm
      TOV,\,T}} \lesssim 2~M_\odot$ for \textit{Category~II}. On the other
  hand, the use of a stiffer EOS can make these differences larger, as
  shown in Ref. \cite{Christian:2017jni}.}.

Twin-star solutions of \textit{Category IV} appear for very low values of
$p_{\rm tr}$ (\ie $p_{\rm tr}\lesssim 25\rm\,MeV\,fm^{-3} \;
[15\rm\,MeV\,fm^{-3}]$), as required in order for the maximum mass of the
normal-neutron-star branch to be below the $1\,M_\odot$ value. Twin stars 
of this category might not exist because the mass in the
normal-neutron-star branch is much lower than the canonical value of
$1.4\,M_\odot$, which should be well described as normal neutron stars, given our
present knowledge of nuclear matter at the expected central densities.

Also clear from Fig. \ref{fig:parameters_rhoqcd} is that the category
that contains the largest number of twin-star solutions is
\textit{Category III}, with ${25\rm\,MeV\,fm^{-3}}\lesssim p_{\rm
  tr}\lesssim 180\rm\,MeV\,fm^{-3}$ \,[${15\rm\,MeV\,fm^{-3}}\lesssim
  p_{\rm tr}\lesssim 160\rm\,MeV\,fm^{-3}$] and the width of the $\Delta
e$ range depending on the model for the EOS. In addition,
\textit{Category III} is certainly the most interesting category from an
astrophysical point of view, as it accommodates twin stars of masses
around the canonical $1.4\,M_\odot$ value.

Finally, we show in Fig.~\ref{fig:parameters_gamma} the $\Delta
e$--$p_{\rm tr}$ parameter space for \texttt{Model-1$^\dagger$} (upper 
panels) and \texttt{Model-2$^\dagger$} (lower panels), with a Maxwell 
and Gibbs phase transition, respectively, that implement for the quark
phase a polytrope $p(\rho)=K_{\rm q}\rho^{\Gamma_{\rm q}}$, combined with
a constant speed of sound parametrization when $c_s^2=1$. In this case,
we find that the parameter space increases with increasing polytropic
index. The higher the polytropic index, the stiffer the EOS and, hence,
the easier it is to find twin-star solutions with $M_{_{\rm
    TOV}}^{^{\uparrow}}\ge 2.0\,M_\odot$. By comparing the $\Delta
e$--$p_{\rm tr}$ parameter space of \texttt{Model-1} and \texttt{Model-2}
EOSs in Fig.~\ref{fig:parameters_rhoqcd} with the corresponding space in
Fig.~\ref{fig:parameters_gamma}, we conclude that EOSs of
\texttt{Model-1} and \texttt{Model-2} are still the most effective in
providing twin-star configurations and masses $\ge 2.0\,M_\odot$.

\subsection{Masses and Radii}
\label{subsec:massradius}

\begin{figure*}[hbtp!]
  \hspace*{3cm}\includegraphics[width=0.85\textwidth]{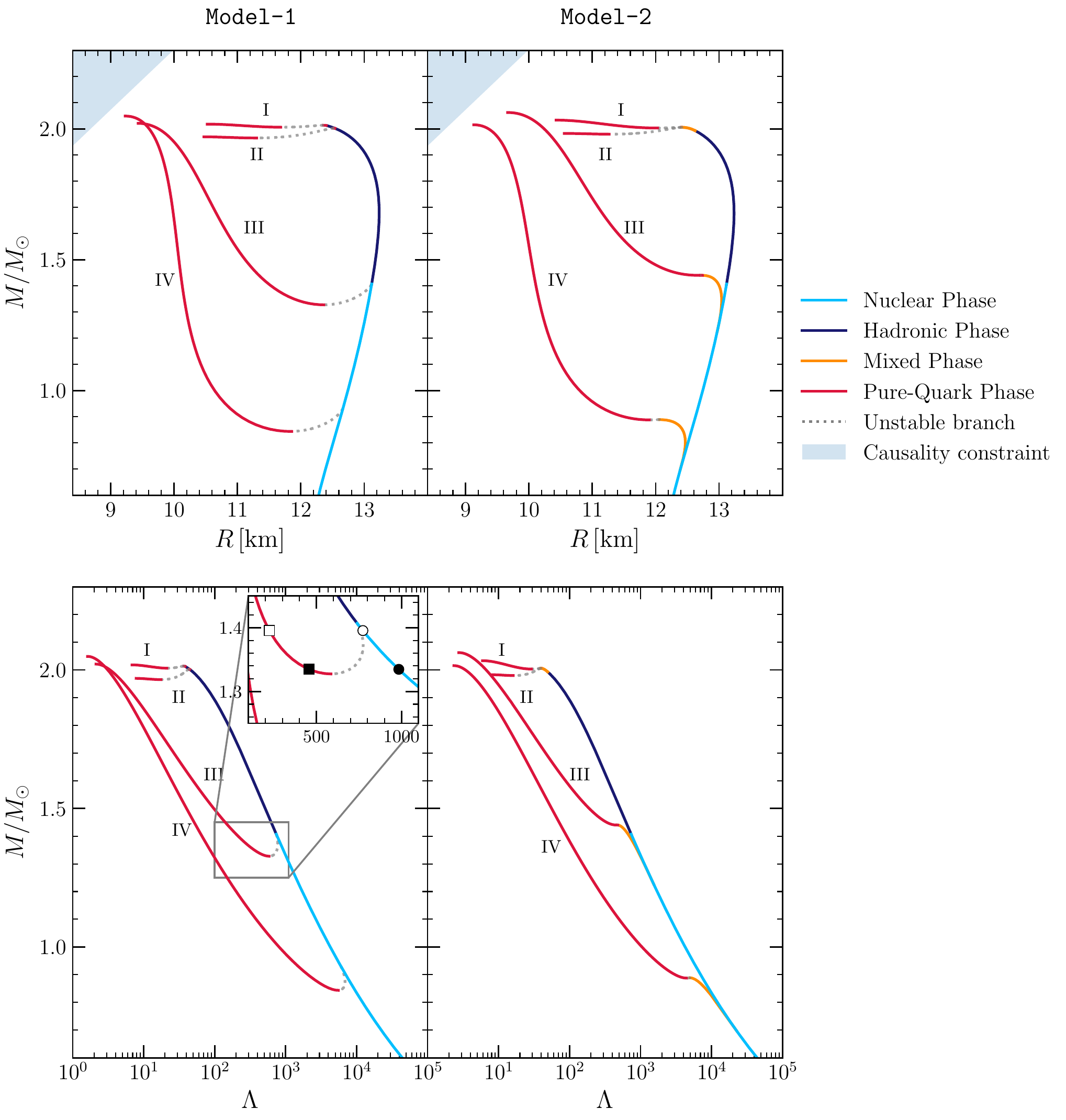}
 \caption{Upper panels: Selected mass--radius relations classified in the
   corresponding twin-star category for \texttt{Model-1} and
   \texttt{Model-2}. The grey dotted lines correspond to the unstable
   regions. The light-blue shaded area marks the causality limit for
   compactness $R\geq 2.94~M$. Lower panels: Dimensionless tidal
   deformability $\Lambda$ of a single neutron star as a function of its
   mass using the same EOSs as in the upper panels. The colouring
   indicates the composition of the innermost region of the neutron star
   (nucleons, nucleons and hyperons, mixed phase and pure-quark phase) at
   a central density $\rho_c$, as seen in Fig.~\ref{fig:eos}. The symbols in the
   inset represent the possible configurations of binaries with masses
   $M_1\gtrsim M_2$ set by the GW170817 chirp mass $\mathcal{M}=1.188
   \,M_\odot$: circles for neutron stars in the hadronic branch and
   squares for hybrid stars in the twin branch; empty symbols for the
   high-mass component of the binary and filled symbols for the low-mass
   one.}
\label{fig:MvsR}
\end{figure*}

A selection of the possible $M$--$R$ relations obtained for
\texttt{Model-1} and \texttt{Model-2} EOSs is displayed in the upper two
panels of Fig.~\ref{fig:MvsR}. Each curve in a given panel shows the
$M$--$R$ relation for a given twin-star category, with the corresponding
values of $\Delta e$ and $p_{\rm tr}$ being provided in the rightmost
upper and lower panels of Fig.~\ref{fig:parameters_rhoqcd}, where the red
circles single out the values plotted in Fig.~\ref{fig:MvsR}. Note that
by using the same colour palette as in Fig.~\ref{fig:eos}, we show with
different colours in the various $M$--$R$ curves the composition of the
innermost region of the star: light blue for neutron stars entirely
composed of nucleonic matter, dark blue if the central pressure is large
enough to allow for the appearance of hyperons, orange if there is an
inner core of mixed matter surrounded by a hadronic- (or nuclear-) matter
mantle, and red for the hybrid star composed of a quark-matter core and a
mantle of hadronic (or nuclear) matter, separated by a mixed-phase region
within \texttt{Model-2}. Dashed lines in grey correspond to unstable
configurations.

\begin{table*}[hbt]
\begin{tabular}{ll|ccc|ccc}
\hline
\multicolumn{2}{c}{\multirow{2}{*}{Model}} & \multicolumn{3}{|c|}{Normal-neutron-star branch} & \multicolumn{3}{c}{Twin branch} \\ \cline{3-8} 
\multicolumn{2}{c|}{} & $M_{_{\rm TOV}}~[M_\odot]$ & $R_{1.4}\rm~[km]$ & $\Lambda_{1.4}$ & $M_{_{\rm TOV,T}}~[M_\odot]$ & $R_{1.4}\rm~[km]$ & $\Lambda_{1.4}$ \\ \hline
\multicolumn{1}{l|}{\multirow{3}{*}{\rotatebox[origin=c]{90}{Maxwell}}} & \texttt{Model-1} & $[1.06,\,2.00)$ & $13.1$ & $760$ & $[2.00,\,2.44]$ & $[10.1,\,12.9]$ & $[69,\,609]$ \\ \cline{2-8} 
\multicolumn{1}{l|}{} & $\rho_{\rm QCD}=5\rho_0$ & $[1.06,\,1.76]$ & $13.1$ & $760$ & $[2.00,\,2.38]$ & $[11.1,\,12.9]$ & $[149,\,609]$ \\
\multicolumn{1}{l|}{} & $\Gamma_{\rm q}=4.0$ & $[1.06,\,2.00)$ & $13.1$ & $760$ & $[2.00,\,2.05]$ & $[11.0,\,12.9]$ & $[145,\,599]$ \\ \hline
\multicolumn{1}{l|}{\multirow{3}{*}{\rotatebox[origin=c]{90}{Gibbs}}} & \texttt{Model-2} & $[1.02,\,2.00)$ & $[12.9,\,13.1]$ & $[679,\,760]$ & $[2.00,\,2.08]$ & $[10.4,\,11.9]$ & $[114,\,295]$ \\ \cline{2-8} 
\multicolumn{1}{l|}{} & $\rho_{\rm QCD}=6\rho_0$ & $[1.72,\,1.99]$ & $13.1$ & $760$ & $[2.00,\,2.02]$ & $-$ & $-$ \\
\multicolumn{1}{l|}{} & $\Gamma_{\rm q}=4.0$ & $[1.84,\,2.00)$ & $13.1$ & $760$ & $[2.00,\,2.03]$ & $-$ & $-$ \\ \hline
\end{tabular}
\caption{Physical properties of the stars of \textit{Category~III}
  obtained within the most representative models described in the text.}
\label{table:properties}
\end{table*}

With these considerations, one can readily appreciate the multiplicity of
possibilities concerning masses, radii and internal structures for each
of the twin stars obtained with our two EOS models. We note that for
\textit{Categories I-II}, the allowed $p_{\rm tr}\gtrsim 160~\rm
MeV\,fm^{-3}$ region seen in Fig.~\ref{fig:parameters_rhoqcd} results in
a nearly flat twin mass--radius branch, as noted in
Ref. \cite{Christian:2017jni}. Also, for values of $p_{\rm tr}$ similar
to those in \textit{Category I}, higher values of $\Delta e$ in
\textit{Category II} (see Fig.~\ref{fig:parameters_rhoqcd}) are
responsible for unstable regions separating the two stable branches that
are larger in \textit{Category II} than in \textit{Category I}. This is
due to the fact that a larger $\Delta e$ produces heavier quark cores,
with the subsequent greater gravitational pull on the nuclear mantle, so
that the twin branch takes longer to stabilize
\cite{Macher:2004vw,Alford:2013aca}.

An interesting quantity to consider across the different twin-star
categories and the EOS models is the radius difference between the two
equal-mass twin stars, $\Delta R$. In Ref.~\cite{Christian:2017jni}
values of $\Delta R$ as large as $4\,{\rm km}$ were claimed to be
possible. This radius difference would allow for the distinction of the
two stars given that a few per cent accuracy might be expected in future
determinations of the radius, either via electromagnetic emissions
\citep{Watts:2016uzu} or via gravitational waves \cite{Bose2017}.
However, we here find that the largest differences are $\Delta R\sim
2.7\,{\rm km}$ and $\Delta R\sim 2.3\,{\rm km}$, that correspond to twin
stars of \textit{Categories IV} and \textit{II}, respectively, in the
\texttt{Model-1} EOSs, and $\Delta R\sim 2.2\,{\rm km}$ for 
\textit{Category II} in \texttt{Model-2}. This is most certainly due to
the different hadronic EOS here, which is softer than that employed in
\cite{Christian:2017jni}. Finally, we note that for \textit{Category
  III}, which is possibly the most interesting case as it can accommodate
twin stars with masses around $1.4\,M_\odot$, the largest difference in
radii is $\Delta R \sim 1.9\,{\rm km}$ and $\Delta R \sim 1.4\,{\rm km}$ in
the case of \texttt{Model-1} and  \texttt{Model-2}, respectively, thus
making it more difficult to distinguish the two types of stars.

\subsection{Tidal deformabilities}
\label{subsec:tidaldef}

The tidal deformability is a property of the EOS that is in principle
measurable via gravitational-wave observations of binary neutron-star
inspirals, as done with the recent GW170817 event 
\cite{TheLIGOScientific:2017qsa}. It is therefore interesting to explore
the behaviour of the tidal deformability for different EOSs that allow
for the appearance of twin stars. This is done in the lower panels of
Fig.~\ref{fig:MvsR}, which report the dimensionless tidal deformability
$\Lambda$ as a function of the mass of the neutron star for the same
selection of EOSs as in the upper panels. From the figure it is clearly
seen that $\Lambda$ spans several orders of magnitude for different EOSs.

With regards to the dimensionless tidal deformability for the reference
star with a mass of $1.4\,M_\odot$, \ie $\Lambda_{1.4}$, we observe a
considerable difference between EOSs that exhibit a phase transition at
low densities (such as \textit{Category IV}) and at high densities
(\textit{Categories I-II}) for the two models considered. More
specifically, a reference $1.4\,M_\odot$ star with a dense core of quark
matter in \textit{Category IV} has $\Lambda_{1.4}$ ranging from a few
tens to a few hundreds, while a $1.4\,M_\odot$ pure hadronic neutron star
in \textit{Categories I-II} has $\Lambda_{1.4}=760$. The differences in
the values of $\Lambda_{1.4}$ can be explained by the different
compactnesses of the stars. We recall, in fact, that $\Lambda\propto
k_2\mathcal{C}^{-5}$, where $\mathcal{C}:=M/R$ is the stellar
compactness and $k_2$  the second tidal Love number.
On the other hand, $k_2\propto\mathcal{C}^{-1}$ in the mass
range of typical neutron stars \cite{Hinderer:2009ca,Zhao:2018nyf}, so
that $\Lambda\propto \mathcal{C}^{-6}$. In the presence of a HQPT,
however, this correlation is expected to be weakened
\cite{Postnikov:2010yn}. At any rate, for the same total mass of
$1.4\,M_\odot$, stars with a quark-matter core have smaller radii and,
hence, larger compactness or, equivalently, smaller values of
$\Lambda_{1.4}$.

When considering the case of \textit{Category III}, we obtain twin stars
with masses around $1.4\,M_\odot$. These configurations have a core of
mixed or pure quark matter with a radius for the star between the radius
for \textit{Categories I-II} and \textit{Category IV}. Thus, the value of
$\Lambda_{1.4}$ lies between the values for $\Lambda_{1.4}$ in
\textit{Categories I-II} and \textit{Category IV}. The inset in the left
panel shows in greater detail the behaviour of $\Lambda(M)$ around the
phase transition for an EOS of \textit{Category III} that is of
particular interest because it holds stars with masses of
$1.365\,M_\odot$ in both branches and will be further discussed in
Section~\ref{subsec:tidaldefbinary}.

We note that using the detection of GW170817,
Ref.~\cite{TheLIGOScientific:2017qsa} derived an upper bound
$\Lambda_{1.4}\leq 800$ (corrected later to $\leq 900$) upon the GW170817
event, that was later on reanalysed to be $300 ^{+420}_{-230}$
\cite{Abbott:2018wiz}.  However we note that this constraint is obtained
by expanding $\Lambda(M)$ linearly about $M=1.4\,M_\odot$, and from
Fig.~\ref{fig:MvsR} we can see that if the twin branch appears at $M\sim
1.4\,M_\odot$ this approach is no longer valid and the upper bound on
$\Lambda_{1.4}$ could be further decreased, as shown in
Refs.~\cite{Most:2018hfd,Paschalidis:2017qmb,Burgio:2018yix}.  In fact,
Ref.~\cite{Most:2018hfd} has shown that the lower limit on
$\Lambda_{1.4}$ is decreased from $\Lambda_{1.4}\ge 375$ to
$\Lambda_{1.4}\ge 265$ at $2$-$\sigma$ level (see the supplemental
material of Ref.~\cite{Most:2018hfd}) when allowing for a phase
transition.

\begin{figure*}[htbp!]
 \centering
 \hspace*{1cm}\includegraphics[width=0.9\textwidth]{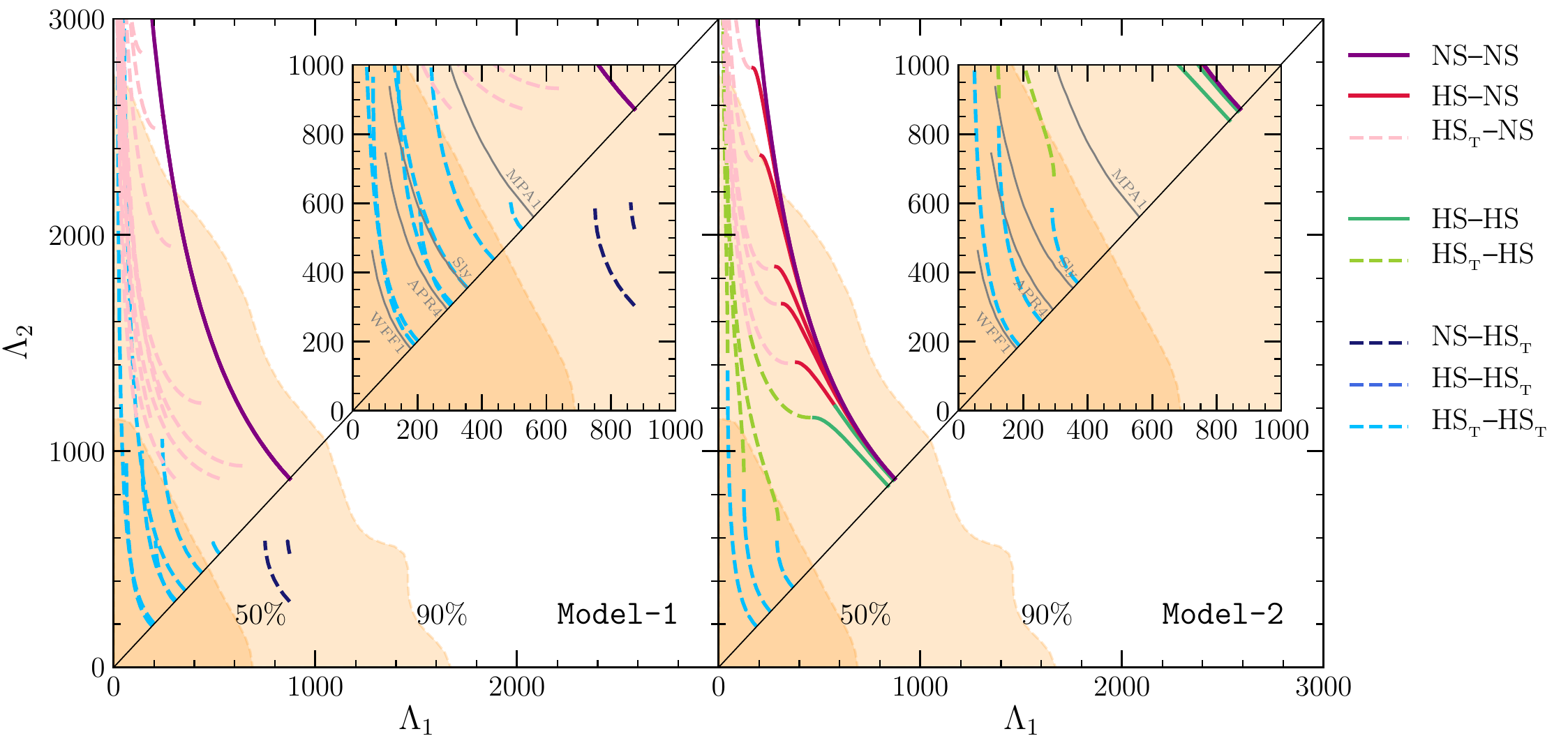}
 \caption{Relation between the tidal deformabilities of the high-mass and
   the low-mass components, $\Lambda_1$ and $\Lambda_2$, of a binary
   neutron star with a chirp mass $\mathcal{M}=1.188\,M_\odot$ for
   \texttt{Model-1} and \texttt{Model-2} EOSs of \textit{Category III}
   only.  The colours are related to the nature of each of the components
   of the $M_1$--$M_2$ binary system: NS for a hadronic or pure neutron
   star, HS for a hybrid star in the normal-neutron-star branch, and
   HS$_{_{\rm T}}$ for a hybrid star in the twin branch, with the first
   label referring to the massive component of the binary ($M_1$) and the
   second to the less massive ($M_2$) separated by a long dash. The
   lines displayed correspond to the EOSs indicated with empty squares in
   Fig.~\ref{fig:parameters_rhoqcd}, each EOS giving a unique 
   ``connection'' among NS, HS and HS$_{_{\rm T}}$ in the 
   $\Lambda_1$--$\Lambda_2$ plane: 0. (NS--NS), 1. (NS--NS, HS--NS,
   HS$_{_{\rm T}}$--NS), 2. (NS--NS, HS$_{_{\rm T}}$--NS), 3. (NS--NS,
   HS$_{_{\rm T}}$--NS, HS$_{_{\rm T}}$--HS$_{_{\rm T}}$, 
   NS--HS$_{_{\rm T}}$), 4. (HS--HS, HS--NS, HS$_{_{\rm T}}$--NS), 5. 
   (HS--HS, HS$_{_{\rm T}}$--HS, HS$_{_{\rm T}}$--NS), 6. 
   (HS$_{_{\rm T}}$--HS$_{_{\rm T}}$, HS$_{_{\rm T}}$--NS), 7. 
   (HS$_{_{\rm T}}$--HS$_{_{\rm T}}$, HS$_{_{\rm T}}$--HS,
   HS$_{_{\rm T}}$--NS), 8. (HS$_{_{\rm T}}$--HS$_{_{\rm T}}$,
   HS$_{_{\rm T}}$--HS), 9. (HS$_{_{\rm T}}$--HS$_{_{\rm T}}$). The
   shaded areas correspond to the 50\% and 90\% credibility regions set
   by GW170817 for a low-spin scenario $|\chi|\leq 0.05$
   \cite{TheLIGOScientific:2017qsa}. The inset also reports for
   comparison the tidal deformabilities of representative nucleonic EOSs
   (grey lines). }
\label{fig:lam1lam2_1.188}
\end{figure*}

In Table \ref{table:properties} we report the range of values for the
maximum mass, radius ($R_{1.4}$) and tidal deformability for a
$1.4\,M_{\odot}$ star for \texttt{Model-1} and \texttt{Model-2} EOSs in
\textit{Category~III}, both in the normal and in the twin
branch. Moreover, for completeness, we also show the ranges for these
quantities coming from four representative EOS models, that have been
computed using the Maxwell or Gibbs construction for the phase transition
and taking into account the two different descriptions of the
quark-matter phase discussed in Section \ref{subsec:parameterspace}. We
observe a larger variance for $R_{1.4}$ and $\Lambda_{1.4},$ for both
\texttt{Model-1} and \texttt{Model-2} EOSs in both branches, which
can be understood in terms of the larger $\Delta e$--$p_{\rm tr}$ space
of parameters (\cf Sec. \ref{subsec:parameterspace}). Thus, as mentioned
before, we will restrict our attention to \texttt{Model-1} and
\texttt{Model-2} EOSs when performing the analysis of the tidal
deformabilities of neutron-star binaries.

\subsection{Tidal deformabilities and GW170817}
\label{subsec:tidaldefbinary}

In order to compare directly with the observational analysis from the
GW170817 event \cite{TheLIGOScientific:2017qsa}, we considered a binary
system with a chirp mass $\mathcal{M} := (M_1M_2)^{3/5}/(M_1+M_2)^{1/5} =
1.188\,M_\odot$ and calculated the tidal deformabilities $\Lambda_1$ and
$\Lambda_2$ of the high-mass $M_1$ and low-mass $M_2$ components,
respectively, plotting the 50\% and 90\% credibility regions\footnote{
  The confidence levels were obtained from the LIGO data analysis with
  EOSs that do not account for twin stars; hence, they serve as a
  reference with this caveat in mind.} for the low-spin scenario
$|\chi|\leq 0.05$ given in
Refs. \cite{TheLIGOScientific:2017qsa,Abbott:2018exr}. This is shown in
Fig.~\ref{fig:lam1lam2_1.188} for selected EOSs with twin stars of
\textit{Category III} only within \texttt{Model-1} and \texttt{Model-2}
EOSs. Lines of different colour and type show different number of branches
and shapes in the $\Lambda_1$--$\Lambda_2$ plane obtained by varying
$M_1\in[1.365,1.8] \,M_\odot$ and $M_2\in[1.0,1.365]\,M_\odot$ (with
fixed $\mathcal{M} =1.188\,M_\odot$). The corresponding values of $\Delta
e$ and $p_{\rm tr}$ of the selected EOSs are indicated with black empty
squares in the rightmost upper and lower panels of
Fig.~\ref{fig:parameters_rhoqcd}.

The different colour lines in Fig.~\ref{fig:lam1lam2_1.188} are related to
the nature of each of the components, with the labels having the
following meaning: NS for a purely hadronic \textit{neutron star}, HS for
a \textit{hybrid star} with a core of mixed and/or quark matter in the
``normal-neutron-star'' branch, and HS$_{_{\rm T}}$ for a \textit{hybrid
  star} with a quark core in the ``twin'' branch, with the first label
referring to the massive component of the binary ($M_1$) and the second
to the less massive ($M_2$) separated by a long dash. Allowing for all
these possibilities, there can be up to eight lines in the
$\Lambda_1$--$\Lambda_2$ plane (see legend of
Fig.~\ref{fig:lam1lam2_1.188}). However, not all of them are produced by
each of the EOS models at a given $\mathcal{M}$.

\begin{figure}[H]
 \centering
     \includegraphics[width=0.47\textwidth]{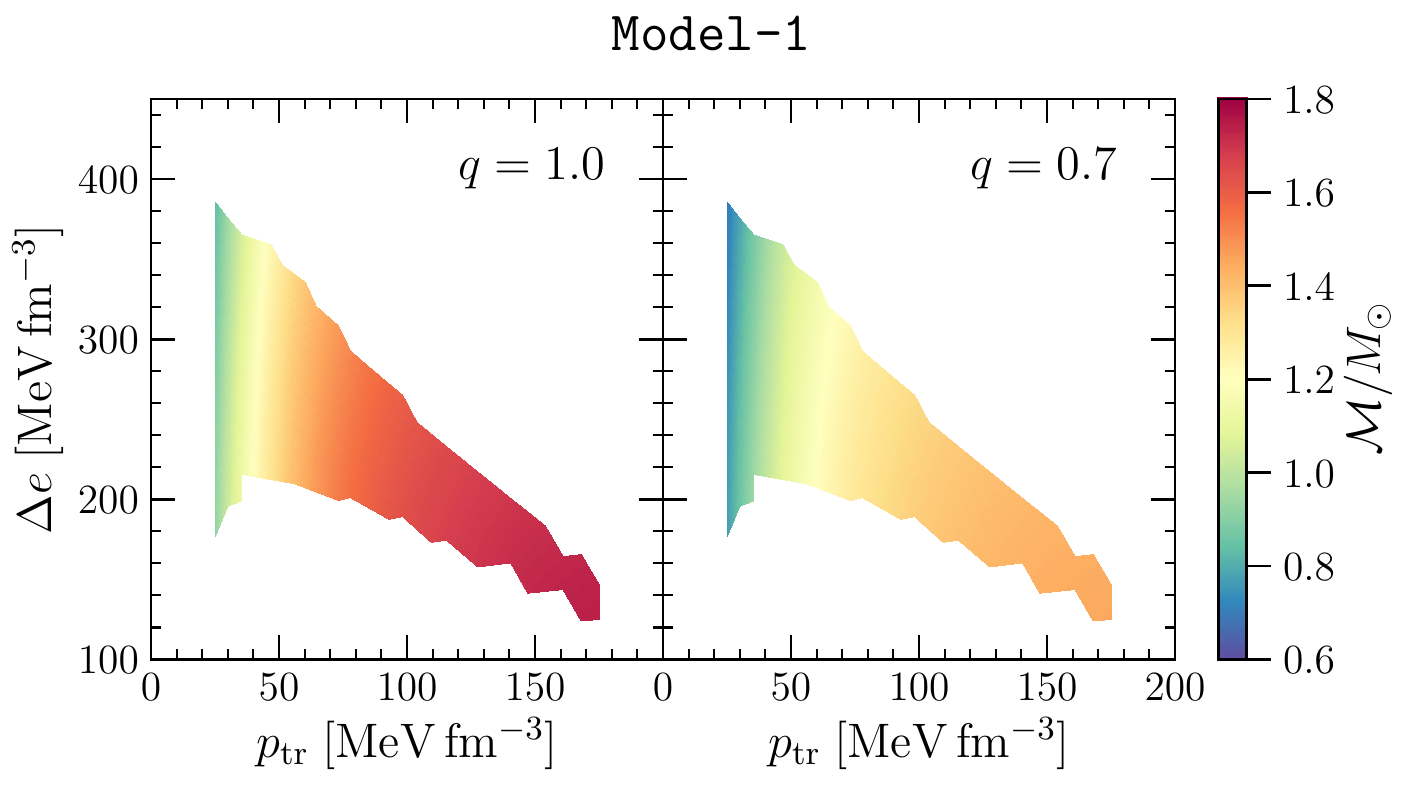}
     \includegraphics[width=0.47\textwidth]{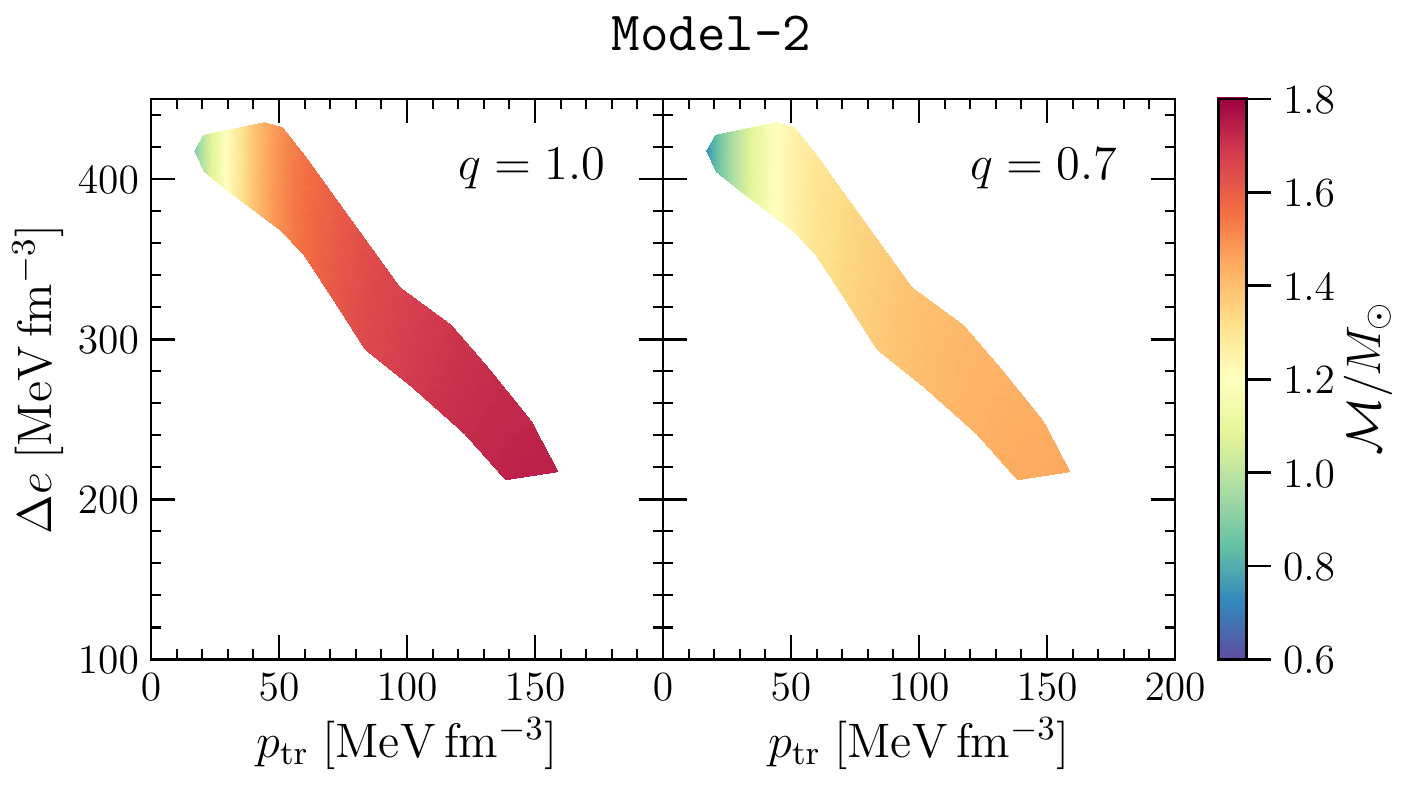}     
     \caption{Upper panels: Minimum value of the chirp mass for which at
   least one of the components of the binary is in the twin branch, shown
   as a function of the values of the transition pressure and energy-density
   jump of the EOSs of \textit{Category III} for the \texttt{Model-1}
   EOSs for $q=1$ (left panel) and $q=0.7$ (right panel). Lower panels:
   Same as upper panels but for the \texttt{Model-2} EOSs. }
\label{fig:properties_chirpm}
\end{figure}

This is reported in Fig. \ref{fig:properties_chirpm}, which shows the
minimum value of the chirp mass for which at least one of the components
of the binary before the merger is in the twin branch as a function of
the values of the transition pressure and energy-density jump of the EOSs of
\textit{Category III} for the \texttt{Model-1} EOSs (upper panels); left 
and right panels refer to mass ratios of $q=1.0$ and $q=0.7$, respectively.
In full similarity, same quantities are shown in the lower panels for the
\texttt{Model-2} EOSs.

As expected, the lower the transition pressure, the lower the minimum
chirp mass required for the HQPT to occur in at least one of the stars in
the binary given that it is easier to populate the twin branch for low
transition pressures. Note that the lowest value for the chirp mass is
$0.6\,M_{\odot}$ and that for $q=1$ higher chirp masses are needed to
have one of the components in the twin branch when compared with the
$q=0.7$ case. This is because for unequal-mass binaries, it is easier for
the high-mass component to be on the twin branch.

For the EOSs that allow for twin stars of \textit{Categories I-II} (not
shown in Fig.~\ref{fig:lam1lam2_1.188}), for which the mass of the twins
is significantly larger than that of the components in GW170817, only the
hadronic part of the EOS is reported for $\Lambda_i\in[0,3000]$ and only
the NS--NS (purple) line is shown.  The opposite limiting case
corresponds to EOSs that allow for twin stars of \textit{Category IV}
(not shown in Fig.~\ref{fig:lam1lam2_1.188}) and hence with very low
masses. For this case, both components of the binary system are located
in the twin branch, producing only lines of the type HS$_{_{\rm
    T}}$--HS$_{_{\rm T}}$ (light blue) in the $\Lambda_1$--$\Lambda_2$
plot, with $\Lambda_1<200$ for both models. On the other hand, in the
case of EOSs with twins stars of \textit{Category III}, the range of
possibilities is larger, mostly due to the existence of twin stars with
masses similar to those of the GW170817 binary ($M_1=M_2=1.365\,M_\odot$
in the equal-mass limit). Indeed, as shown in
Fig.~\ref{fig:lam1lam2_1.188}, \texttt{Model-1} and \texttt{Model-2} EOSs
show clear differences with regards to the number of possible
scenarios. 

The general considerations made above can be made more specific starting,
in particular, from EOSs of \texttt{Model-1} (left panel of
Fig.~\ref{fig:lam1lam2_1.188}). 

In this case, for an EOS with a HQPT at
high transition pressure (\ie with $p_{\rm tr}\gtrsim
80\rm\,MeV\,fm^{-3}$ in Fig.~\ref{fig:parameters_rhoqcd}, which corresponds
to $\rho_{\rm tr}\gtrsim3\,\rho_0$), only the
NS--NS sequence (purple line) is found. If the phase transition takes
place at lower densities ($2.2\,\rho_0\lesssim \rho_{\rm tr}\lesssim 3.0\,\rho_0$,
\ie $40{\rm\,MeV\,fm^{-3}}\lesssim p_{\rm tr}\lesssim 80\rm\,MeV\,fm^{-3}$), 
the twin branch contains hybrid
stars with a mass that is low enough to hold the high-mass component of
the binary. In this case, when $M_1\approx M_2$ both stars are in the
normal-neutron-star branch, but as $M_1$ is increased (and $M_2$
decreased to keep $\mathcal{M}$ constant) it jumps to the twin branch and
the NS--NS sequence (purple line) connects with a HS$_{_{\rm T}}$--NS
(pink lines) line. 
On the other hand, the normal-neutron-star branch of EOSs with low 
transition pressure ($p_{\rm tr}\lesssim 25\rm\,MeV\,fm^{-3}$ in 
Fig.~\ref{fig:parameters_rhoqcd}, \ie $\rho_{\rm tr}\lesssim 1.9\,\rho_0$) 
cannot support any of the components of the binary and the only allowed
configuration is HS$_{_{\rm T}}$--HS$_{_{\rm T}}$ (light blue lines). 
Larger values of $p_{\rm tr}$ ($30{\rm\,MeV\,fm^{-3}}\lesssim p_{\rm tr}
\lesssim 40\rm\,MeV\,fm^{-3}$, \ie $2.0\,\rho_0\lesssim \rho_{\rm tr}\lesssim
2.2\,\rho_0$) allow the low-mass component to be a neutron star. This situation
corresponds to having the two components in the twin branch when their masses
are equal and the low-mass star jumps to the normal-neutron-star branch
as $M_2$ is decreased, giving a HS$_{_{\rm T}}$--NS (pink lines) line connecting with 
the HS$_{_{\rm T}}$--HS$_{_{\rm T}}$ sequence (light blue lines).

There is in addition a particular case in which the value
$M_1=M_2=1.365\,M_\odot$ is contained within the range of masses of the
twin stars produced. This is indeed what happens for some EOSs in
\texttt{Model-1}, as it can be appreciated in the inset of the third
panel in Fig.~\ref{fig:MvsR}. In this case, we can have both components
of the binary in the normal-neutron-star branch (marked as {\large
  $\circ$} and {\large $\bullet$} in Fig.~\ref{fig:MvsR}), both in the
twin branch ({\footnotesize $\square$} and {\footnotesize $\blacksquare$}
in Fig.~\ref{fig:MvsR}); alternatively, we can have the high-mass star in
the twin branch and the low-mass star in the normal-neutron-star branch
({\footnotesize $\square$} and {\large $\bullet$} in Fig.~\ref{fig:MvsR})
and also the high-mass star in the normal-neutron-star branch and the
low-mass star in the twin branch ({\large $\circ$} and {\footnotesize
  $\blacksquare$} in Fig.~\ref{fig:MvsR}). These configurations are
marked in the left panel of Fig.~\ref{fig:lam1lam2_1.188} respectively
as: NS--NS (purple lines), HS$_{_{\rm T}}$--HS$_{_{\rm T}}$ (dashed
light-blue lines), HS$_{_{\rm T}}$--NS (dashed pink lines) and
NS--HS$_{_{\rm T}}$ (dashed dark-blue lines).

Also noticeable is that this last extra NS--HS$_{_{\rm T}}$ sequences
(dashed dark-blue lines) appear in the otherwise empty
$\Lambda_1>\Lambda_2$ region. For EOSs not producing twin stars and given
that $M_1>M_2$, one has $\mathcal{C}_1 > \mathcal{C}_2$ and, hence,
$\Lambda_1<\Lambda_2$, so it is usually not possible to have solutions in
the $\Lambda_1>\Lambda_2$ area. However, this does not hold for EOSs
giving rise to twin stars, since in this case the high-mass star can be
less compact than the low-mass one, as seen in the inset of
Fig.~\ref{fig:MvsR} for the {\large $\circ$} and {\footnotesize
  $\blacksquare$} cases. This type of pairs, where the heavier star is
also less compact, has been named \textit{``rising-twins''} pair in
Ref. \cite{Schertler:2000xq}; by definition, therefore, rising twins can
only appear with EOSs that allow for twin stars and their existence is not
allowed by any other kind of EOS of compact stars. In summary, if the
EOS allows for rising twins of masses $M_1$ and $M_2<M_1$, tied together
by a given value of the chirp mass $\mathcal{M}$, there must be a line in
the $\Lambda_1>\Lambda_2$ side of the plot. Indeed, in
Ref.~\cite{Christian:2018jyd} it was suggested that this is the case so
long as $0 < (M_1-M_2)/(R_1-R_2)< M_1/R_1$.

As a consequence, since only EOSs producing twin stars can access the
region with $\Lambda_1>\Lambda_2$ for $M_1>M_2$, any experimental
indication that the binary occupies this region of the
$\Lambda_1$--$\Lambda_2$ space would be a strong evidence for the
existence of twin stars. Other EOSs with a HQPT but not generating twin
stars would show similar lines in the $\Lambda_1< \Lambda_2$ region as
those displayed in the left panel of Fig.~\ref{fig:lam1lam2_1.188}. In
this case, one might expect all the lines of stable configurations
connected to one another, but the $\Lambda_1>\Lambda_2$ would be
unattainable. This situation was analysed in
Ref. \cite{Sieniawska:2018zzj} for polytropic EOSs with a CSS
parametrization of the quark phase and different values of the
energy-density jump. However, to find such a signature in the
$\Lambda_1$--$\Lambda_2$ plot is very challenging as it requires a
modeling of the LIGO/Virgo data that includes a phase transition and 
more accurate measurements of the component masses, since in our models
the twin stars have similar masses in this region.

A similar discussion can be made for the \texttt{Model-2} EOSs and is
shown in the right panel of Fig.~\ref{fig:lam1lam2_1.188}. The variety of
lines (colours) increases because the probability of having a HS in the
normal-neutron-star branch is higher than for \texttt{Model-1}. In the
Maxwell construction, for combinations of phase transition parameters
below the Seidov line [see Eq.~\eqref{eq:Seidov}], the
normal-neutron-star branch is composed of a hybrid segment connected to
the purely-hadronic segment whose length is typically quite small and
hence hard to capture in the right panel of
Fig.~\ref{fig:lam1lam2_1.188}. On the other hand, in the Gibbs
construction, hybrid stars with a core of hadron-quark mixed phase can be
found in a relevant portion of the normal-neutron-branch, as shown in
Fig.~\ref{fig:MvsR}.  Several previous works have also studied the
possibility of interpreting the GW170817 event as the coalescence of pure
neutron stars and hybrid stars \cite{Paschalidis:2017qmb, Nandi:2017rhy,
  Alvarez-Castillo:2018pve, Burgio:2018yix, Sieniawska:2018zzj,
  Li:2018ayl, Gomes:2018eiv, Christian:2018jyd,Han:2018mtj}, although
only some of them considered the possibility of twin stars
\cite{Paschalidis:2017qmb, Alvarez-Castillo:2018pve, Burgio:2018yix,
  Sieniawska:2018zzj, Christian:2018jyd}. In
Ref. \cite{Paschalidis:2017qmb}, in particular, NS--NS and HS$_{_{\rm
    T}}$--NS merger combinations were considered. It was then shown that
a HQPT can soften the EOS making it compatible, even for a stiff hadronic
EOS, with the GW170817 observations. In particular, the authors found
that GW170817 is consistent with the coalescence of a HS$_{_{\rm T}}$--NS
binary.

Similarly, in Ref. \cite{Alvarez-Castillo:2018pve}, the GW170817 event
was interpreted as the merger of either a HS$_{_{\rm T}}$--NS or a 
HS$_{_{\rm T}}$--HS$_{_{\rm T}}$ binary and actually disfavoured a 
NS--NS scenario. This was mostly due to the stiffness of the hadronic EOS
employed, that made a neutron-star merger incompatible with the 
compactness expected from GW170817. More recently, 
Ref. \cite{Christian:2018jyd} has interpreted the GW170817 event as the
merger scenario of either a NS--NS, a NS--HS$_{_{\rm T}}$, or 
HS$_{_{\rm T}}$--HS$_{_{\rm T}}$ binary, where all three merger scenarios
can be potentially plausible within a single EOS. This finding is in
agreement with the conclusions presented here, although in our study we 
have performed a more detailed analysis of the merger scenarios in which 
we have also varied the parameters characterizing the phase transition 
(\ie $\Delta e$ and $p_{\rm tr}$), while considering both Maxwell and 
Gibbs constructions for the phase transition, \ie having a significant 
number of HS configurations in addition to NS and HS$_{_{\rm T}}$ 
configurations.

Finally, Ref. \cite{Han:2018mtj} has very recently explored the
sensitivity of the tidal deformability to the properties of a sharp HQPT,
not necessarily producing twin stars, finding that a smoothing of the
transition will not have distinguishable effects. In our case, when
twin-star solutions and masses above $2\,M_\odot$ are produced, we find
however clear differences in masses, radii and tidal deformabilities when
comparing our \texttt{Model-1} and \texttt{Model-2} EOSs in
Figs.~\ref{fig:parameters_rhoqcd}, \ref{fig:MvsR} and
\ref{fig:lam1lam2_1.188}. This is due to the ``smoothing'' of the mixed
phase between the Maxwell and Gibbs constructions, which is different
from that of the rapid crossover transition in Ref.~\cite{Han:2018mtj},
and that leads to the rather different behaviour of the speed of sound in
the two cases.

\begin{figure*}[htb]
 \centering
 \includegraphics[width=0.95\textwidth]{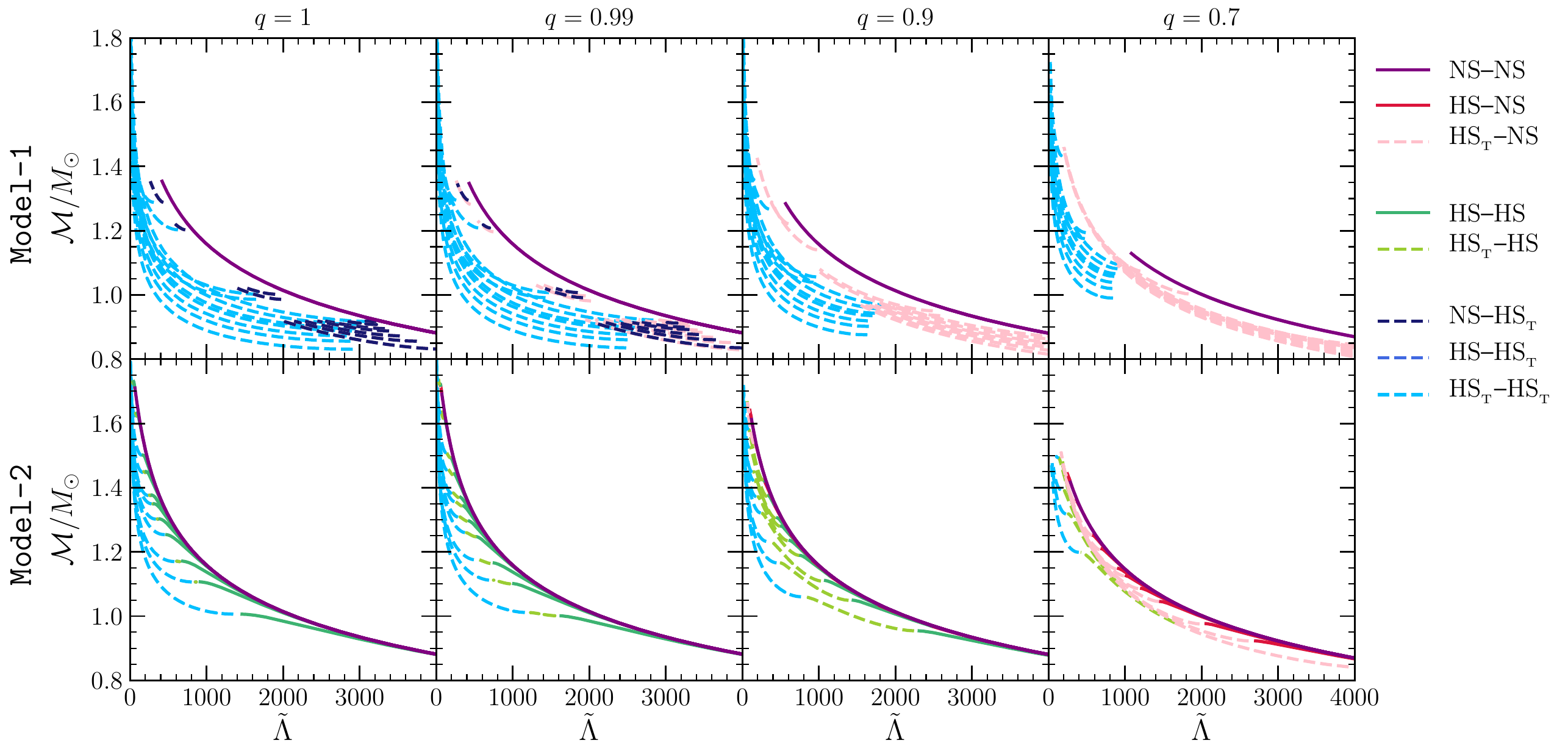} 
 \caption{Relation between the chirp mass, $\mathcal{M}$, and the 
  dimensionless tidal deformability, $\tilde\Lambda$, of binary systems 
  with mass ratios $q=1,0.99,0.9,0.7$ for the same \texttt{Model-1} 
  (upper panels) and \texttt{Model-2} (lower panels) EOSs of 
  \textit{Category III} shown in \ref{fig:lam1lam2_1.188}}.
\label{fig:chirpM_Lambdatilde}
\end{figure*}

In order to distinguish the different kinds of compact star merger
scenarios that are compatible with future gravitational wave events, a 
more promising tool would be to analyse the chirp mass, $\mathcal{M}$,
as a function of the weighted dimensionless tidal deformability
$\tilde \Lambda$ of a neutron-star binary \cite{Flanagan:2007ix,
  Favata:2013rwa, TheLIGOScientific:2017qsa}
\begin{eqnarray}\label{eq:lambdatilde}
\tilde \Lambda& := &\frac{8}{13} [(1+7 \eta -31 \eta^2) (\Lambda_1 + \Lambda_2) \nonumber  \\
 &&+ \sqrt{1-4 \eta}(1+ 9 \eta -11 \eta^2)(\Lambda_1 -\Lambda_2)\nonumber \\
 &=& \frac{16}{13M^5}[(M_1+12M_2)M_1^4\Lambda_1+(M_2+12M_1)M_2^4\Lambda_2]\,,\nonumber \\
\end{eqnarray}
where $\eta:=M_1 M_2/M^2$ is the symmetric mass ratio, $M:=M_1+M_2$ is
the total mass of the binary, and where, in the equal-mass case,
$\tilde{\Lambda}=\Lambda$.

Figure \ref{fig:chirpM_Lambdatilde} displays the relation
$\mathcal{M}$--$\tilde\Lambda$ for different values of the mass ratio,
$q$, for the selected \texttt{Model-1} (upper panels) and
\texttt{Model-2} (lower panels) EOSs with twin stars of \textit{Category
  III} shown in the $\Lambda_1$--$\Lambda_2$ plot of
Fig.~\ref{fig:lam1lam2_1.188} and using the same colour palette to refer
to the various merger scenarios. We can see that the merger of rising
twins happens for mass ratios $q\lesssim 1$. However, rising twins cannot
be identified in Fig.~\ref{fig:chirpM_Lambdatilde} because, due to the
symmetry of $\tilde\Lambda$ with respect to the two components of the
binary [see Eq.~\eqref{eq:lambdatilde}], for $q=1$ the HS$_{_{\rm
    T}}$--NS sequence (dashed pink line) overlaps exactly with the
NS--HS$_{_{\rm T}}$ rising twins (dashed dark-blue line) in the leftmost
upper panel, and the HS$_{_{\rm T}}$--HS sequence (dashed light-green
line) lies on top of the HS--HS$_{_{\rm T}}$ rising twins (dashed
medium-blue line) in the leftmost lower panel. With a small asymmetry
($q=0.99$) these lines do not exactly overlap each other but still lie in
the same region of the $\mathcal{M}$--$\tilde\Lambda$ plot. We also note
that the rising twins extend over a larger range of $\mathcal{M}$ and $q$
in \texttt{Model-1} than in \texttt{Model-2} because the larger unstable
branches in the $M$--$R$ relation obtained with the Maxwell construction
of the HQPT allow for broader ranges of twin stars (see
Fig.~\ref{fig:MvsR}) than the Gibbs construction. Also from
Fig.~\ref{fig:chirpM_Lambdatilde} it is clear that within our
description, the possibility of having a merger of two hybrid stars in
the twin branch (\ie HS$_{_{\rm T}}$--HS$_{_{\rm T}}$) diminishes with
decreasing values of $q$ in favor of either the HS$_{_{\rm T}}$--NS
scenario in the case of \texttt{Model-1} or the HS$_{_{\rm T}}$--NS and
HS$_{_{\rm T}}$--HS scenarios in the case of \texttt{Model-2}. Indeed at
large asymmetries (\eg $q=0.7$) the merger of two hybrid stars with a
chirp mass $\mathcal{M}\lesssim 1\,M_\odot$ is ruled out in our models
because the twin branch cannot hold the low-mass star. Therefore, the
analysis above reveals that within our models the merger scenario can
be readily determined given a measure of the chirp mass and the weighted
dimensionless tidal deformability.

\subsection{Constraining twin stars with GW170817}
\label{subsec:TS_GW170817}
\begin{figure*}[htb]
 \centering
 \includegraphics[width=0.495\textwidth]{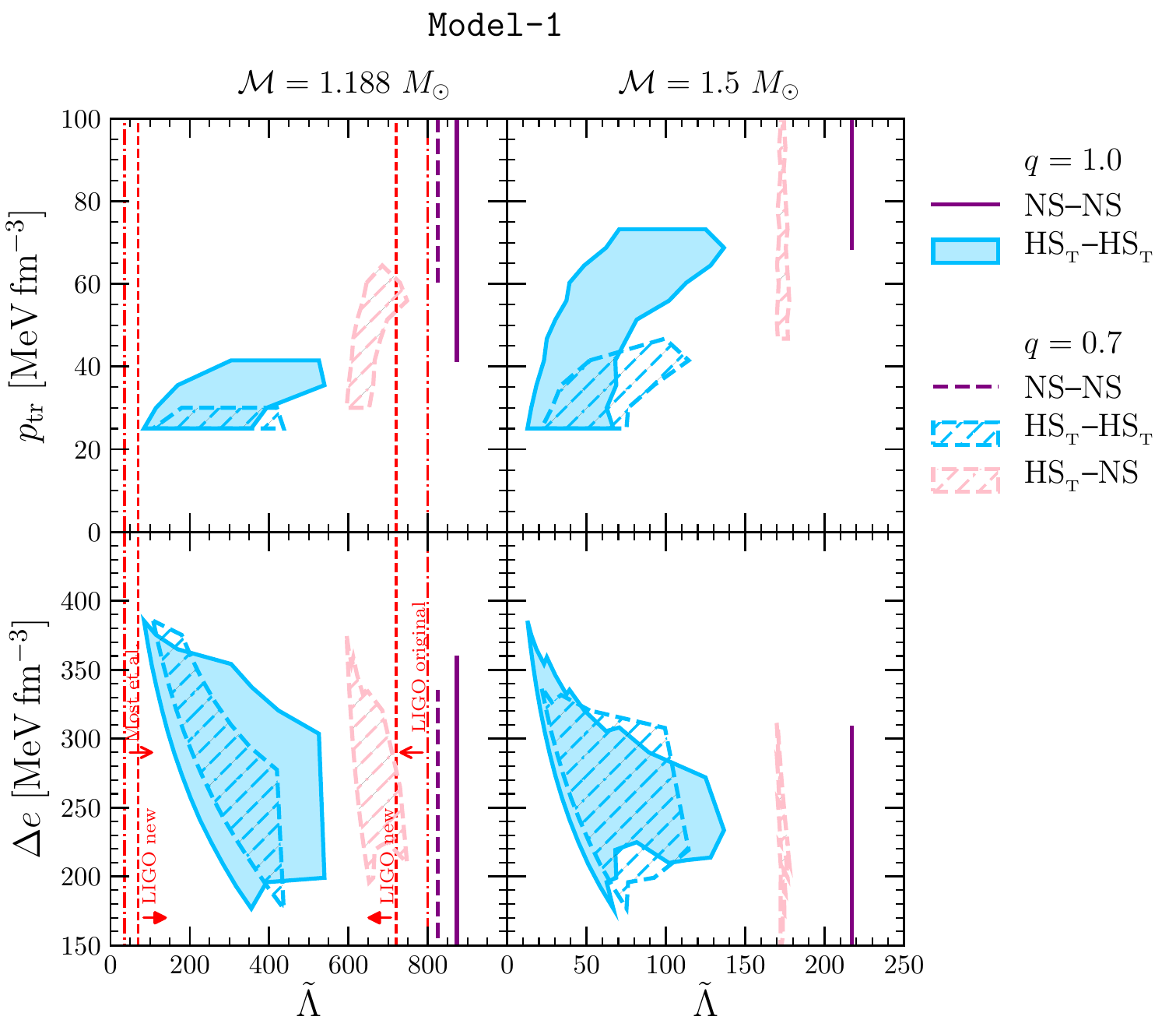} 
 \includegraphics[width=0.495\textwidth]{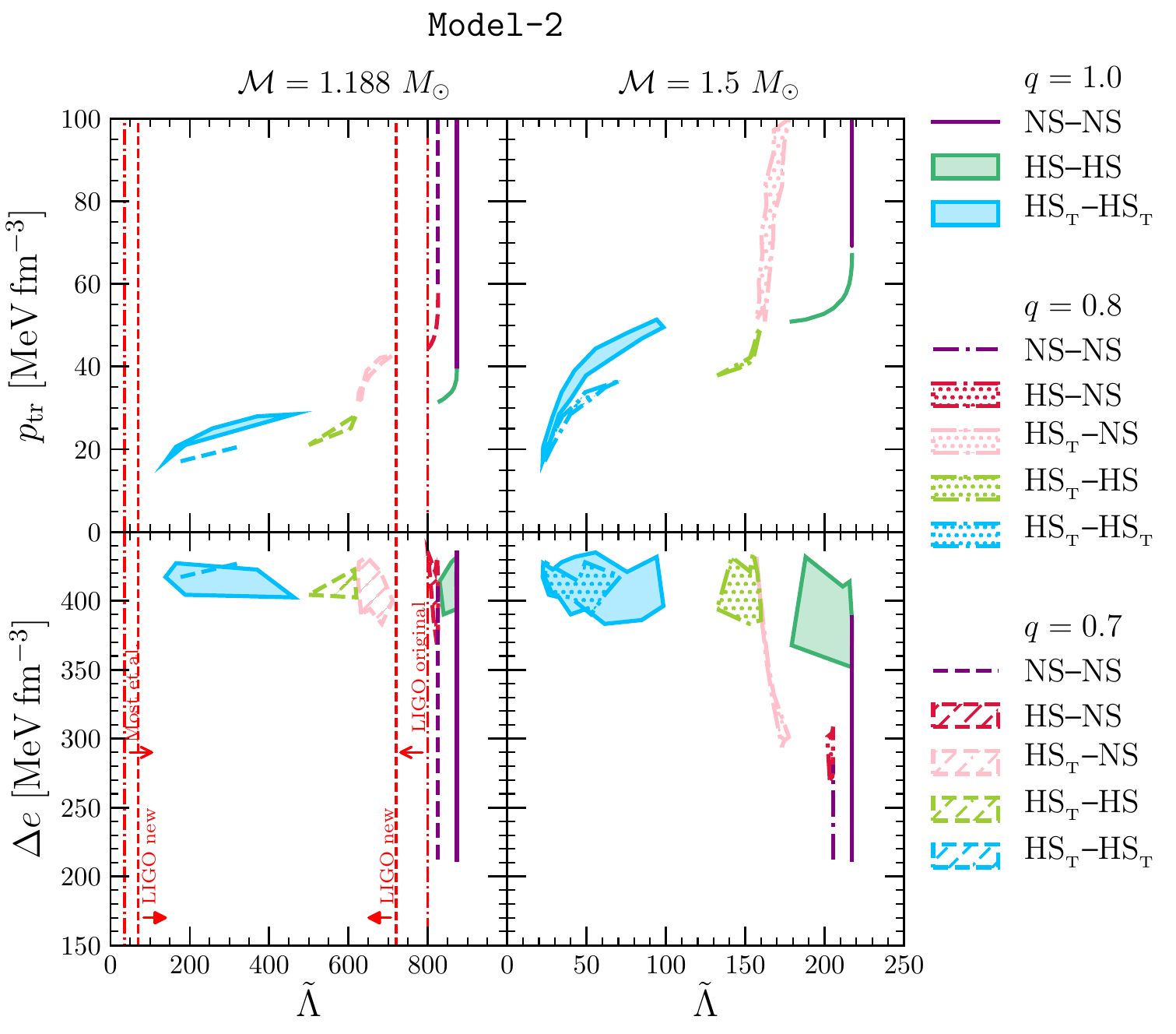} 
 \caption{Left plot: Transition pressure $p_{\rm tr}$ (upper
   panels) and energy-density jump $\Delta e$ (lower panels) as a
   function of the $\tilde{\Lambda}$ for \texttt{Model-1} EOSs and twins
   of $\textit{Category~III}$. The left (right) panels correspond to a
   chirp mass $\mathcal{M}=1.188\,M_\odot$
   ($\mathcal{M}=1.5\,M_\odot$). Shaded (striped) regions with solid
   (dashed) contours show the case for $q=M_2/M_1=1$ ($q=0.7$). The
   vertical lines stand for LIGO-Advanced Virgo upper limit of
   $\tilde\Lambda =800$ \cite{TheLIGOScientific:2017qsa}, the improved
   LIGO-Advanced Virgo analysis of $70 < \tilde\Lambda < 720$
   \cite{Abbott:2018wiz} and Most et al. \cite{Most:2018hfd} lower
   estimate for EOSs with phase transition $\Lambda_{1.4} >
   35.5$. Right plot: Same as the left plot but for
   \texttt{Model-2} EOSs. Note that for $\mathcal{M}=1.5\,M_\odot$ we
   consider shaded (dotted) regions with solid (dashed-dotted) contours
   for $q=M_2/M_1=1$ ($q=0.8$).  
 }
\label{fig:parameters_vs_lambda_G_1}
\end{figure*}

\begin{figure*}[htb]
 \centering
 \includegraphics[width=0.495\textwidth]{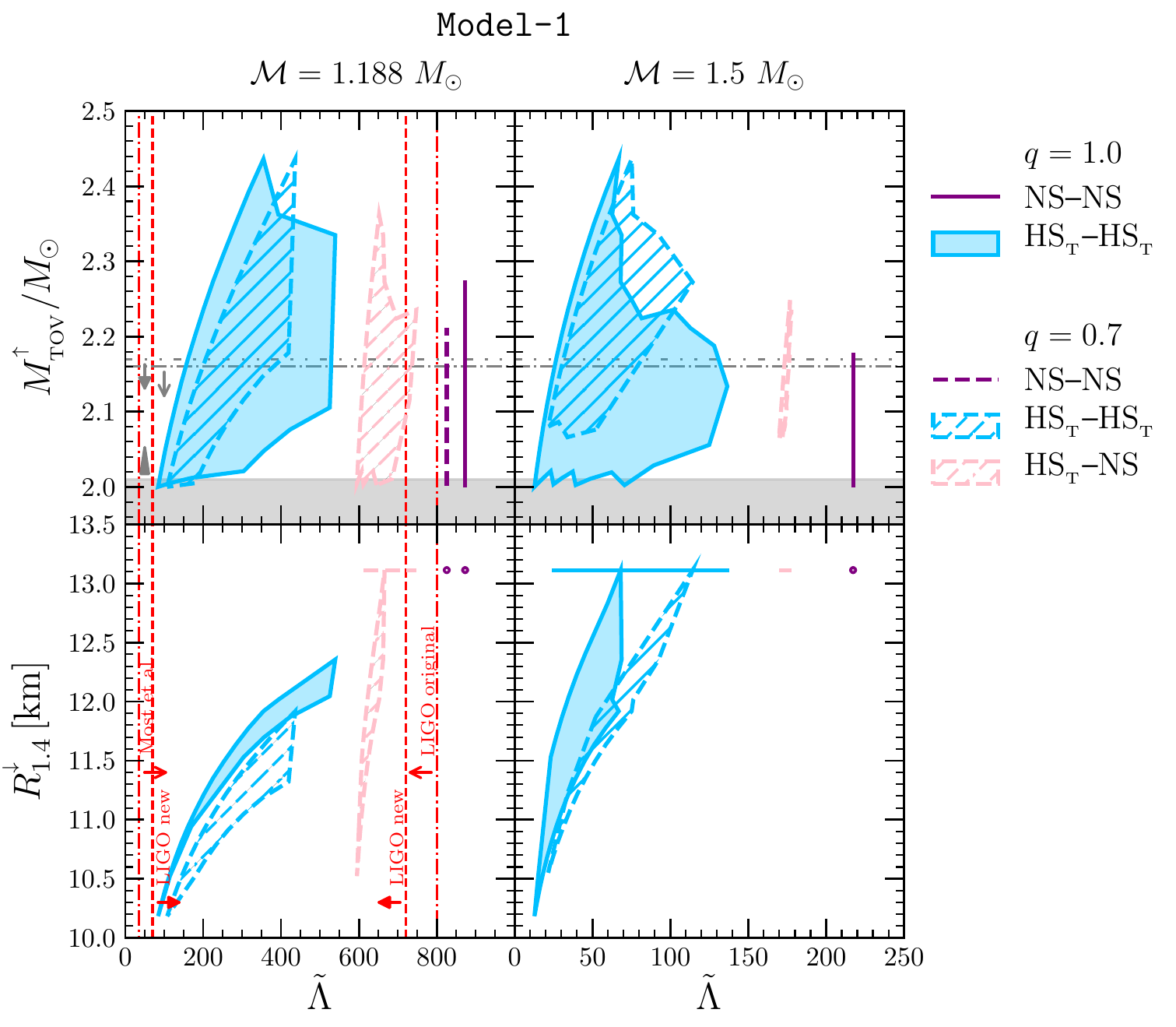} 
 \includegraphics[width=0.495\textwidth]{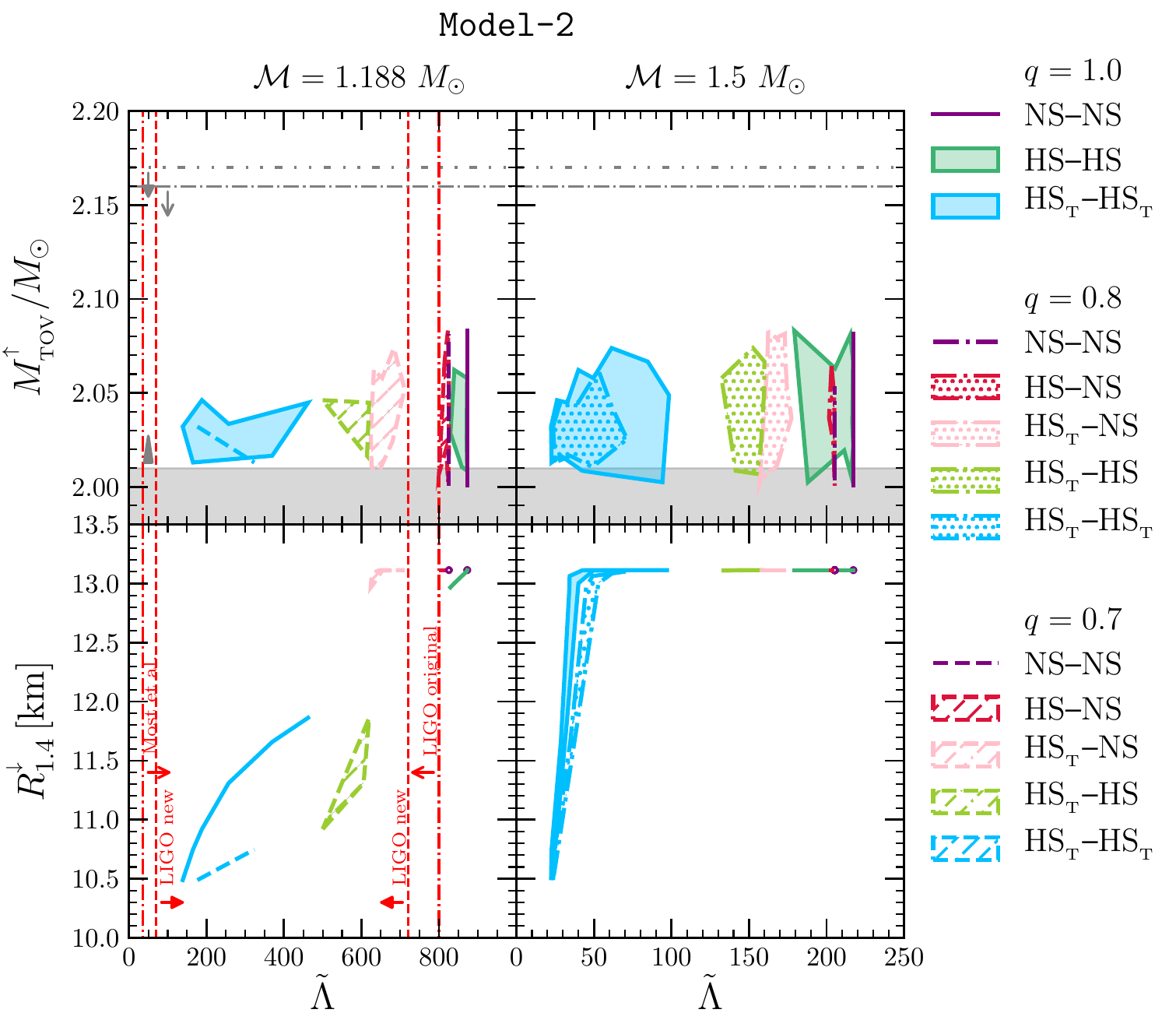} 
 \caption{
  Left plot: Maximum mass (upper panels) and radius of a
   $1.4\,M_\odot$ star (lower panels) as a function of the weighted
   $\tilde{\Lambda}$ for the same cases as in the left plot of Fig.~\ref{fig:parameters_vs_lambda_G_1}.
   Right plot: Maximum mass (upper panels) and radius of a
   $1.4\,M_\odot$ star (lower panels) as a function of the weighted
   $\tilde{\Lambda}$ for the same cases as in the right plot of Fig.~\ref{fig:parameters_vs_lambda_G_1}. In
   these plots, together with the constraints on tidal deformability,
   we display a lower horizontal band coming from the lower limit of
   $2\,M_{\odot}$ observations \cite{Demorest:2010bx,Antoniadis:2013pzd}
   as well as recent constraints on the maximum mass of $\sim
   2.16$--$2.17\,M_{\odot}$ from multi-messenger observations of GW170817
   \cite{Margalit:2017dij,Rezzolla:2017aly}.}
\label{fig:parameters_vs_lambda_G_2}
\end{figure*}

The properties of the phase transition, which are contained in the two
free parameters $\Delta e$ and $p_{\rm tr}$, can be constrained using the
observational information on tidal deformabilities of the GW170817
event. Moreover, given an allowed space of parameters for $\Delta e$ and
$p_{\rm tr}$, the predictions for the maximum mass and the reference
radius $R_{1.4}$ can be further restricted by taking into account the
limits set on the maximum mass and reference radius after the detection
of GW170817.

In what follows we discuss how to constrain twin-star models with
GW170817 and we start by studying the values of $\Delta e$ and $p_{\rm
  tr}$ as a function of $\tilde\Lambda$.

Figure~\ref{fig:parameters_vs_lambda_G_1} reports the ranges in $\Delta e$
and $p_{\rm tr}$ for the \texttt{Model-1} (left plot) and
\texttt{Model-2} (right plot) EOSs yielding twin stars of
$\textit{Category~III}$, as a function of $\tilde \Lambda$ for a chirp
mass $\mathcal{M}=1.188\,M_\odot$ (left panels of each plot). In addition, we also
show results for a higher chirp mass of $\mathcal{M}=1.5\,M_\odot$ (right
panels of each plot), so as to have access to larger $M_1$ and $M_2$ masses closer to
the $2\,M_{\odot}$ limit.

For the \texttt{Model-1} EOSs, in particular, we use two different values
of the mass ratio $q:=M_2/M_1=0.7$ (\ie $M_1=1.64\,M_\odot$, $M_2=1.14\,
M_\odot$ for $\mathcal{M}=1.188\,M_\odot$, and $M_1=2.07\,M_\odot$,
$M_2=1.45\,M_\odot$ for $\mathcal{M}=1.5\,M_\odot$) and $q=1$ (\ie
$M_1=M_2=1.36\,M_\odot$ for $\mathcal{M}=1.188\,M_\odot$ and
$M_1=M_2=1.72 \,M_\odot$ for $\mathcal{M}=1.5\,M_\odot$), which
correspond to the constraints set by the analysis of the LIGO/Virgo data
in Refs. \cite{TheLIGOScientific:2017qsa, Tews:2018chv}. For the
\texttt{Model-2} EOSs, instead, we use $q=0.7,1$ for
$\mathcal{M}=1.188\,M_\odot$, whereas we take $q=0.8,1$ for
$\mathcal{M}=1.5\,M_\odot$ (note that a maximum mass of $2\,M_{\odot}$ is
reached easier for a \texttt{Model-1} EOS than for a \texttt{Model-2}
EOS, as seen in Fig.~\ref{fig:parameters_rhoqcd}). This is due to the
presence of the Gibbs mixed phase that softens the EOS for the transition
region (by contrast, the Maxwell construction leads to the stiffest EOS
parametrization in the quark phase). By increasing $q$ to $0.8$, we have
access to a value of $M_1=1.93\,M_\odot$, slightly below $2\,M_{\odot}$
and thus easier to be found within \texttt{Model-2}. In the case of
$\mathcal{M}=1.188\,M_\odot$, we also show with vertical lines the
LIGO/Virgo upper limit of $\tilde\Lambda =800$
\cite{TheLIGOScientific:2017qsa}, as well as the improved analysis of
Ref. \cite{Abbott:2018wiz}, which gives of $70 < \tilde\Lambda <
720$. We also show the lowest value of $\Lambda_{1.4}$ for a star with
a phase transition, \ie $\Lambda_{1.4} > 35.5$ (at $2$-$\sigma$ level)
of Ref. \cite{Most:2018hfd}. We note that our results are overall 
consistent with the previous results of Refs. \cite{Tews:2018chv,
Han:2018mtj,Zhao:2018nyf}.

In summary, Fig.~\ref{fig:parameters_vs_lambda_G_1} shows that, in
agreement with the results reported in Refs. \cite{Paschalidis:2017qmb,
  Alvarez-Castillo:2018pve, Christian:2018jyd}, a HQPT softens the EOS
and expands the parameter space of EOSs that are compatible with the
GW170817 event, allowing for HS$_{_{\rm T}}$--HS$_{_{\rm T}}$
configurations, while HS$_{_{\rm T}}$--NS solutions for $q=0.7$ are also
permitted.  Such constraints also allow for HS$_{_{\rm T}}$--HS
configurations at $q=0.7$ for a \texttt{Model-2} EOS.  The global
parameters of the HQPT ($p_{\rm tr}$ and $\Delta e$) are thus constrained
by GW170817 to be in the range
\begin{align}
  &\bullet \texttt{Model-1}  \nonumber \\
  & \quad p_{\rm tr}\in [25,65] \,
  \rm MeV\,fm^{-3}, & \Delta e \in [175,395] \, \rm MeV\,fm^{-3}\,,
  \nonumber \\
  \nonumber \\
  &\bullet \texttt{Model-2} \nonumber\\
  & \quad p_{\rm tr}\in [15,45]
  \, \rm MeV\,fm^{-3}, & \Delta e \in [380,435] \, \rm MeV\,fm^{-3}\,.
  \nonumber
\end{align}
Note that for a larger $\mathcal{M}=1.5\,M_\odot$, a larger range of
$\Delta e$ and $p_{\rm tr}$ parameters is found for both models, with
lower values of $\tilde \Lambda$ up to $\sim 220$; since these specific
values correspond to the NS--NS configuration, they obviously depend on
the specific hadronic EOS considered.

In a similar manner, the plots of Fig.~\ref{fig:parameters_vs_lambda_G_2}
display the maximum mass (upper panels) and minimum radius for a
$1.4\,M_{\odot}$ star (lower panels) for the \texttt{Model-1} (left plot)
and \texttt{Model-2} (right plot) EOSs as a function of $\tilde{\Lambda}$
for the same NS and HS/HS$_{_{\rm T}}$ configurations and at the same
mass ratios in the plots of Fig.~\ref{fig:parameters_vs_lambda_G_1}. Note
that as in Fig. \ref{fig:parameters_vs_lambda_G_1}, the left panels for
each plot in Fig. \ref{fig:parameters_vs_lambda_G_2} refer to
$\mathcal{M}=1.188\,M_\odot$, while the right ones to
$\mathcal{M}=1.5\,M_\odot$.

Together with the previous constraints on the tidal deformability shown
in Fig.~\ref{fig:parameters_vs_lambda_G_1}, we display in
Fig.~\ref{fig:parameters_vs_lambda_G_2} the excluded range of masses up
to the $2\,M_{\odot}$ limit coming from $2\,M_{\odot}$ observations
\cite{Demorest:2010bx, Antoniadis:2013pzd, Arzoumanian:2017puf}, as well
as recent constraints on the maximum mass of $\sim
2.16$--$2.17\,M_{\odot}$ from multi-messenger observations of GW170817
\cite{Margalit:2017dij, Rezzolla:2017aly}. We note that for the
\texttt{Model-2} EOSs, and $q=0.7$ and $\mathcal{M}=1.188\,M_\odot$, the
HS$_{_{\rm T}}$--HS$_{_{\rm T}}$, HS$_{_{\rm T}}$--HS and HS$_{_{\rm
    T}}$--NS configurations satisfy the maximum-mass constraints for the
whole range of values of $\Delta e$ and $p_{\rm tr}$. On the other hand,
for the \texttt{Model-1} EOSs (again with $q=0.7$ and
$\mathcal{M}=1.188\,M_\odot$), the HS$_{_{\rm T}}$--HS$_{_{\rm T}}$ and
HS$_{_{\rm T}}$--NS solutions satisfy the constraint for less than
$~50\%$ of the parameter space, further requiring the energy-density jump
to be $\Delta e\in[245,395]\rm\,MeV\,fm^{-3}$.  This is due to the fact
that the maximum mass increases as $\Delta e$ decreases, as long as
$p_{\rm tr}$ is not too high (see also Fig.~2 of
\cite{Alford:2015dpa}). As a result, imposing a $M_{_{\rm
    TOV}}^{^\uparrow} < 2.16-2.17$ rules out small $\Delta e$ values.  A
similar behaviour for a $\mathcal{M}=1.5\,M_\odot$ is seen, although in
this case we cannot impose any constraint on $\tilde{\Lambda}$.


\begin{table}[]
  \setlength{\tabcolsep}{-0.2em}
\begin{tabular}{lllrcl}
  \hline
\multicolumn{2}{l}{Reference} & \phantom{MTHR} & & $R_{i}\,{\rm [km]}$ \\ \hline
\multicolumn{2}{l}{\textit{Without a phase transition}}     & &             &          & \\
\multicolumn{2}{l}{Bauswein et al. \cite{Bauswein:2017vtn}} & & $10.68^{+0.15}_{-0.03}\leq$ & $R_{1.6}$ &  \\
\multicolumn{2}{l}{Most et al.     \cite{Most:2018hfd}}     & & $12.00\leq$ & $R_{1.4}$ & $\leq 13.45$ \\
\multicolumn{2}{l}{Burgio et al.   \cite{Burgio:2018yix}}   & & $11.8\leq$  & $R_{1.5}$ & $\leq 13.1$ \\
\multicolumn{2}{l}{Tews et al.     \cite{Tews:2018chv}}     & & $11.3\leq$  & $R_{1.4}$ & $\leq 13.6$ \\
\multicolumn{2}{l}{De et al.       \cite{De:2018uhw}}       & & $8.9\leq$   & $R_{1.4}$ & $\leq 13.2$ \\
\multicolumn{2}{l}{LIGO/Virgo      \cite{Abbott:2018exr}}   & & $10.5\leq$  & $R_{1.4}$ & $\leq 13.3$ \\ 
\multicolumn{2}{l}{Koeppel et al.  \cite{Koeppel2019}}      & & $10.9\leq$  & $R_{1.4}$ &             \\ 
\hline                                                          
\multicolumn{2}{l}{\textit{With a phase transition}}        & &             &          & \\
\multicolumn{2}{l}{Annala et al.   \cite{Annala:2017llu}}   & &             & $R_{1.4}$ & $\leq 13.6$ \\
\multicolumn{2}{l}{Most et al.     \cite{Most:2018hfd}}     & & $8.53\leq$  & $R_{1.4}$ & $\leq 13.74$ \\
\multicolumn{2}{l}{Burgio et al.   \cite{Burgio:2018yix}}   & &             & $R_{1.5}$ & $=10.7$ \\
\multicolumn{2}{l}{Tews et al.     \cite{Tews:2018chv}}     & & $9.0\leq$   & $R_{1.4}$ & $\leq 13.6$ \\ 
\hline
\multicolumn{2}{l}{\textit{This work}}                      & &             &          & \\
\multicolumn{2}{l}{NS}                                      & &             & $R_{1.4}$ & $=13.11$ \\
HS                  & \texttt{Model-2}			    & & $12.9\leq$  & $R_{1.4}$ & $\leq 13.11$ \\
HS$_{_{\rm T}}\ \ \ $ & \texttt{Model-1} 			    & & $10.1\leq$  & $R_{1.4}$ & $\leq 12.9$ \\
HS$_{_{\rm T}}\ \ \ $ & \texttt{Model-2} 			    & & $10.4\leq$  & $R_{1.4}$ & $\leq 11.9$ \\
\hline
\end{tabular}
\caption{Constraints on the radius of neutron stars from  GW170817 for
 models without a phase transition (top), works considering the
 possibility of a transition to quark matter (middle) and for EOSs of
 \textit{Category III} in the present work (bottom).}
\label{table:radius}
\end{table}

\begin{figure*}
 \centering
     \includegraphics[width=0.8\textwidth]{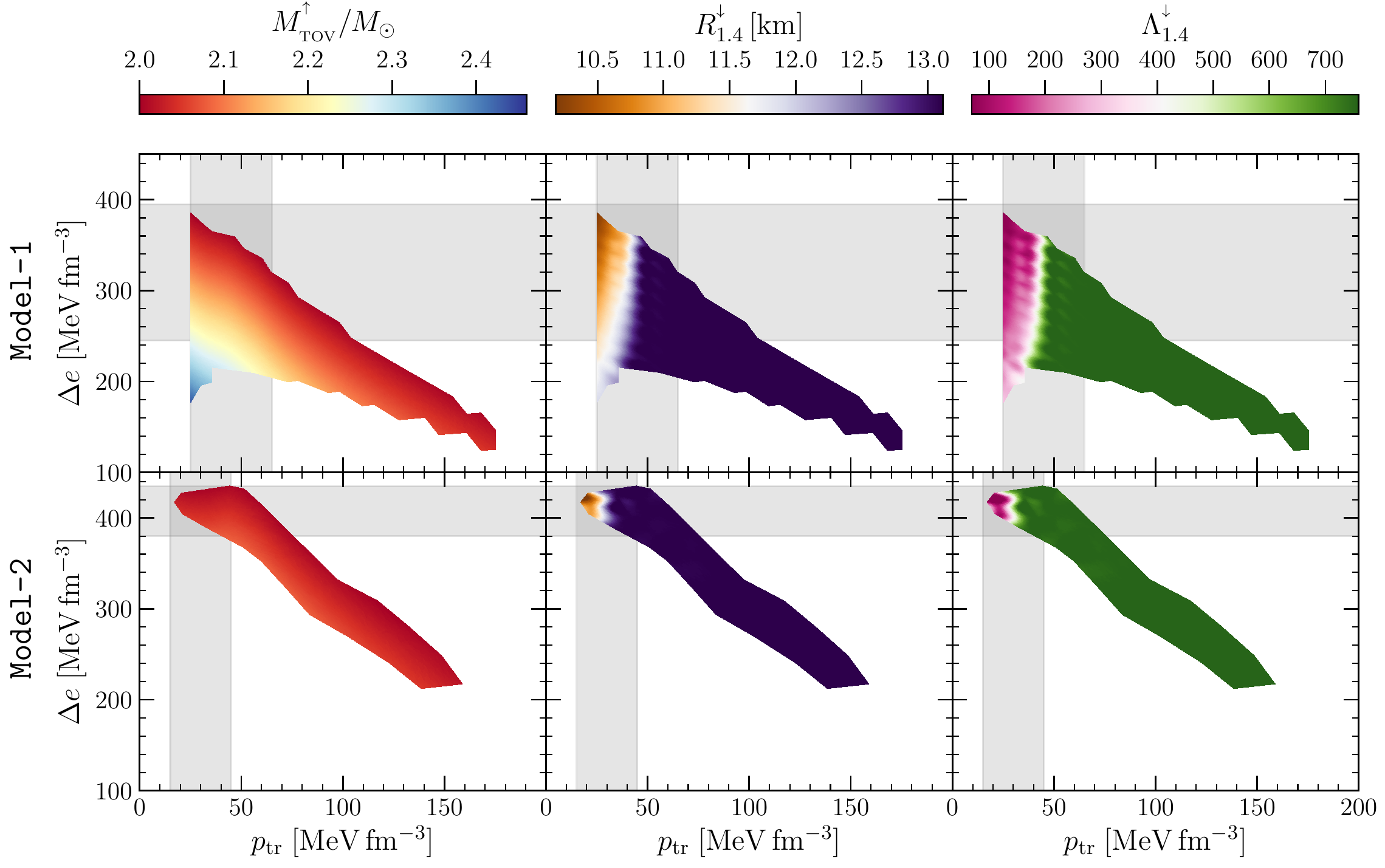}
 \caption{Maximum mass (left panels), minimum radius of a $1.4\,M_\odot$
   star (middle panels) and minimum tidal deformability of a
   $1.4\,M_\odot$ star (right panels) as a function of the values of the
   transition pressure and energy-density jump of twin stars of
   \textit{Category III} for the \texttt{Model-1} (upper panels) and
   \texttt{Model-2} (lower panels) EOSs. The shaded areas correspond to
   the parameter space allowed by the GW170817 constraints on
   ${\Lambda}_{1.4}^{^\downarrow}$ and $M_{_{\rm TOV}}^{^\uparrow}$ (see details in the
   text).}
\label{fig:properties}
\end{figure*}

As for the limits on the radius of $1.4\, M_{\odot}$, several works have
reported values for stars of a given mass \cite{Annala:2017llu,
  Most:2018hfd, Abbott:2018exr, De:2018uhw, Tews:2018chv, Kumar:2017wqp,
  Fattoyev:2017jql, Malik:2018zcf, Lim:2018bkq, Raithel:2018ncd,
  Burgio:2018yix} and we have collected them in Table~\ref{table:radius}.
Also, we recall that in Ref.~\cite{Sieniawska:2018zzj} it was found that
the minimal radius that can be produced on a twin branch lies between
$9.5$ and $10.5\,{\rm km}$. This result was obtained using a set of
relativistic polytropes and a quark bag model for the phase transition
with a Maxwell construction.
  
When concentrating on the predictions of this work, and obviously for the
hadronic EOS considered here, we note that for the allowed configurations
of type HS$_{_{\rm T}}$--HS$_{_{\rm T}}$ and HS$_{_{\rm T}}$--NS with
mass ratio $q=0.7$ and for the \texttt{Model-1} and \texttt{Model-2}
EOSs, we find that $10< R_{1.4}/{\rm km}\lesssim 13$ and that $10<
R_{1.4}/{\rm km}\lesssim 12.5$ when considering only HS$_{_{\rm
    T}}$--HS$_{_{\rm T}}$ solutions.
Similarly good agreements with previous investigations can be found when
considering HS$_{_{\rm T}}$--HS binaries with $q=0.7$ and the
\texttt{Model-2} EOSs. On the other hand, larger radii closer to
$13\,{\rm km}$ are reached when $\mathcal{M}=1.5\,M_\odot$ for HS$_{_{\rm
    T}}$--HS$_{_{\rm T}}$ configurations for both \texttt{Model-1} and
\texttt{Model-2} EOSs.

A summary plot of our results is given in Fig.~\ref{fig:properties},
where we show the maximum mass, $M_{_{\rm TOV}} ^{^{\uparrow}}$, the
minimum radius of a $1.4\,M_\odot$ star, $R_{1.4} ^{^{\downarrow}}$ , and
the minimum tidal deformability for the same star,
$\Lambda_{1.4}^{^{\downarrow}}$, as a function of the values of the
transition pressure and energy density of twin stars of \textit{Category
  III} for the \texttt{Model-1} (upper panels) and \texttt{Model-2} EOSs
(lower panels). With shaded areas we indicate the parameter space allowed
by the GW170817 as analysed in Fig.~\ref{fig:parameters_vs_lambda_G_1};
once again, the value of $R_{1.4}$ relative to NSs depends on the
hadronic EOS considered here.
  
Overall, we find that $M_{_{\rm TOV}}^{^{\uparrow}} \simeq
2$--$2.1\,M_\odot$ are commonly generated for both models, whereas
$R_{1.4}^{^{\downarrow}} \simeq 13 \, {\rm km}$ and
$\Lambda_{1.4}^{^{\downarrow}} \simeq 700$ are also produced for almost
the entire range in the $(p_{\rm tr}, \Delta e)$ space. The lowest values
of $R_{1.4}^{^{\downarrow}} \simeq 10 \, {\rm km}$ and
$\Lambda_{1.4}^{^{\downarrow}} \simeq 100$ are produced for transition
pressures well below $p_{\rm tr}= 50\, {\rm MeV\, fm^{-3}}$, \ie
$\rho_{\rm tr}=2.4\,\rho_0$. If such small radii and tidal deformabilities
given by our hadronic EOS are confirmed by future measurements, the HQPT
with a Maxwell construction would imply that hyperons exist in a very
narrow region of the interiors of neutron stars. In the case of the FSU2H
model \cite{Tolos:2016hhl, Tolos:2017lgv}, only the $\Lambda$ particle
would be present, since it appears at $\rho=2.2\,\rho_0$ (see
Fig.~\ref{fig:eos}). In the case of a Gibbs construction of the HQPT,
hyperons would still exist in the mixed phase before entering the pure
deconfined quark phase and the fractions of these particles would depend
on both the hadronic and quark models at these densities. Hence, if
future detections of gravitational waves from LIGO/Virgo determine values
of $\Lambda_{1.4} \lesssim 400$ and, at the same time, chirp masses
$\mathcal{M} \lesssim 1.2M_\odot$, our modeling reveals that these can be
interpreted in terms of a HQPT with a low transition pressure taking place
during the inspiral. Otherwise, it will be difficult to distinguish
during the inspiral whether one of the components of the binary is a HS,
as a HQPT with transition pressures above $50\,{\rm MeV \, fm^{-3}}$
might be indistinguishable from no phase transition.

\section{Conclusions and Outlook}
\label{sec:conclusions}

We have performed an extensive and detailed analysis of the features of
the hadron-quark phase transition that is needed in order to obtain
twin-star configurations and enforcing, at the same time, the constraint
on the minimum value of the maximum mass, \ie $M_{_{\rm TOV}} \gtrsim
2\,M_{\odot}$ and the information on multi-messenger observation of the
GW170817 event. In our analysis we have employed two general EOS models
for the neutron-star matter, \ie \texttt{Model-1} and \texttt{Model-2}
EOSs, that share the same description for the hadronic EOS, but take into
account a parametrization of the hadron--quark phase transition assuming
either a Maxwell or a Gibbs construction, combined with a constant speed
of sound (CSS) parametrization for the quark phase.

The parameter space of the phase transition, which is set by the
energy-density jump and transition pressure, $\Delta e$ and $p_{\rm tr}$,
has been explored systematically and the twin-star solutions found have
been classified according to \textit{Categories I-IV}
\cite{Christian:2017jni}. We find that the largest number of twin-star
solutions that satisfy the $2\,M_{\odot}$ constraint is \textit{Category
  III}, with masses in the normal-neutron-star branch of
$1$--$2\,M_{\odot}$ and maximum masses of the twin-star branch $M_{_{\rm
    TOV,T}} \gtrsim$ $2\,M_{\odot}$. This category is potentially the
most interesting one, as it accommodates twin stars with masses around
the canonical value of $1.4\,M_{\odot}$. The masses, radii and tidal
deformabilities have been thoroughly studied for the different categories
and parameter sets, showing that, when twin-star solutions appear, the
tidal deformability also displays two distinct branches having the same
mass. This behaviour, which is in agreement with what is found in
Refs. \cite{Burgio:2018yix, Sieniawska:2018zzj, Christian:2018jyd}, is
radically different from what is shown for pure neutron stars and could
be used as a signature for the existence of twin stars.

Making use of the large space of solutions found, we have considered the
evidence for the existence of EOSs with a HQPT and thus originating
twin-star solutions. In particular, we have exploited the 
weighted tidal deformability and chirp mass,
as deduced from the recent binary neutron-star
merger event GW170817. In this way, we have found that the presence of a
phase transition is not excluded by the observational data and that, in
addition to standard NS--NS binaries, also binaries of the type
HS$_{_{\rm T}}$--HS$_{_{\rm T}}$ and HS$_{_{\rm T}}$--NS (for
\texttt{Model-1} and \texttt{Model-2} EOSs) and HS$_{_{\rm T}}$--HS (for
\texttt{Model-2} EOSs) are allowed, in principle. In addition, we have
used the multi-messenger astronomical observations associated with
GW170817, namely, the new predictions on the maximum mass, to set
constraints on the values of $\Delta e$ and $p_{\rm tr}$, as well as on
the radius of a reference model with a mass of $1.4\,M_{\odot}$.

Interestingly, the time of occurrence of the HQPT in a binary system of
compact stars will depend on the total mass of the binary and on the
global properties of the HQPT (\ie $\Delta e$ and $p_{\rm tr}$). For
example, in the case of a binary with chirp mass $\mathcal{M} =
1.188~M_\odot$ (as for GW170817), and assuming that the hadronic part of
the EOS is given by the FSU2H model \cite{Tolos:2016hhl, Tolos:2017lgv},
the phase transition takes place (at least for one of the two stars)
already in the inspiral phase. In particular, the lower the transition
pressure, the lower the minimum chirp mass for the HQPT to occur in
pre-merger NSs. The lowest chirp mass leading to the appearance of a
phase transition is $0.6\,M_{\odot}$ and, quite generically, higher chirp
masses are needed in the equal-mass case to obtain a phase
transition. Indeed, our results show that future gravitational-wave
detections with chirp masses $\mathcal{M} \lesssim 1.2M_\odot$ and, at
the same time, tidal deformabilities of $\Lambda_{1.4} \lesssim 400$, can
be interpreted as due to a HQPT with a low transition pressure taking
place in the inspiral phase. Because these precise values depend on the
chosen hadronic EOS, we have also considered the hadronic FSU2 model,
which represents the baseline of the FSU2H model employed in this work
and is stiffer around saturation density, giving rise to higher values of
$R_{1.4}$ and $\Lambda_{1.4}$ for the hadronic phase. At the same time,
because of the similar stiffness at high densities, the values of the
$\Lambda_{1.4}$ and chirp mass where the phase transition takes place are
similar to both models as well as the twin-star parameter space. The
dependence of our results on the chosen hadronic EOS will be the subject
of our future work.

On the other hand, if the central pressure of either of the two stars is
below $p_{\rm tr}$ during the inspiral, then a measurement of the tidal
deformabilities cannot contain information on the properties of the
structure of the HQPT. For such cases, the HQPT will take place during
the post-merger evolution of the merger remnant, giving rise to a variety
of interesting phenomena. For instance, the rearrangement of the angular
momentum in the remnant as a result of the formation of a quark-core
could be accompanied by a prompt burst of neutrinos followed by a
gamma-ray burst \cite{Cheng1996PRL, Bombaci2000ApJ,
  Mishustin2003PhLB}. Furthermore, the $f_2$-frequency peak of the
gravitational-wave signal \cite{Takami:2014, takami2015spectral,
  Rezzolla2016} would change rapidly due to the sudden speed up of the
differentially rotating remnant \cite{Hanauske2018EPJWC,
  Hanauske2018JApA}.

Preliminary investigations in this direction have already been made and
the consequences of the appearance of the HQPT after the merger and its
impact on the spectral properties of the emitted gravitational waves have
been recently discussed in Refs. \cite{Most:2018eaw,
  Bauswein:2018bma}. Although the two studies have employed
different temperature-dependent EOSs which include a strong HQPT but do
not allow for twin-star solutions, both reach the conclusion that the
impact of the phase transition might be measurable with future
gravitational-wave detections \cite{Most:2018eaw,
  Bauswein:2018bma}. Such measurement, together with those
performed by X- and gamma-ray space missions such as NICER
\cite{2014SPIE.9144E..20A},
eXTP \cite{Watts:2018iom} and THESEUS \cite{Amati2018}, have the
potential of providing essential information to clarify whether a HQPT
should indeed be accounted for a binary neutron-star merger.

\begin{acknowledgments}

We thank J\"urgen Schaffner-Bielich, Armen Sedrakian and Fiorella Burgio
for valuable discussions. Support comes  from ``PHAROS'',
COST Action CA16214; the LOEWE-Program in HIC for FAIR; the European
Union's Horizon 2020 Research and Innovation Programme (Grant 671698)
(call FETHPC-1-2014, project ExaHyPE); the Deutsche
Forschungsgemeinschaft (DFG, German Research Foundation) -Project number
315477589 -TRR 211 and  the Heisenberg Programme under the Project number 383452331; the Spanish Ministerio de
Ciencia, Innovacion y Universidades Project No. MDM-2014-0369 of ICCUB (Unidad de Excelencia Mar\'{\i}a
de Maeztu) and with additional European FEDER funds under Contract
No. FIS2017-87534-P and also the FPA2016-81114-P Grant; the Spanish Ministerio de Educaci\'on, Cultura y
Deporte (MECD) under the fellowship FPU17/04910; the Generalitat de
Catalunya under Contract No. 2014SGR-401 and Contract No. 2017SGR-247; and the fellowship 2018 FI\_B
00234 from the Secretaria d'Universitats i Recerca del Departament
d'Empresa i Coneixement. M.H. acknowledges support from Frankfurt Institute for Advanced Studies
(FIAS) and the Institute for Theoretical Physics (ITP) at the Goethe
University in Frankfurt.

\end{acknowledgments}

\bibliography{neue_bib}

\begin{thebibliography}{127}%
\makeatletter
\providecommand \@ifxundefined [1]{%
 \@ifx{#1\undefined}
}%
\providecommand \@ifnum [1]{%
 \ifnum #1\expandafter \@firstoftwo
 \else \expandafter \@secondoftwo
 \fi
}%
\providecommand \@ifx [1]{%
 \ifx #1\expandafter \@firstoftwo
 \else \expandafter \@secondoftwo
 \fi
}%
\providecommand \natexlab [1]{#1}%
\providecommand \enquote  [1]{``#1''}%
\providecommand \bibnamefont  [1]{#1}%
\providecommand \bibfnamefont [1]{#1}%
\providecommand \citenamefont [1]{#1}%
\providecommand \href@noop [0]{\@secondoftwo}%
\providecommand \href [0]{\begingroup \@sanitize@url \@href}%
\providecommand \@href[1]{\@@startlink{#1}\@@href}%
\providecommand \@@href[1]{\endgroup#1\@@endlink}%
\providecommand \@sanitize@url [0]{\catcode `\\12\catcode `\$12\catcode
  `\&12\catcode `\#12\catcode `\^12\catcode `\_12\catcode `\%12\relax}%
\providecommand \@@startlink[1]{}%
\providecommand \@@endlink[0]{}%
\providecommand \url  [0]{\begingroup\@sanitize@url \@url }%
\providecommand \@url [1]{\endgroup\@href {#1}{\urlprefix }}%
\providecommand \urlprefix  [0]{URL }%
\providecommand \Eprint [0]{\href }%
\providecommand \doibase [0]{http://dx.doi.org/}%
\providecommand \selectlanguage [0]{\@gobble}%
\providecommand \bibinfo  [0]{\@secondoftwo}%
\providecommand \bibfield  [0]{\@secondoftwo}%
\providecommand \translation [1]{[#1]}%
\providecommand \BibitemOpen [0]{}%
\providecommand \bibitemStop [0]{}%
\providecommand \bibitemNoStop [0]{.\EOS\space}%
\providecommand \EOS [0]{\spacefactor3000\relax}%
\providecommand \BibitemShut  [1]{\csname bibitem#1\endcsname}%
\let\auto@bib@innerbib\@empty
\bibitem [{\citenamefont {Itoh}(1970)}]{Itoh:1970uw}%
  \BibitemOpen
  \bibfield  {author} {\bibinfo {author} {\bibfnamefont {N.}~\bibnamefont
  {Itoh}},\ }\href {\doibase 10.1143/PTP.44.291} {\bibfield  {journal}
  {\bibinfo  {journal} {Prog.Theor.Phys.}\ }\textbf {\bibinfo {volume} {44}},\
  \bibinfo {pages} {291} (\bibinfo {year} {1970})}\BibitemShut {NoStop}%
\bibitem [{\citenamefont {{Cheng}}\ and\ \citenamefont
  {{Dai}}(1996)}]{Cheng1996PRL}%
  \BibitemOpen
  \bibfield  {author} {\bibinfo {author} {\bibfnamefont {K.~S.}\ \bibnamefont
  {{Cheng}}}\ and\ \bibinfo {author} {\bibfnamefont {Z.~G.}\ \bibnamefont
  {{Dai}}},\ }\href {\doibase 10.1103/PhysRevLett.77.1210} {\bibfield
  {journal} {\bibinfo  {journal} {Phys. Rev. Lett.}\ }\textbf {\bibinfo
  {volume} {77}},\ \bibinfo {pages} {1210} (\bibinfo {year} {1996})},\ \Eprint
  {http://arxiv.org/abs/astro-ph/9510073} {arXiv:astro-ph/9510073 [astro-ph]}
  \BibitemShut {NoStop}%
\bibitem [{\citenamefont {{Bombaci}}\ and\ \citenamefont
  {{Datta}}(2000)}]{Bombaci2000ApJ}%
  \BibitemOpen
  \bibfield  {author} {\bibinfo {author} {\bibfnamefont {I.}~\bibnamefont
  {{Bombaci}}}\ and\ \bibinfo {author} {\bibfnamefont {B.}~\bibnamefont
  {{Datta}}},\ }\href {\doibase 10.1086/312497} {\bibfield  {journal} {\bibinfo
   {journal} {Astrophys. J.}\ }\textbf {\bibinfo {volume} {530}},\ \bibinfo
  {pages} {L69} (\bibinfo {year} {2000})},\ \Eprint
  {http://arxiv.org/abs/astro-ph/0001478} {arXiv:astro-ph/0001478 [astro-ph]}
  \BibitemShut {NoStop}%
\bibitem [{\citenamefont {{Hanauske}}\ \emph {et~al.}(2001)\citenamefont
  {{Hanauske}}, \citenamefont {{Satarov}}, \citenamefont {{Mishustin}},
  \citenamefont {{St{\"o}cker}},\ and\ \citenamefont
  {{Greiner}}}]{Hanauske2001PRD}%
  \BibitemOpen
  \bibfield  {author} {\bibinfo {author} {\bibfnamefont {M.}~\bibnamefont
  {{Hanauske}}}, \bibinfo {author} {\bibfnamefont {L.~M.}\ \bibnamefont
  {{Satarov}}}, \bibinfo {author} {\bibfnamefont {I.~N.}\ \bibnamefont
  {{Mishustin}}}, \bibinfo {author} {\bibfnamefont {H.}~\bibnamefont
  {{St{\"o}cker}}}, \ and\ \bibinfo {author} {\bibfnamefont {W.}~\bibnamefont
  {{Greiner}}},\ }\href {\doibase 10.1103/PhysRevD.64.043005} {\bibfield
  {journal} {\bibinfo  {journal} {Phys. Rev. D}\ }\textbf {\bibinfo {volume}
  {64}},\ \bibinfo {eid} {043005} (\bibinfo {year} {2001})},\ \Eprint
  {http://arxiv.org/abs/astro-ph/0101267} {arXiv:astro-ph/0101267 [astro-ph]}
  \BibitemShut {NoStop}%
\bibitem [{\citenamefont {Weber}(2005)}]{Weber:2004kj}%
  \BibitemOpen
  \bibfield  {author} {\bibinfo {author} {\bibfnamefont {F.}~\bibnamefont
  {Weber}},\ }\href@noop {} {\bibfield  {journal} {\bibinfo  {journal} {Prog.
  Part. Nucl. Phys.}\ }\textbf {\bibinfo {volume} {54}},\ \bibinfo {pages}
  {193} (\bibinfo {year} {2005})},\ \Eprint
  {http://arxiv.org/abs/astro-ph/0407155} {astro-ph/0407155} \BibitemShut
  {NoStop}%
\bibitem [{\citenamefont {{Zacchi}}\ \emph {et~al.}(2015)\citenamefont
  {{Zacchi}}, \citenamefont {{Stiele}},\ and\ \citenamefont
  {{Schaffner-Bielich}}}]{Zacchi2015PRD}%
  \BibitemOpen
  \bibfield  {author} {\bibinfo {author} {\bibfnamefont {A.}~\bibnamefont
  {{Zacchi}}}, \bibinfo {author} {\bibfnamefont {R.}~\bibnamefont {{Stiele}}},
  \ and\ \bibinfo {author} {\bibfnamefont {J.}~\bibnamefont
  {{Schaffner-Bielich}}},\ }\href {\doibase 10.1103/PhysRevD.92.045022}
  {\bibfield  {journal} {\bibinfo  {journal} {Phys. Rev. D}\ }\textbf {\bibinfo
  {volume} {92}},\ \bibinfo {eid} {045022} (\bibinfo {year}
  {2015})}\BibitemShut {NoStop}%
\bibitem [{\citenamefont {{Pal}}\ \emph {et~al.}(1999)\citenamefont {{Pal}},
  \citenamefont {{Hanauske}}, \citenamefont {{Zakout}}, \citenamefont
  {{St{\"o}cker}},\ and\ \citenamefont {{Greiner}}}]{Pal1999PRC}%
  \BibitemOpen
  \bibfield  {author} {\bibinfo {author} {\bibfnamefont {S.}~\bibnamefont
  {{Pal}}}, \bibinfo {author} {\bibfnamefont {M.}~\bibnamefont {{Hanauske}}},
  \bibinfo {author} {\bibfnamefont {I.}~\bibnamefont {{Zakout}}}, \bibinfo
  {author} {\bibfnamefont {H.}~\bibnamefont {{St{\"o}cker}}}, \ and\ \bibinfo
  {author} {\bibfnamefont {W.}~\bibnamefont {{Greiner}}},\ }\href {\doibase
  10.1103/PhysRevC.60.015802} {\bibfield  {journal} {\bibinfo  {journal}
  {Physical Review C}\ }\textbf {\bibinfo {volume} {60}},\ \bibinfo {eid}
  {015802} (\bibinfo {year} {1999})},\ \Eprint
  {http://arxiv.org/abs/astro-ph/9905010} {arXiv:astro-ph/9905010 [astro-ph]}
  \BibitemShut {NoStop}%
\bibitem [{\citenamefont {{Hanauske}}\ \emph {et~al.}(2000)\citenamefont
  {{Hanauske}} \emph {et~al.}}]{Hanauske2000ApJ}%
  \BibitemOpen
  \bibfield  {author} {\bibinfo {author} {\bibfnamefont {M.}~\bibnamefont
  {{Hanauske}}} \emph {et~al.},\ }\href {\doibase 10.1086/309052} {\bibfield
  {journal} {\bibinfo  {journal} {Astrophys. J.}\ }\textbf {\bibinfo {volume}
  {537}},\ \bibinfo {pages} {958} (\bibinfo {year} {2000})},\ \Eprint
  {http://arxiv.org/abs/astro-ph/9909052} {arXiv:astro-ph/9909052 [astro-ph]}
  \BibitemShut {NoStop}%
\bibitem [{\citenamefont {{Pal}}\ \emph {et~al.}(2000)\citenamefont {{Pal}},
  \citenamefont {{Bandyopadhyay}},\ and\ \citenamefont
  {{Greiner}}}]{Pal2000NuPhA}%
  \BibitemOpen
  \bibfield  {author} {\bibinfo {author} {\bibfnamefont {S.}~\bibnamefont
  {{Pal}}}, \bibinfo {author} {\bibfnamefont {D.}~\bibnamefont
  {{Bandyopadhyay}}}, \ and\ \bibinfo {author} {\bibfnamefont {W.}~\bibnamefont
  {{Greiner}}},\ }\href {\doibase 10.1016/S0375-9474(00)00175-5} {\bibfield
  {journal} {\bibinfo  {journal} {Nuclear Physics A}\ }\textbf {\bibinfo
  {volume} {674}},\ \bibinfo {pages} {553} (\bibinfo {year} {2000})},\ \Eprint
  {http://arxiv.org/abs/astro-ph/0001039} {arXiv:astro-ph/0001039 [astro-ph]}
  \BibitemShut {NoStop}%
\bibitem [{\citenamefont {Lattimer}\ and\ \citenamefont
  {Prakash}(2007)}]{Lattimer:2006xb}%
  \BibitemOpen
  \bibfield  {author} {\bibinfo {author} {\bibfnamefont {J.~M.}\ \bibnamefont
  {Lattimer}}\ and\ \bibinfo {author} {\bibfnamefont {M.}~\bibnamefont
  {Prakash}},\ }\href {\doibase 10.1016/j.physrep.2007.02.003} {\bibfield
  {journal} {\bibinfo  {journal} {Phys. Rept.}\ }\textbf {\bibinfo {volume}
  {442}},\ \bibinfo {pages} {109} (\bibinfo {year} {2007})},\ \Eprint
  {http://arxiv.org/abs/astro-ph/0612440} {arXiv:astro-ph/0612440 [astro-ph]}
  \BibitemShut {NoStop}%
\bibitem [{\citenamefont {{Heiselberg}}\ \emph {et~al.}(1993)\citenamefont
  {{Heiselberg}}, \citenamefont {{Pethick}},\ and\ \citenamefont
  {{Staubo}}}]{Heiselberg1993PRL}%
  \BibitemOpen
  \bibfield  {author} {\bibinfo {author} {\bibfnamefont {H.}~\bibnamefont
  {{Heiselberg}}}, \bibinfo {author} {\bibfnamefont {C.~J.}\ \bibnamefont
  {{Pethick}}}, \ and\ \bibinfo {author} {\bibfnamefont {E.~F.}\ \bibnamefont
  {{Staubo}}},\ }\href {\doibase 10.1103/PhysRevLett.70.1355} {\bibfield
  {journal} {\bibinfo  {journal} {Phys. Rev. Lett.}\ }\textbf {\bibinfo
  {volume} {70}},\ \bibinfo {pages} {1355} (\bibinfo {year}
  {1993})}\BibitemShut {NoStop}%
\bibitem [{\citenamefont {Oestgaard}(1994)}]{Oestgaard:1994gy}%
  \BibitemOpen
  \bibfield  {author} {\bibinfo {author} {\bibfnamefont {E.}~\bibnamefont
  {Oestgaard}},\ }\bibfield  {booktitle} {\emph {\bibinfo {booktitle} {{Phys.
  Rep. 242 (1994) 313-332}}},\ }\href {\doibase 10.1016/0370-1573(94)90166-X}
  {\bibfield  {journal} {\bibinfo  {journal} {Phys. Rept.}\ }\textbf {\bibinfo
  {volume} {242}},\ \bibinfo {pages} {313} (\bibinfo {year}
  {1994})}\BibitemShut {NoStop}%
\bibitem [{\citenamefont {{Schertler}}\ \emph {et~al.}(1999)\citenamefont
  {{Schertler}}, \citenamefont {{Leupold}},\ and\ \citenamefont
  {{Schaffner-Bielich}}}]{Schertler1999PRC}%
  \BibitemOpen
  \bibfield  {author} {\bibinfo {author} {\bibfnamefont {K.}~\bibnamefont
  {{Schertler}}}, \bibinfo {author} {\bibfnamefont {S.}~\bibnamefont
  {{Leupold}}}, \ and\ \bibinfo {author} {\bibfnamefont {J.}~\bibnamefont
  {{Schaffner-Bielich}}},\ }\href {\doibase 10.1103/PhysRevC.60.025801}
  {\bibfield  {journal} {\bibinfo  {journal} {Physical Review C}\ }\textbf
  {\bibinfo {volume} {60}},\ \bibinfo {eid} {025801} (\bibinfo {year}
  {1999})},\ \Eprint {http://arxiv.org/abs/astro-ph/9901152}
  {arXiv:astro-ph/9901152 [astro-ph]} \BibitemShut {NoStop}%
\bibitem [{\citenamefont {{Steiner}}\ \emph {et~al.}(2000)\citenamefont
  {{Steiner}}, \citenamefont {{Prakash}},\ and\ \citenamefont
  {{Lattimer}}}]{Steiner2000PLB}%
  \BibitemOpen
  \bibfield  {author} {\bibinfo {author} {\bibfnamefont {A.~W.}\ \bibnamefont
  {{Steiner}}}, \bibinfo {author} {\bibfnamefont {M.}~\bibnamefont
  {{Prakash}}}, \ and\ \bibinfo {author} {\bibfnamefont {J.~M.}\ \bibnamefont
  {{Lattimer}}},\ }\href {\doibase 10.1016/S0370-2693(00)00780-2} {\bibfield
  {journal} {\bibinfo  {journal} {Physics Letters B}\ }\textbf {\bibinfo
  {volume} {486}},\ \bibinfo {pages} {239} (\bibinfo {year} {2000})},\ \Eprint
  {http://arxiv.org/abs/nucl-th/0003066} {arXiv:nucl-th/0003066 [nucl-th]}
  \BibitemShut {NoStop}%
\bibitem [{\citenamefont {{Hanauske}}\ and\ \citenamefont
  {{Greiner}}(2001)}]{Hanauske2001GRG}%
  \BibitemOpen
  \bibfield  {author} {\bibinfo {author} {\bibfnamefont {M.}~\bibnamefont
  {{Hanauske}}}\ and\ \bibinfo {author} {\bibfnamefont {W.}~\bibnamefont
  {{Greiner}}},\ }\href {\doibase 10.1023/A:1010291321139} {\bibfield
  {journal} {\bibinfo  {journal} {General Relativity and Gravitation}\ }\textbf
  {\bibinfo {volume} {33}},\ \bibinfo {pages} {739} (\bibinfo {year}
  {2001})}\BibitemShut {NoStop}%
\bibitem [{\citenamefont {{Mishustin}}\ \emph {et~al.}(2003)\citenamefont
  {{Mishustin}} \emph {et~al.}}]{Mishustin2003PhLB}%
  \BibitemOpen
  \bibfield  {author} {\bibinfo {author} {\bibfnamefont {I.~N.}\ \bibnamefont
  {{Mishustin}}} \emph {et~al.},\ }\href {\doibase
  10.1016/S0370-2693(02)03108-8} {\bibfield  {journal} {\bibinfo  {journal}
  {Physics Letters B}\ }\textbf {\bibinfo {volume} {552}},\ \bibinfo {pages}
  {1} (\bibinfo {year} {2003})},\ \Eprint {http://arxiv.org/abs/hep-ph/0210422}
  {arXiv:hep-ph/0210422 [hep-ph]} \BibitemShut {NoStop}%
\bibitem [{\citenamefont {{Banik}}\ \emph {et~al.}(2004)\citenamefont
  {{Banik}}, \citenamefont {{Hanauske}}, \citenamefont {{Bandyopadhyay}},\ and\
  \citenamefont {{Greiner}}}]{Banik2004PRD}%
  \BibitemOpen
  \bibfield  {author} {\bibinfo {author} {\bibfnamefont {S.}~\bibnamefont
  {{Banik}}}, \bibinfo {author} {\bibfnamefont {M.}~\bibnamefont {{Hanauske}}},
  \bibinfo {author} {\bibfnamefont {D.}~\bibnamefont {{Bandyopadhyay}}}, \ and\
  \bibinfo {author} {\bibfnamefont {W.}~\bibnamefont {{Greiner}}},\ }\href
  {\doibase 10.1103/PhysRevD.70.123004} {\bibfield  {journal} {\bibinfo
  {journal} {Phys. Rev. D}\ }\textbf {\bibinfo {volume} {70}},\ \bibinfo {eid}
  {123004} (\bibinfo {year} {2004})},\ \Eprint
  {http://arxiv.org/abs/astro-ph/0406315} {arXiv:astro-ph/0406315 [astro-ph]}
  \BibitemShut {NoStop}%
\bibitem [{\citenamefont {{Zacchi}}\ \emph {et~al.}(2016)\citenamefont
  {{Zacchi}}, \citenamefont {{Hanauske}},\ and\ \citenamefont
  {{Schaffner-Bielich}}}]{Zacchi2016PRD}%
  \BibitemOpen
  \bibfield  {author} {\bibinfo {author} {\bibfnamefont {A.}~\bibnamefont
  {{Zacchi}}}, \bibinfo {author} {\bibfnamefont {M.}~\bibnamefont
  {{Hanauske}}}, \ and\ \bibinfo {author} {\bibfnamefont {J.}~\bibnamefont
  {{Schaffner-Bielich}}},\ }\href {\doibase 10.1103/PhysRevD.93.065011}
  {\bibfield  {journal} {\bibinfo  {journal} {Phys. Rev. D}\ }\textbf {\bibinfo
  {volume} {93}},\ \bibinfo {eid} {065011} (\bibinfo {year}
  {2016})}\BibitemShut {NoStop}%
\bibitem [{\citenamefont {{Weber}}(2017)}]{WeberBook2017}%
  \BibitemOpen
  \bibfield  {author} {\bibinfo {author} {\bibfnamefont {F.}~\bibnamefont
  {{Weber}}},\ }\href@noop {} {\emph {\bibinfo {title} {Pulsars as
  Astrophysical Laboratories for Nuclear and Particle Physics. Series: Series
  in High Energy Physics, Cosmology and Gravitation, ISBN: 978-0-7503-0332-3.
  Institute of Physics Publishing (Bristol and Philadelphia), Edited by Weber,
  Fridolin}}}\ (\bibinfo {year} {2017})\BibitemShut {NoStop}%
\bibitem [{\citenamefont {Roark}\ \emph {et~al.}(2018)\citenamefont {Roark},
  \citenamefont {Du}, \citenamefont {Constantinou}, \citenamefont {Dexheimer},
  \citenamefont {Steiner},\ and\ \citenamefont {Stone}}]{Roark:2018boj}%
  \BibitemOpen
  \bibfield  {author} {\bibinfo {author} {\bibfnamefont {J.}~\bibnamefont
  {Roark}}, \bibinfo {author} {\bibfnamefont {X.}~\bibnamefont {Du}}, \bibinfo
  {author} {\bibfnamefont {C.}~\bibnamefont {Constantinou}}, \bibinfo {author}
  {\bibfnamefont {V.}~\bibnamefont {Dexheimer}}, \bibinfo {author}
  {\bibfnamefont {A.~W.}\ \bibnamefont {Steiner}}, \ and\ \bibinfo {author}
  {\bibfnamefont {J.~R.}\ \bibnamefont {Stone}},\ }\href@noop {} {\  (\bibinfo
  {year} {2018})},\ \Eprint {http://arxiv.org/abs/1812.08157} {arXiv:1812.08157
  [astro-ph.HE]} \BibitemShut {NoStop}%
\bibitem [{\citenamefont {Demorest}\ \emph {et~al.}(2010)\citenamefont
  {Demorest}, \citenamefont {Pennucci}, \citenamefont {Ransom}, \citenamefont
  {Roberts},\ and\ \citenamefont {Hessels}}]{Demorest:2010bx}%
  \BibitemOpen
  \bibfield  {author} {\bibinfo {author} {\bibfnamefont {P.}~\bibnamefont
  {Demorest}}, \bibinfo {author} {\bibfnamefont {T.}~\bibnamefont {Pennucci}},
  \bibinfo {author} {\bibfnamefont {S.}~\bibnamefont {Ransom}}, \bibinfo
  {author} {\bibfnamefont {M.}~\bibnamefont {Roberts}}, \ and\ \bibinfo
  {author} {\bibfnamefont {J.}~\bibnamefont {Hessels}},\ }\href {\doibase
  10.1038/nature09466} {\bibfield  {journal} {\bibinfo  {journal} {Nature}\
  }\textbf {\bibinfo {volume} {467}},\ \bibinfo {pages} {1081} (\bibinfo {year}
  {2010})},\ \Eprint {http://arxiv.org/abs/1010.5788} {arXiv:1010.5788
  [astro-ph.HE]} \BibitemShut {NoStop}%
\bibitem [{\citenamefont {Antoniadis}\ \emph {et~al.}(2013)\citenamefont
  {Antoniadis} \emph {et~al.}}]{Antoniadis:2013pzd}%
  \BibitemOpen
  \bibfield  {author} {\bibinfo {author} {\bibfnamefont {J.}~\bibnamefont
  {Antoniadis}} \emph {et~al.},\ }\href {\doibase 10.1126/science.1233232}
  {\bibfield  {journal} {\bibinfo  {journal} {Science}\ }\textbf {\bibinfo
  {volume} {340}},\ \bibinfo {pages} {6131} (\bibinfo {year} {2013})},\ \Eprint
  {http://arxiv.org/abs/1304.6875} {arXiv:1304.6875 [astro-ph.HE]} \BibitemShut
  {NoStop}%
\bibitem [{\citenamefont {Arzoumanian}\ \emph {et~al.}(2018)\citenamefont
  {Arzoumanian} \emph {et~al.}}]{Arzoumanian:2017puf}%
  \BibitemOpen
  \bibfield  {author} {\bibinfo {author} {\bibfnamefont {Z.}~\bibnamefont
  {Arzoumanian}} \emph {et~al.} (\bibinfo {collaboration} {NANOGrav}),\ }\href
  {\doibase 10.3847/1538-4365/aab5b0} {\bibfield  {journal} {\bibinfo
  {journal} {Astrophys. J. Suppl.}\ }\textbf {\bibinfo {volume} {235}},\
  \bibinfo {pages} {37} (\bibinfo {year} {2018})},\ \Eprint
  {http://arxiv.org/abs/1801.01837} {arXiv:1801.01837 [astro-ph.HE]}
  \BibitemShut {NoStop}%
\bibitem [{\citenamefont {Verbiest}\ \emph {et~al.}(2008)\citenamefont
  {Verbiest} \emph {et~al.}}]{Verbiest:2008gy}%
  \BibitemOpen
  \bibfield  {author} {\bibinfo {author} {\bibfnamefont {J.~P.~W.}\
  \bibnamefont {Verbiest}} \emph {et~al.},\ }\href {\doibase 10.1086/529576}
  {\bibfield  {journal} {\bibinfo  {journal} {Astrophys. J.}\ }\textbf
  {\bibinfo {volume} {679}},\ \bibinfo {pages} {675} (\bibinfo {year}
  {2008})},\ \Eprint {http://arxiv.org/abs/0801.2589} {arXiv:0801.2589
  [astro-ph]} \BibitemShut {NoStop}%
\bibitem [{\citenamefont {\"Ozel}\ \emph {et~al.}(2010)\citenamefont {\"Ozel},
  \citenamefont {Baym},\ and\ \citenamefont {Guver}}]{Ozel:2010fw}%
  \BibitemOpen
  \bibfield  {author} {\bibinfo {author} {\bibfnamefont {F.}~\bibnamefont
  {\"Ozel}}, \bibinfo {author} {\bibfnamefont {G.}~\bibnamefont {Baym}}, \ and\
  \bibinfo {author} {\bibfnamefont {T.}~\bibnamefont {Guver}},\ }\href
  {\doibase 10.1103/PhysRevD.82.101301} {\bibfield  {journal} {\bibinfo
  {journal} {Phys. Rev.}\ }\textbf {\bibinfo {volume} {D82}},\ \bibinfo {pages}
  {101301} (\bibinfo {year} {2010})},\ \Eprint {http://arxiv.org/abs/1002.3153}
  {arXiv:1002.3153 [astro-ph.HE]} \BibitemShut {NoStop}%
\bibitem [{\citenamefont {Suleimanov}\ \emph {et~al.}(2011)\citenamefont
  {Suleimanov}, \citenamefont {Poutanen}, \citenamefont {Revnivtsev},\ and\
  \citenamefont {Werner}}]{Suleimanov:2010th}%
  \BibitemOpen
  \bibfield  {author} {\bibinfo {author} {\bibfnamefont {V.}~\bibnamefont
  {Suleimanov}}, \bibinfo {author} {\bibfnamefont {J.}~\bibnamefont
  {Poutanen}}, \bibinfo {author} {\bibfnamefont {M.}~\bibnamefont
  {Revnivtsev}}, \ and\ \bibinfo {author} {\bibfnamefont {K.}~\bibnamefont
  {Werner}},\ }\href {\doibase 10.1088/0004-637X/742/2/122} {\bibfield
  {journal} {\bibinfo  {journal} {Astrophys. J.}\ }\textbf {\bibinfo {volume}
  {742}},\ \bibinfo {pages} {122} (\bibinfo {year} {2011})},\ \Eprint
  {http://arxiv.org/abs/1004.4871} {arXiv:1004.4871 [astro-ph.HE]} \BibitemShut
  {NoStop}%
\bibitem [{\citenamefont {Lattimer}\ and\ \citenamefont
  {Lim}(2013)}]{Lattimer:2012xj}%
  \BibitemOpen
  \bibfield  {author} {\bibinfo {author} {\bibfnamefont {J.~M.}\ \bibnamefont
  {Lattimer}}\ and\ \bibinfo {author} {\bibfnamefont {Y.}~\bibnamefont {Lim}},\
  }\href {\doibase 10.1088/0004-637X/771/1/51} {\bibfield  {journal} {\bibinfo
  {journal} {Astrophys. J.}\ }\textbf {\bibinfo {volume} {771}},\ \bibinfo
  {pages} {51} (\bibinfo {year} {2013})},\ \Eprint
  {http://arxiv.org/abs/1203.4286} {arXiv:1203.4286 [nucl-th]} \BibitemShut
  {NoStop}%
\bibitem [{\citenamefont {Steiner}\ \emph {et~al.}(2013)\citenamefont
  {Steiner}, \citenamefont {Lattimer},\ and\ \citenamefont
  {Brown}}]{Steiner:2012xt}%
  \BibitemOpen
  \bibfield  {author} {\bibinfo {author} {\bibfnamefont {A.~W.}\ \bibnamefont
  {Steiner}}, \bibinfo {author} {\bibfnamefont {J.~M.}\ \bibnamefont
  {Lattimer}}, \ and\ \bibinfo {author} {\bibfnamefont {E.~F.}\ \bibnamefont
  {Brown}},\ }\href {\doibase 10.1088/2041-8205/765/1/L5} {\bibfield  {journal}
  {\bibinfo  {journal} {Astrophys.J.}\ }\textbf {\bibinfo {volume} {765}},\
  \bibinfo {pages} {L5} (\bibinfo {year} {2013})},\ \Eprint
  {http://arxiv.org/abs/1205.6871} {arXiv:1205.6871 [nucl-th]} \BibitemShut
  {NoStop}%
\bibitem [{\citenamefont {Bogdanov}(2013)}]{Bogdanov:2012md}%
  \BibitemOpen
  \bibfield  {author} {\bibinfo {author} {\bibfnamefont {S.}~\bibnamefont
  {Bogdanov}},\ }\href {\doibase 10.1088/0004-637X/762/2/96} {\bibfield
  {journal} {\bibinfo  {journal} {Astrophys.J.}\ }\textbf {\bibinfo {volume}
  {762}},\ \bibinfo {pages} {96} (\bibinfo {year} {2013})},\ \Eprint
  {http://arxiv.org/abs/1211.6113} {arXiv:1211.6113 [astro-ph.HE]} \BibitemShut
  {NoStop}%
\bibitem [{\citenamefont {Guver}\ and\ \citenamefont
  {\"Ozel}(2013)}]{Guver:2013xa}%
  \BibitemOpen
  \bibfield  {author} {\bibinfo {author} {\bibfnamefont {T.}~\bibnamefont
  {Guver}}\ and\ \bibinfo {author} {\bibfnamefont {F.}~\bibnamefont {\"Ozel}},\
  }\href {\doibase 10.1088/2041-8205/765/1/L1} {\bibfield  {journal} {\bibinfo
  {journal} {Astrophys. J.}\ }\textbf {\bibinfo {volume} {765}},\ \bibinfo
  {pages} {L1} (\bibinfo {year} {2013})},\ \Eprint
  {http://arxiv.org/abs/1301.0831} {arXiv:1301.0831 [astro-ph.HE]} \BibitemShut
  {NoStop}%
\bibitem [{\citenamefont {Guillot}\ \emph {et~al.}(2013)\citenamefont
  {Guillot}, \citenamefont {Servillat}, \citenamefont {Webb},\ and\
  \citenamefont {Rutledge}}]{Guillot:2013wu}%
  \BibitemOpen
  \bibfield  {author} {\bibinfo {author} {\bibfnamefont {S.}~\bibnamefont
  {Guillot}}, \bibinfo {author} {\bibfnamefont {M.}~\bibnamefont {Servillat}},
  \bibinfo {author} {\bibfnamefont {N.~A.}\ \bibnamefont {Webb}}, \ and\
  \bibinfo {author} {\bibfnamefont {R.~E.}\ \bibnamefont {Rutledge}},\ }\href
  {\doibase 10.1088/0004-637X/772/1/7} {\bibfield  {journal} {\bibinfo
  {journal} {Astrophys. J.}\ }\textbf {\bibinfo {volume} {772}},\ \bibinfo
  {pages} {7} (\bibinfo {year} {2013})},\ \Eprint
  {http://arxiv.org/abs/1302.0023} {arXiv:1302.0023 [astro-ph.HE]} \BibitemShut
  {NoStop}%
\bibitem [{\citenamefont {Lattimer}\ and\ \citenamefont
  {Steiner}(2014)}]{Lattimer:2013hma}%
  \BibitemOpen
  \bibfield  {author} {\bibinfo {author} {\bibfnamefont {J.~M.}\ \bibnamefont
  {Lattimer}}\ and\ \bibinfo {author} {\bibfnamefont {A.~W.}\ \bibnamefont
  {Steiner}},\ }\href {\doibase 10.1088/0004-637X/784/2/123} {\bibfield
  {journal} {\bibinfo  {journal} {Astrophys. J.}\ }\textbf {\bibinfo {volume}
  {784}},\ \bibinfo {pages} {123} (\bibinfo {year} {2014})},\ \Eprint
  {http://arxiv.org/abs/1305.3242} {arXiv:1305.3242 [astro-ph.HE]} \BibitemShut
  {NoStop}%
\bibitem [{\citenamefont {Poutanen}\ \emph {et~al.}(2014)\citenamefont
  {Poutanen} \emph {et~al.}}]{Poutanen:2014xqa}%
  \BibitemOpen
  \bibfield  {author} {\bibinfo {author} {\bibfnamefont {J.}~\bibnamefont
  {Poutanen}} \emph {et~al.},\ }\href {\doibase 10.1093/mnras/stu1139}
  {\bibfield  {journal} {\bibinfo  {journal} {Mon. Not. Roy. Astron. Soc.}\
  }\textbf {\bibinfo {volume} {442}},\ \bibinfo {pages} {3777} (\bibinfo {year}
  {2014})},\ \Eprint {http://arxiv.org/abs/1405.2663} {arXiv:1405.2663
  [astro-ph.HE]} \BibitemShut {NoStop}%
\bibitem [{\citenamefont {Heinke}\ \emph {et~al.}(2014)\citenamefont {Heinke}
  \emph {et~al.}}]{Heinke:2014xaa}%
  \BibitemOpen
  \bibfield  {author} {\bibinfo {author} {\bibfnamefont {C.~O.}\ \bibnamefont
  {Heinke}} \emph {et~al.},\ }\href {\doibase 10.1093/mnras/stu1449} {\bibfield
   {journal} {\bibinfo  {journal} {Mon. Not. Roy. Astron. Soc.}\ }\textbf
  {\bibinfo {volume} {444}},\ \bibinfo {pages} {443} (\bibinfo {year}
  {2014})},\ \Eprint {http://arxiv.org/abs/1406.1497} {arXiv:1406.1497
  [astro-ph.HE]} \BibitemShut {NoStop}%
\bibitem [{\citenamefont {Guillot}\ and\ \citenamefont
  {Rutledge}(2014)}]{Guillot:2014lla}%
  \BibitemOpen
  \bibfield  {author} {\bibinfo {author} {\bibfnamefont {S.}~\bibnamefont
  {Guillot}}\ and\ \bibinfo {author} {\bibfnamefont {R.~E.}\ \bibnamefont
  {Rutledge}},\ }\href {\doibase 10.1088/2041-8205/796/1/L3} {\bibfield
  {journal} {\bibinfo  {journal} {Astrophys. J.}\ }\textbf {\bibinfo {volume}
  {796}},\ \bibinfo {pages} {L3} (\bibinfo {year} {2014})},\ \Eprint
  {http://arxiv.org/abs/1409.4306} {arXiv:1409.4306 [astro-ph.HE]} \BibitemShut
  {NoStop}%
\bibitem [{\citenamefont {\"Ozel}\ \emph {et~al.}(2016)\citenamefont {\"Ozel}
  \emph {et~al.}}]{Ozel:2015fia}%
  \BibitemOpen
  \bibfield  {author} {\bibinfo {author} {\bibfnamefont {F.}~\bibnamefont
  {\"Ozel}} \emph {et~al.},\ }\href {\doibase 10.3847/0004-637X/820/1/28}
  {\bibfield  {journal} {\bibinfo  {journal} {Astrophys. J.}\ }\textbf
  {\bibinfo {volume} {820}},\ \bibinfo {pages} {28} (\bibinfo {year} {2016})},\
  \Eprint {http://arxiv.org/abs/1505.05155} {arXiv:1505.05155 [astro-ph.HE]}
  \BibitemShut {NoStop}%
\bibitem [{\citenamefont {\"Ozel}\ and\ \citenamefont
  {Psaltis}(2015)}]{Ozel:2015gia}%
  \BibitemOpen
  \bibfield  {author} {\bibinfo {author} {\bibfnamefont {F.}~\bibnamefont
  {\"Ozel}}\ and\ \bibinfo {author} {\bibfnamefont {D.}~\bibnamefont
  {Psaltis}},\ }\href {\doibase 10.1088/0004-637X/810/2/135} {\bibfield
  {journal} {\bibinfo  {journal} {Astrophys. J.}\ }\textbf {\bibinfo {volume}
  {810}},\ \bibinfo {pages} {135} (\bibinfo {year} {2015})},\ \Eprint
  {http://arxiv.org/abs/1505.05156} {arXiv:1505.05156 [astro-ph.HE]}
  \BibitemShut {NoStop}%
\bibitem [{\citenamefont {Lattimer}\ and\ \citenamefont
  {Prakash}(2016)}]{Lattimer:2015nhk}%
  \BibitemOpen
  \bibfield  {author} {\bibinfo {author} {\bibfnamefont {J.~M.}\ \bibnamefont
  {Lattimer}}\ and\ \bibinfo {author} {\bibfnamefont {M.}~\bibnamefont
  {Prakash}},\ }\href {\doibase 10.1016/j.physrep.2015.12.005} {\bibfield
  {journal} {\bibinfo  {journal} {Phys. Rept.}\ }\textbf {\bibinfo {volume}
  {621}},\ \bibinfo {pages} {127} (\bibinfo {year} {2016})},\ \Eprint
  {http://arxiv.org/abs/1512.07820} {arXiv:1512.07820 [astro-ph.SR]}
  \BibitemShut {NoStop}%
\bibitem [{\citenamefont {\"Ozel}\ and\ \citenamefont
  {Freire}(2016)}]{Ozel:2016oaf}%
  \BibitemOpen
  \bibfield  {author} {\bibinfo {author} {\bibfnamefont {F.}~\bibnamefont
  {\"Ozel}}\ and\ \bibinfo {author} {\bibfnamefont {P.}~\bibnamefont
  {Freire}},\ }\href {\doibase 10.1146/annurev-astro-081915-023322} {\bibfield
  {journal} {\bibinfo  {journal} {Annu. Rev. Astron. Astrophys.}\ }\textbf
  {\bibinfo {volume} {54}},\ \bibinfo {pages} {401} (\bibinfo {year} {2016})},\
  \Eprint {http://arxiv.org/abs/1603.02698} {arXiv:1603.02698 [astro-ph.HE]}
  \BibitemShut {NoStop}%
\bibitem [{\citenamefont {Steiner}\ \emph {et~al.}(2018)\citenamefont {Steiner}
  \emph {et~al.}}]{Steiner:2017vmg}%
  \BibitemOpen
  \bibfield  {author} {\bibinfo {author} {\bibfnamefont {A.~W.}\ \bibnamefont
  {Steiner}} \emph {et~al.},\ }\href {\doibase 10.1093/mnras/sty215} {\bibfield
   {journal} {\bibinfo  {journal} {Mon. Not. Roy. Astron. Soc.}\ }\textbf
  {\bibinfo {volume} {476}},\ \bibinfo {pages} {421} (\bibinfo {year}
  {2018})},\ \Eprint {http://arxiv.org/abs/1709.05013} {arXiv:1709.05013
  [astro-ph.HE]} \BibitemShut {NoStop}%
\bibitem [{\citenamefont {{Arzoumanian}}\ \emph {et~al.}(2014)\citenamefont
  {{Arzoumanian}} \emph {et~al.}}]{2014SPIE.9144E..20A}%
  \BibitemOpen
  \bibfield  {author} {\bibinfo {author} {\bibfnamefont {Z.}~\bibnamefont
  {{Arzoumanian}}} \emph {et~al.},\ }in\ \href {\doibase 10.1117/12.2056811}
  {\emph {\bibinfo {booktitle} {Space Telescopes and Instrumentation 2014:
  Ultraviolet to Gamma Ray}}},\ \bibinfo {series} {Proceedings of the SPIE},
  Vol.\ \bibinfo {volume} {9144}\ (\bibinfo {year} {2014})\ p.\ \bibinfo
  {pages} {914420}\BibitemShut {NoStop}%
\bibitem [{\citenamefont {Watts}\ \emph {et~al.}(2019)\citenamefont {Watts}
  \emph {et~al.}}]{Watts:2018iom}%
  \BibitemOpen
  \bibfield  {author} {\bibinfo {author} {\bibfnamefont {A.~L.}\ \bibnamefont
  {Watts}} \emph {et~al.},\ }\href {\doibase 10.1007/s11433-017-9188-4}
  {\bibfield  {journal} {\bibinfo  {journal} {Sci. China Phys. Mech. Astron.}\
  }\textbf {\bibinfo {volume} {62}},\ \bibinfo {pages} {29503} (\bibinfo {year}
  {2019})}\BibitemShut {NoStop}%
\bibitem [{\citenamefont {Margalit}\ and\ \citenamefont
  {Metzger}(2017)}]{Margalit:2017dij}%
  \BibitemOpen
  \bibfield  {author} {\bibinfo {author} {\bibfnamefont {B.}~\bibnamefont
  {Margalit}}\ and\ \bibinfo {author} {\bibfnamefont {B.~D.}\ \bibnamefont
  {Metzger}},\ }\href {\doibase 10.3847/2041-8213/aa991c} {\bibfield  {journal}
  {\bibinfo  {journal} {Astrophys. J.}\ }\textbf {\bibinfo {volume} {850}},\
  \bibinfo {pages} {L19} (\bibinfo {year} {2017})},\ \Eprint
  {http://arxiv.org/abs/1710.05938} {arXiv:1710.05938 [astro-ph.HE]}
  \BibitemShut {NoStop}%
\bibitem [{\citenamefont {Bauswein}\ \emph {et~al.}(2017)\citenamefont
  {Bauswein}, \citenamefont {Just}, \citenamefont {Janka},\ and\ \citenamefont
  {Stergioulas}}]{Bauswein:2017vtn}%
  \BibitemOpen
  \bibfield  {author} {\bibinfo {author} {\bibfnamefont {A.}~\bibnamefont
  {Bauswein}}, \bibinfo {author} {\bibfnamefont {O.}~\bibnamefont {Just}},
  \bibinfo {author} {\bibfnamefont {H.-T.}\ \bibnamefont {Janka}}, \ and\
  \bibinfo {author} {\bibfnamefont {N.}~\bibnamefont {Stergioulas}},\ }\href
  {\doibase 10.3847/2041-8213/aa9994} {\bibfield  {journal} {\bibinfo
  {journal} {Astrophys. J.}\ }\textbf {\bibinfo {volume} {850}},\ \bibinfo
  {pages} {L34} (\bibinfo {year} {2017})},\ \Eprint
  {http://arxiv.org/abs/1710.06843} {arXiv:1710.06843 [astro-ph.HE]}
  \BibitemShut {NoStop}%
\bibitem [{\citenamefont {Shibata}\ \emph {et~al.}(2017)\citenamefont {Shibata}
  \emph {et~al.}}]{Shibata:2017xdx}%
  \BibitemOpen
  \bibfield  {author} {\bibinfo {author} {\bibfnamefont {M.}~\bibnamefont
  {Shibata}} \emph {et~al.},\ }\href {\doibase 10.1103/PhysRevD.96.123012}
  {\bibfield  {journal} {\bibinfo  {journal} {Phys. Rev.}\ }\textbf {\bibinfo
  {volume} {D96}},\ \bibinfo {pages} {123012} (\bibinfo {year} {2017})},\
  \Eprint {http://arxiv.org/abs/1710.07579} {arXiv:1710.07579 [astro-ph.HE]}
  \BibitemShut {NoStop}%
\bibitem [{\citenamefont {Rezzolla}\ \emph {et~al.}(2018)\citenamefont
  {Rezzolla}, \citenamefont {Most},\ and\ \citenamefont
  {Weih}}]{Rezzolla:2017aly}%
  \BibitemOpen
  \bibfield  {author} {\bibinfo {author} {\bibfnamefont {L.}~\bibnamefont
  {Rezzolla}}, \bibinfo {author} {\bibfnamefont {E.~R.}\ \bibnamefont {Most}},
  \ and\ \bibinfo {author} {\bibfnamefont {L.~R.}\ \bibnamefont {Weih}},\
  }\href {\doibase 10.3847/2041-8213/aaa401} {\bibfield  {journal} {\bibinfo
  {journal} {Astrophys. J.}\ }\textbf {\bibinfo {volume} {852}},\ \bibinfo
  {pages} {L25} (\bibinfo {year} {2018})},\ \bibinfo {note} {[Astrophys. J.
  Lett.852,L25(2018)]},\ \Eprint {http://arxiv.org/abs/1711.00314}
  {arXiv:1711.00314 [astro-ph.HE]} \BibitemShut {NoStop}%
\bibitem [{\citenamefont {Ruiz}\ \emph {et~al.}(2018)\citenamefont {Ruiz},
  \citenamefont {Shapiro},\ and\ \citenamefont {Tsokaros}}]{Ruiz:2017due}%
  \BibitemOpen
  \bibfield  {author} {\bibinfo {author} {\bibfnamefont {M.}~\bibnamefont
  {Ruiz}}, \bibinfo {author} {\bibfnamefont {S.~L.}\ \bibnamefont {Shapiro}}, \
  and\ \bibinfo {author} {\bibfnamefont {A.}~\bibnamefont {Tsokaros}},\ }\href
  {\doibase 10.1103/PhysRevD.97.021501} {\bibfield  {journal} {\bibinfo
  {journal} {Phys. Rev.}\ }\textbf {\bibinfo {volume} {D97}},\ \bibinfo {pages}
  {021501} (\bibinfo {year} {2018})},\ \Eprint
  {http://arxiv.org/abs/1711.00473} {arXiv:1711.00473 [astro-ph.HE]}
  \BibitemShut {NoStop}%
\bibitem [{\citenamefont {Annala}\ \emph {et~al.}(2018)\citenamefont {Annala},
  \citenamefont {Gorda}, \citenamefont {Kurkela},\ and\ \citenamefont
  {Vuorinen}}]{Annala:2017llu}%
  \BibitemOpen
  \bibfield  {author} {\bibinfo {author} {\bibfnamefont {E.}~\bibnamefont
  {Annala}}, \bibinfo {author} {\bibfnamefont {T.}~\bibnamefont {Gorda}},
  \bibinfo {author} {\bibfnamefont {A.}~\bibnamefont {Kurkela}}, \ and\
  \bibinfo {author} {\bibfnamefont {A.}~\bibnamefont {Vuorinen}},\ }\href
  {\doibase 10.1103/PhysRevLett.120.172703} {\bibfield  {journal} {\bibinfo
  {journal} {Phys. Rev. Lett.}\ }\textbf {\bibinfo {volume} {120}},\ \bibinfo
  {pages} {172703} (\bibinfo {year} {2018})},\ \Eprint
  {http://arxiv.org/abs/1711.02644} {arXiv:1711.02644 [astro-ph.HE]}
  \BibitemShut {NoStop}%
\bibitem [{\citenamefont {Abbott}\ \emph {et~al.}(2017)\citenamefont {Abbott}
  \emph {et~al.}}]{TheLIGOScientific:2017qsa}%
  \BibitemOpen
  \bibfield  {author} {\bibinfo {author} {\bibfnamefont {B.}~\bibnamefont
  {Abbott}} \emph {et~al.} (\bibinfo {collaboration} {Virgo, LIGO
  Scientific}),\ }\href {\doibase 10.1103/PhysRevLett.119.161101} {\bibfield
  {journal} {\bibinfo  {journal} {Phys. Rev. Lett.}\ }\textbf {\bibinfo
  {volume} {119}},\ \bibinfo {pages} {161101} (\bibinfo {year} {2017})},\
  \Eprint {http://arxiv.org/abs/1710.05832} {arXiv:1710.05832 [gr-qc]}
  \BibitemShut {NoStop}%
\bibitem [{\citenamefont {Abbott}\ \emph {et~al.}(2019)\citenamefont {Abbott}
  \emph {et~al.}}]{Abbott:2018wiz}%
  \BibitemOpen
  \bibfield  {author} {\bibinfo {author} {\bibfnamefont {B.~P.}\ \bibnamefont
  {Abbott}} \emph {et~al.} (\bibinfo {collaboration} {LIGO Scientific,
  Virgo}),\ }\href {\doibase 10.1103/PhysRevX.9.011001} {\bibfield  {journal}
  {\bibinfo  {journal} {Phys. Rev.}\ }\textbf {\bibinfo {volume} {X9}},\
  \bibinfo {pages} {011001} (\bibinfo {year} {2019})},\ \Eprint
  {http://arxiv.org/abs/1805.11579} {arXiv:1805.11579 [gr-qc]} \BibitemShut
  {NoStop}%
\bibitem [{\citenamefont {Kumar}\ \emph {et~al.}(2018)\citenamefont {Kumar},
  \citenamefont {Agrawal},\ and\ \citenamefont {Patra}}]{Kumar:2017wqp}%
  \BibitemOpen
  \bibfield  {author} {\bibinfo {author} {\bibfnamefont {B.}~\bibnamefont
  {Kumar}}, \bibinfo {author} {\bibfnamefont {B.~K.}\ \bibnamefont {Agrawal}},
  \ and\ \bibinfo {author} {\bibfnamefont {S.~K.}\ \bibnamefont {Patra}},\
  }\href {\doibase 10.1103/PhysRevC.97.045806} {\bibfield  {journal} {\bibinfo
  {journal} {Phys. Rev.}\ }\textbf {\bibinfo {volume} {C97}},\ \bibinfo {pages}
  {045806} (\bibinfo {year} {2018})},\ \Eprint
  {http://arxiv.org/abs/1711.04940} {arXiv:1711.04940 [nucl-th]} \BibitemShut
  {NoStop}%
\bibitem [{\citenamefont {Fattoyev}\ \emph {et~al.}(2018)\citenamefont
  {Fattoyev}, \citenamefont {Piekarewicz},\ and\ \citenamefont
  {Horowitz}}]{Fattoyev:2017jql}%
  \BibitemOpen
  \bibfield  {author} {\bibinfo {author} {\bibfnamefont {F.~J.}\ \bibnamefont
  {Fattoyev}}, \bibinfo {author} {\bibfnamefont {J.}~\bibnamefont
  {Piekarewicz}}, \ and\ \bibinfo {author} {\bibfnamefont {C.~J.}\ \bibnamefont
  {Horowitz}},\ }\href {\doibase 10.1103/PhysRevLett.120.172702} {\bibfield
  {journal} {\bibinfo  {journal} {Phys. Rev. Lett.}\ }\textbf {\bibinfo
  {volume} {120}},\ \bibinfo {pages} {172702} (\bibinfo {year} {2018})},\
  \Eprint {http://arxiv.org/abs/1711.06615} {arXiv:1711.06615 [nucl-th]}
  \BibitemShut {NoStop}%
\bibitem [{\citenamefont {Most}\ \emph
  {et~al.}(2018{\natexlab{a}})\citenamefont {Most}, \citenamefont {Weih},
  \citenamefont {Rezzolla},\ and\ \citenamefont
  {Schaffner-Bielich}}]{Most:2018hfd}%
  \BibitemOpen
  \bibfield  {author} {\bibinfo {author} {\bibfnamefont {E.~R.}\ \bibnamefont
  {Most}}, \bibinfo {author} {\bibfnamefont {L.~R.}\ \bibnamefont {Weih}},
  \bibinfo {author} {\bibfnamefont {L.}~\bibnamefont {Rezzolla}}, \ and\
  \bibinfo {author} {\bibfnamefont {J.}~\bibnamefont {Schaffner-Bielich}},\
  }\href {\doibase 10.1103/PhysRevLett.120.261103} {\bibfield  {journal}
  {\bibinfo  {journal} {Phys. Rev. Lett.}\ }\textbf {\bibinfo {volume} {120}},\
  \bibinfo {pages} {261103} (\bibinfo {year} {2018}{\natexlab{a}})},\ \Eprint
  {http://arxiv.org/abs/1803.00549} {arXiv:1803.00549 [gr-qc]} \BibitemShut
  {NoStop}%
\bibitem [{\citenamefont {Lim}\ and\ \citenamefont {Holt}(2018)}]{Lim:2018bkq}%
  \BibitemOpen
  \bibfield  {author} {\bibinfo {author} {\bibfnamefont {Y.}~\bibnamefont
  {Lim}}\ and\ \bibinfo {author} {\bibfnamefont {J.~W.}\ \bibnamefont {Holt}},\
  }\href {\doibase 10.1103/PhysRevLett.121.062701} {\bibfield  {journal}
  {\bibinfo  {journal} {Phys. Rev. Lett.}\ }\textbf {\bibinfo {volume} {121}},\
  \bibinfo {pages} {062701} (\bibinfo {year} {2018})},\ \Eprint
  {http://arxiv.org/abs/1803.02803} {arXiv:1803.02803 [nucl-th]} \BibitemShut
  {NoStop}%
\bibitem [{\citenamefont {Raithel}\ \emph {et~al.}(2018)\citenamefont
  {Raithel}, \citenamefont {{\"O}zel},\ and\ \citenamefont
  {Psaltis}}]{Raithel:2018ncd}%
  \BibitemOpen
  \bibfield  {author} {\bibinfo {author} {\bibfnamefont {C.}~\bibnamefont
  {Raithel}}, \bibinfo {author} {\bibfnamefont {F.}~\bibnamefont {{\"O}zel}}, \
  and\ \bibinfo {author} {\bibfnamefont {D.}~\bibnamefont {Psaltis}},\ }\href
  {\doibase 10.3847/2041-8213/aabcbf} {\bibfield  {journal} {\bibinfo
  {journal} {Astrophys. J.}\ }\textbf {\bibinfo {volume} {857}},\ \bibinfo
  {pages} {L23} (\bibinfo {year} {2018})},\ \Eprint
  {http://arxiv.org/abs/1803.07687} {arXiv:1803.07687 [astro-ph.HE]}
  \BibitemShut {NoStop}%
\bibitem [{\citenamefont {Burgio}\ \emph {et~al.}(2018)\citenamefont {Burgio},
  \citenamefont {Drago}, \citenamefont {Pagliara}, \citenamefont {Schulze},\
  and\ \citenamefont {Wei}}]{Burgio:2018yix}%
  \BibitemOpen
  \bibfield  {author} {\bibinfo {author} {\bibfnamefont {G.~F.}\ \bibnamefont
  {Burgio}}, \bibinfo {author} {\bibfnamefont {A.}~\bibnamefont {Drago}},
  \bibinfo {author} {\bibfnamefont {G.}~\bibnamefont {Pagliara}}, \bibinfo
  {author} {\bibfnamefont {H.~J.}\ \bibnamefont {Schulze}}, \ and\ \bibinfo
  {author} {\bibfnamefont {J.~B.}\ \bibnamefont {Wei}},\ }\href {\doibase
  10.3847/1538-4357/aac6ee} {\bibfield  {journal} {\bibinfo  {journal}
  {Astrophys. J.}\ }\textbf {\bibinfo {volume} {860}},\ \bibinfo {pages} {139}
  (\bibinfo {year} {2018})},\ \Eprint {http://arxiv.org/abs/1803.09696}
  {arXiv:1803.09696 [astro-ph.HE]} \BibitemShut {NoStop}%
\bibitem [{\citenamefont {Tews}\ \emph
  {et~al.}(2018{\natexlab{a}})\citenamefont {Tews}, \citenamefont {Margueron},\
  and\ \citenamefont {Reddy}}]{Tews:2018chv}%
  \BibitemOpen
  \bibfield  {author} {\bibinfo {author} {\bibfnamefont {I.}~\bibnamefont
  {Tews}}, \bibinfo {author} {\bibfnamefont {J.}~\bibnamefont {Margueron}}, \
  and\ \bibinfo {author} {\bibfnamefont {S.}~\bibnamefont {Reddy}},\ }\href
  {\doibase 10.1103/PhysRevC.98.045804} {\bibfield  {journal} {\bibinfo
  {journal} {Phys. Rev.}\ }\textbf {\bibinfo {volume} {C98}},\ \bibinfo {pages}
  {045804} (\bibinfo {year} {2018}{\natexlab{a}})},\ \Eprint
  {http://arxiv.org/abs/1804.02783} {arXiv:1804.02783 [nucl-th]} \BibitemShut
  {NoStop}%
\bibitem [{\citenamefont {De}\ \emph {et~al.}(2018)\citenamefont {De} \emph
  {et~al.}}]{De:2018uhw}%
  \BibitemOpen
  \bibfield  {author} {\bibinfo {author} {\bibfnamefont {S.}~\bibnamefont {De}}
  \emph {et~al.},\ }\href {\doibase 10.1103/PhysRevLett.121.091102} {\bibfield
  {journal} {\bibinfo  {journal} {Phys. Rev. Lett.}\ }\textbf {\bibinfo
  {volume} {121}},\ \bibinfo {pages} {091102} (\bibinfo {year} {2018})},\
  \Eprint {http://arxiv.org/abs/1804.08583} {arXiv:1804.08583 [astro-ph.HE]}
  \BibitemShut {NoStop}%
\bibitem [{\citenamefont {Abbott}\ \emph {et~al.}(2018)\citenamefont {Abbott}
  \emph {et~al.}}]{Abbott:2018exr}%
  \BibitemOpen
  \bibfield  {author} {\bibinfo {author} {\bibfnamefont {B.~P.}\ \bibnamefont
  {Abbott}} \emph {et~al.} (\bibinfo {collaboration} {LIGO Scientific,
  Virgo}),\ }\href {\doibase 10.1103/PhysRevLett.121.161101} {\bibfield
  {journal} {\bibinfo  {journal} {Phys. Rev. Lett.}\ }\textbf {\bibinfo
  {volume} {121}},\ \bibinfo {pages} {161101} (\bibinfo {year} {2018})},\
  \Eprint {http://arxiv.org/abs/1805.11581} {arXiv:1805.11581 [gr-qc]}
  \BibitemShut {NoStop}%
\bibitem [{\citenamefont {Malik}\ \emph {et~al.}(2018)\citenamefont {Malik},
  \citenamefont {Alam}, \citenamefont {Fortin}, \citenamefont {Providência},
  \citenamefont {Agrawal}, \citenamefont {Jha}, \citenamefont {Kumar},\ and\
  \citenamefont {Patra}}]{Malik:2018zcf}%
  \BibitemOpen
  \bibfield  {author} {\bibinfo {author} {\bibfnamefont {T.}~\bibnamefont
  {Malik}}, \bibinfo {author} {\bibfnamefont {N.}~\bibnamefont {Alam}},
  \bibinfo {author} {\bibfnamefont {M.}~\bibnamefont {Fortin}}, \bibinfo
  {author} {\bibfnamefont {C.}~\bibnamefont {Providência}}, \bibinfo {author}
  {\bibfnamefont {B.~K.}\ \bibnamefont {Agrawal}}, \bibinfo {author}
  {\bibfnamefont {T.~K.}\ \bibnamefont {Jha}}, \bibinfo {author} {\bibfnamefont
  {B.}~\bibnamefont {Kumar}}, \ and\ \bibinfo {author} {\bibfnamefont {S.~K.}\
  \bibnamefont {Patra}},\ }\href {\doibase 10.1103/PhysRevC.98.035804}
  {\bibfield  {journal} {\bibinfo  {journal} {Phys. Rev.}\ }\textbf {\bibinfo
  {volume} {C98}},\ \bibinfo {pages} {035804} (\bibinfo {year} {2018})},\
  \Eprint {http://arxiv.org/abs/1805.11963} {arXiv:1805.11963 [nucl-th]}
  \BibitemShut {NoStop}%
\bibitem [{\citenamefont {Hinderer}\ \emph {et~al.}(2010)\citenamefont
  {Hinderer}, \citenamefont {Lackey}, \citenamefont {Lang},\ and\ \citenamefont
  {Read}}]{Hinderer:2009ca}%
  \BibitemOpen
  \bibfield  {author} {\bibinfo {author} {\bibfnamefont {T.}~\bibnamefont
  {Hinderer}}, \bibinfo {author} {\bibfnamefont {B.~D.}\ \bibnamefont
  {Lackey}}, \bibinfo {author} {\bibfnamefont {R.~N.}\ \bibnamefont {Lang}}, \
  and\ \bibinfo {author} {\bibfnamefont {J.~S.}\ \bibnamefont {Read}},\ }\href
  {\doibase 10.1103/PhysRevD.81.123016} {\bibfield  {journal} {\bibinfo
  {journal} {Phys. Rev.}\ }\textbf {\bibinfo {volume} {D81}},\ \bibinfo {pages}
  {123016} (\bibinfo {year} {2010})},\ \Eprint {http://arxiv.org/abs/0911.3535}
  {arXiv:0911.3535 [astro-ph.HE]} \BibitemShut {NoStop}%
\bibitem [{\citenamefont {Paschalidis}\ \emph {et~al.}(2018)\citenamefont
  {Paschalidis}, \citenamefont {Yagi}, \citenamefont {Alvarez-Castillo},
  \citenamefont {Blaschke},\ and\ \citenamefont
  {Sedrakian}}]{Paschalidis:2017qmb}%
  \BibitemOpen
  \bibfield  {author} {\bibinfo {author} {\bibfnamefont {V.}~\bibnamefont
  {Paschalidis}}, \bibinfo {author} {\bibfnamefont {K.}~\bibnamefont {Yagi}},
  \bibinfo {author} {\bibfnamefont {D.}~\bibnamefont {Alvarez-Castillo}},
  \bibinfo {author} {\bibfnamefont {D.~B.}\ \bibnamefont {Blaschke}}, \ and\
  \bibinfo {author} {\bibfnamefont {A.}~\bibnamefont {Sedrakian}},\ }\href
  {\doibase 10.1103/PhysRevD.97.084038} {\bibfield  {journal} {\bibinfo
  {journal} {Phys. Rev.}\ }\textbf {\bibinfo {volume} {D97}},\ \bibinfo {pages}
  {084038} (\bibinfo {year} {2018})},\ \Eprint
  {http://arxiv.org/abs/1712.00451} {arXiv:1712.00451 [astro-ph.HE]}
  \BibitemShut {NoStop}%
\bibitem [{\citenamefont {Nandi}\ and\ \citenamefont
  {Char}(2018)}]{Nandi:2017rhy}%
  \BibitemOpen
  \bibfield  {author} {\bibinfo {author} {\bibfnamefont {R.}~\bibnamefont
  {Nandi}}\ and\ \bibinfo {author} {\bibfnamefont {P.}~\bibnamefont {Char}},\
  }\href {\doibase 10.3847/1538-4357/aab78c} {\bibfield  {journal} {\bibinfo
  {journal} {Astrophys. J.}\ }\textbf {\bibinfo {volume} {857}},\ \bibinfo
  {pages} {12} (\bibinfo {year} {2018})},\ \Eprint
  {http://arxiv.org/abs/1712.08094} {arXiv:1712.08094 [astro-ph.HE]}
  \BibitemShut {NoStop}%
\bibitem [{\citenamefont {Alvarez-Castillo}\ \emph {et~al.}(2018)\citenamefont
  {Alvarez-Castillo}, \citenamefont {Blaschke}, \citenamefont {Grunfeld},\ and\
  \citenamefont {Pagura}}]{Alvarez-Castillo:2018pve}%
  \BibitemOpen
  \bibfield  {author} {\bibinfo {author} {\bibfnamefont {D.~E.}\ \bibnamefont
  {Alvarez-Castillo}}, \bibinfo {author} {\bibfnamefont {D.~B.}\ \bibnamefont
  {Blaschke}}, \bibinfo {author} {\bibfnamefont {A.~G.}\ \bibnamefont
  {Grunfeld}}, \ and\ \bibinfo {author} {\bibfnamefont {V.~P.}\ \bibnamefont
  {Pagura}},\ }\href@noop {} {\  (\bibinfo {year} {2018})},\ \Eprint
  {http://arxiv.org/abs/1805.04105} {arXiv:1805.04105 [hep-ph]} \BibitemShut
  {NoStop}%
\bibitem [{\citenamefont {Gomes}\ \emph {et~al.}(2018)\citenamefont {Gomes},
  \citenamefont {Char},\ and\ \citenamefont {Schramm}}]{Gomes:2018eiv}%
  \BibitemOpen
  \bibfield  {author} {\bibinfo {author} {\bibfnamefont {R.~O.}\ \bibnamefont
  {Gomes}}, \bibinfo {author} {\bibfnamefont {P.}~\bibnamefont {Char}}, \ and\
  \bibinfo {author} {\bibfnamefont {S.}~\bibnamefont {Schramm}},\ }\href@noop
  {} {\  (\bibinfo {year} {2018})},\ \Eprint {http://arxiv.org/abs/1806.04763}
  {arXiv:1806.04763 [nucl-th]} \BibitemShut {NoStop}%
\bibitem [{\citenamefont {Sieniawska}\ \emph {et~al.}(2018)\citenamefont
  {Sieniawska}, \citenamefont {Turczanski}, \citenamefont {Bejger},\ and\
  \citenamefont {Zdunik}}]{Sieniawska:2018zzj}%
  \BibitemOpen
  \bibfield  {author} {\bibinfo {author} {\bibfnamefont {M.}~\bibnamefont
  {Sieniawska}}, \bibinfo {author} {\bibfnamefont {W.}~\bibnamefont
  {Turczanski}}, \bibinfo {author} {\bibfnamefont {M.}~\bibnamefont {Bejger}},
  \ and\ \bibinfo {author} {\bibfnamefont {J.~L.}\ \bibnamefont {Zdunik}},\
  }\href@noop {} {\  (\bibinfo {year} {2018})},\ \Eprint
  {http://arxiv.org/abs/1807.11581} {arXiv:1807.11581 [astro-ph.HE]}
  \BibitemShut {NoStop}%
\bibitem [{\citenamefont {Li}\ \emph {et~al.}(2018)\citenamefont {Li},
  \citenamefont {Yan}, \citenamefont {Geng}, \citenamefont {Huang},\ and\
  \citenamefont {Zong}}]{Li:2018ayl}%
  \BibitemOpen
  \bibfield  {author} {\bibinfo {author} {\bibfnamefont {C.-M.}\ \bibnamefont
  {Li}}, \bibinfo {author} {\bibfnamefont {Y.}~\bibnamefont {Yan}}, \bibinfo
  {author} {\bibfnamefont {J.-J.}\ \bibnamefont {Geng}}, \bibinfo {author}
  {\bibfnamefont {Y.-F.}\ \bibnamefont {Huang}}, \ and\ \bibinfo {author}
  {\bibfnamefont {H.-S.}\ \bibnamefont {Zong}},\ }\href {\doibase
  10.1103/PhysRevD.98.083013} {\bibfield  {journal} {\bibinfo  {journal} {Phys.
  Rev.}\ }\textbf {\bibinfo {volume} {D98}},\ \bibinfo {pages} {083013}
  (\bibinfo {year} {2018})},\ \Eprint {http://arxiv.org/abs/1808.02601}
  {arXiv:1808.02601 [nucl-th]} \BibitemShut {NoStop}%
\bibitem [{\citenamefont {Christian}\ \emph {et~al.}(2019)\citenamefont
  {Christian}, \citenamefont {Zacchi},\ and\ \citenamefont
  {Schaffner-Bielich}}]{Christian:2018jyd}%
  \BibitemOpen
  \bibfield  {author} {\bibinfo {author} {\bibfnamefont {J.-E.}\ \bibnamefont
  {Christian}}, \bibinfo {author} {\bibfnamefont {A.}~\bibnamefont {Zacchi}}, \
  and\ \bibinfo {author} {\bibfnamefont {J.}~\bibnamefont
  {Schaffner-Bielich}},\ }\href {\doibase 10.1103/PhysRevD.99.023009}
  {\bibfield  {journal} {\bibinfo  {journal} {Phys. Rev.}\ }\textbf {\bibinfo
  {volume} {D99}},\ \bibinfo {pages} {023009} (\bibinfo {year} {2019})},\
  \Eprint {http://arxiv.org/abs/1809.03333} {arXiv:1809.03333 [astro-ph.HE]}
  \BibitemShut {NoStop}%
\bibitem [{\citenamefont {Han}\ and\ \citenamefont
  {Steiner}(2018)}]{Han:2018mtj}%
  \BibitemOpen
  \bibfield  {author} {\bibinfo {author} {\bibfnamefont {S.}~\bibnamefont
  {Han}}\ and\ \bibinfo {author} {\bibfnamefont {A.~W.}\ \bibnamefont
  {Steiner}},\ }\href@noop {} {\  (\bibinfo {year} {2018})},\ \Eprint
  {http://arxiv.org/abs/1810.10967} {arXiv:1810.10967 [nucl-th]} \BibitemShut
  {NoStop}%
\bibitem [{\citenamefont {Gerlach}(1968)}]{Gerlach:1968zz}%
  \BibitemOpen
  \bibfield  {author} {\bibinfo {author} {\bibfnamefont {U.~H.}\ \bibnamefont
  {Gerlach}},\ }\href {\doibase 10.1103/PhysRev.172.1325} {\bibfield  {journal}
  {\bibinfo  {journal} {Phys.Rev.}\ }\textbf {\bibinfo {volume} {172}},\
  \bibinfo {pages} {1325} (\bibinfo {year} {1968})}\BibitemShut {NoStop}%
\bibitem [{\citenamefont {K\"ampfer}(1981)}]{Kampfer:1981yr}%
  \BibitemOpen
  \bibfield  {author} {\bibinfo {author} {\bibfnamefont {B.}~\bibnamefont
  {K\"ampfer}},\ }\href {\doibase 10.1088/0305-4470/14/11/009} {\bibfield
  {journal} {\bibinfo  {journal} {J.Phys.}\ }\textbf {\bibinfo {volume}
  {A14}},\ \bibinfo {pages} {L471} (\bibinfo {year} {1981})}\BibitemShut
  {NoStop}%
\bibitem [{\citenamefont {Glendenning}\ and\ \citenamefont
  {Kettner}(2000)}]{Glendenning:1998ag}%
  \BibitemOpen
  \bibfield  {author} {\bibinfo {author} {\bibfnamefont {N.~K.}\ \bibnamefont
  {Glendenning}}\ and\ \bibinfo {author} {\bibfnamefont {C.}~\bibnamefont
  {Kettner}},\ }\href@noop {} {\bibfield  {journal} {\bibinfo  {journal}
  {Astron. Astrophys.}\ }\textbf {\bibinfo {volume} {353}},\ \bibinfo {pages}
  {L9} (\bibinfo {year} {2000})},\ \Eprint
  {http://arxiv.org/abs/astro-ph/9807155} {astro-ph/9807155} \BibitemShut
  {NoStop}%
\bibitem [{\citenamefont {Tolos}\ \emph
  {et~al.}(2017{\natexlab{a}})\citenamefont {Tolos}, \citenamefont
  {Centelles},\ and\ \citenamefont {Ramos}}]{Tolos:2016hhl}%
  \BibitemOpen
  \bibfield  {author} {\bibinfo {author} {\bibfnamefont {L.}~\bibnamefont
  {Tolos}}, \bibinfo {author} {\bibfnamefont {M.}~\bibnamefont {Centelles}}, \
  and\ \bibinfo {author} {\bibfnamefont {A.}~\bibnamefont {Ramos}},\ }\href
  {\doibase 10.3847/1538-4357/834/1/3} {\bibfield  {journal} {\bibinfo
  {journal} {Astrophys. J.}\ }\textbf {\bibinfo {volume} {834}},\ \bibinfo
  {pages} {3} (\bibinfo {year} {2017}{\natexlab{a}})},\ \Eprint
  {http://arxiv.org/abs/1610.00919} {arXiv:1610.00919 [astro-ph.HE]}
  \BibitemShut {NoStop}%
\bibitem [{\citenamefont {Tolos}\ \emph
  {et~al.}(2017{\natexlab{b}})\citenamefont {Tolos}, \citenamefont
  {Centelles},\ and\ \citenamefont {Ramos}}]{Tolos:2017lgv}%
  \BibitemOpen
  \bibfield  {author} {\bibinfo {author} {\bibfnamefont {L.}~\bibnamefont
  {Tolos}}, \bibinfo {author} {\bibfnamefont {M.}~\bibnamefont {Centelles}}, \
  and\ \bibinfo {author} {\bibfnamefont {A.}~\bibnamefont {Ramos}},\ }\href
  {\doibase 10.1017/pasa.2017.60} {\bibfield  {journal} {\bibinfo  {journal}
  {Publ. Astron. Soc. Austral.}\ }\textbf {\bibinfo {volume} {34}},\ \bibinfo
  {pages} {e065} (\bibinfo {year} {2017}{\natexlab{b}})},\ \Eprint
  {http://arxiv.org/abs/1708.08681} {arXiv:1708.08681 [astro-ph.HE]}
  \BibitemShut {NoStop}%
\bibitem [{\citenamefont {Sharma}\ \emph {et~al.}(2015)\citenamefont {Sharma},
  \citenamefont {Centelles}, \citenamefont {Viñas}, \citenamefont {Baldo},\
  and\ \citenamefont {Burgio}}]{Sharma:2015bna}%
  \BibitemOpen
  \bibfield  {author} {\bibinfo {author} {\bibfnamefont {B.~K.}\ \bibnamefont
  {Sharma}}, \bibinfo {author} {\bibfnamefont {M.}~\bibnamefont {Centelles}},
  \bibinfo {author} {\bibfnamefont {X.}~\bibnamefont {Viñas}}, \bibinfo
  {author} {\bibfnamefont {M.}~\bibnamefont {Baldo}}, \ and\ \bibinfo {author}
  {\bibfnamefont {G.~F.}\ \bibnamefont {Burgio}},\ }\href {\doibase
  10.1051/0004-6361/201526642} {\bibfield  {journal} {\bibinfo  {journal}
  {Astron. Astrophys.}\ }\textbf {\bibinfo {volume} {584}},\ \bibinfo {pages}
  {A103} (\bibinfo {year} {2015})},\ \Eprint {http://arxiv.org/abs/1506.00375}
  {arXiv:1506.00375 [nucl-th]} \BibitemShut {NoStop}%
\bibitem [{\citenamefont {Chen}\ and\ \citenamefont
  {Piekarewicz}(2014)}]{Chen:2014sca}%
  \BibitemOpen
  \bibfield  {author} {\bibinfo {author} {\bibfnamefont {W.-C.}\ \bibnamefont
  {Chen}}\ and\ \bibinfo {author} {\bibfnamefont {J.}~\bibnamefont
  {Piekarewicz}},\ }\href {\doibase 10.1103/PhysRevC.90.044305} {\bibfield
  {journal} {\bibinfo  {journal} {Phys. Rev.}\ }\textbf {\bibinfo {volume}
  {C90}},\ \bibinfo {pages} {044305} (\bibinfo {year} {2014})},\ \Eprint
  {http://arxiv.org/abs/1408.4159} {arXiv:1408.4159 [nucl-th]} \BibitemShut
  {NoStop}%
\bibitem [{\citenamefont {Dover}\ and\ \citenamefont
  {Gal}(1983)}]{Dover:1982ng}%
  \BibitemOpen
  \bibfield  {author} {\bibinfo {author} {\bibfnamefont {C.~B.}\ \bibnamefont
  {Dover}}\ and\ \bibinfo {author} {\bibfnamefont {A.}~\bibnamefont {Gal}},\
  }\href {\doibase 10.1016/0003-4916(83)90036-2} {\bibfield  {journal}
  {\bibinfo  {journal} {Annals Phys.}\ }\textbf {\bibinfo {volume} {146}},\
  \bibinfo {pages} {309} (\bibinfo {year} {1983})}\BibitemShut {NoStop}%
\bibitem [{\citenamefont {Millener}\ \emph {et~al.}(1988)\citenamefont
  {Millener}, \citenamefont {Dover},\ and\ \citenamefont
  {Gal}}]{Millener:1988hp}%
  \BibitemOpen
  \bibfield  {author} {\bibinfo {author} {\bibfnamefont {D.~J.}\ \bibnamefont
  {Millener}}, \bibinfo {author} {\bibfnamefont {C.~B.}\ \bibnamefont {Dover}},
  \ and\ \bibinfo {author} {\bibfnamefont {A.}~\bibnamefont {Gal}},\ }\href
  {\doibase 10.1103/PhysRevC.38.2700} {\bibfield  {journal} {\bibinfo
  {journal} {Phys. Rev.}\ }\textbf {\bibinfo {volume} {C38}},\ \bibinfo {pages}
  {2700} (\bibinfo {year} {1988})}\BibitemShut {NoStop}%
\bibitem [{\citenamefont {Fukuda}\ \emph {et~al.}(1998)\citenamefont {Fukuda}
  \emph {et~al.}}]{Fukuda:1998bi}%
  \BibitemOpen
  \bibfield  {author} {\bibinfo {author} {\bibfnamefont {T.}~\bibnamefont
  {Fukuda}} \emph {et~al.} (\bibinfo {collaboration} {E224}),\ }\href {\doibase
  10.1103/PhysRevC.58.1306} {\bibfield  {journal} {\bibinfo  {journal} {Phys.
  Rev.}\ }\textbf {\bibinfo {volume} {C58}},\ \bibinfo {pages} {1306} (\bibinfo
  {year} {1998})}\BibitemShut {NoStop}%
\bibitem [{\citenamefont {Khaustov}\ \emph {et~al.}(2000)\citenamefont
  {Khaustov} \emph {et~al.}}]{Khaustov:1999bz}%
  \BibitemOpen
  \bibfield  {author} {\bibinfo {author} {\bibfnamefont {P.}~\bibnamefont
  {Khaustov}} \emph {et~al.} (\bibinfo {collaboration} {AGS E885}),\ }\href
  {\doibase 10.1103/PhysRevC.61.054603} {\bibfield  {journal} {\bibinfo
  {journal} {Phys. Rev.}\ }\textbf {\bibinfo {volume} {C61}},\ \bibinfo {pages}
  {054603} (\bibinfo {year} {2000})},\ \Eprint
  {http://arxiv.org/abs/nucl-ex/9912007} {arXiv:nucl-ex/9912007 [nucl-ex]}
  \BibitemShut {NoStop}%
\bibitem [{\citenamefont {Noumi}\ \emph {et~al.}(2002)\citenamefont {Noumi}
  \emph {et~al.}}]{Noumi:2001tx}%
  \BibitemOpen
  \bibfield  {author} {\bibinfo {author} {\bibfnamefont {H.}~\bibnamefont
  {Noumi}} \emph {et~al.},\ }\href {\doibase 10.1103/PhysRevLett.89.072301}
  {\bibfield  {journal} {\bibinfo  {journal} {Phys. Rev. Lett.}\ }\textbf
  {\bibinfo {volume} {89}},\ \bibinfo {pages} {072301} (\bibinfo {year}
  {2002})},\ \bibinfo {note} {[Erratum: Phys. Rev.
  Lett.90,049902(2003)]}\BibitemShut {NoStop}%
\bibitem [{\citenamefont {Harada}\ and\ \citenamefont
  {Hirabayashi}(2006)}]{Harada:2006yj}%
  \BibitemOpen
  \bibfield  {author} {\bibinfo {author} {\bibfnamefont {T.}~\bibnamefont
  {Harada}}\ and\ \bibinfo {author} {\bibfnamefont {Y.}~\bibnamefont
  {Hirabayashi}},\ }\href {\doibase 10.1016/j.nuclphysa.2005.12.018} {\bibfield
   {journal} {\bibinfo  {journal} {Nucl. Phys.}\ }\textbf {\bibinfo {volume}
  {A767}},\ \bibinfo {pages} {206} (\bibinfo {year} {2006})}\BibitemShut
  {NoStop}%
\bibitem [{\citenamefont {Kohno}\ \emph {et~al.}(2006)\citenamefont {Kohno},
  \citenamefont {Fujiwara}, \citenamefont {Watanabe}, \citenamefont {Ogata},\
  and\ \citenamefont {Kawai}}]{Kohno:2006iq}%
  \BibitemOpen
  \bibfield  {author} {\bibinfo {author} {\bibfnamefont {M.}~\bibnamefont
  {Kohno}}, \bibinfo {author} {\bibfnamefont {Y.}~\bibnamefont {Fujiwara}},
  \bibinfo {author} {\bibfnamefont {Y.}~\bibnamefont {Watanabe}}, \bibinfo
  {author} {\bibfnamefont {K.}~\bibnamefont {Ogata}}, \ and\ \bibinfo {author}
  {\bibfnamefont {M.}~\bibnamefont {Kawai}},\ }\href {\doibase
  10.1103/PhysRevC.74.064613} {\bibfield  {journal} {\bibinfo  {journal} {Phys.
  Rev.}\ }\textbf {\bibinfo {volume} {C74}},\ \bibinfo {pages} {064613}
  (\bibinfo {year} {2006})},\ \Eprint {http://arxiv.org/abs/nucl-th/0611080}
  {arXiv:nucl-th/0611080 [nucl-th]} \BibitemShut {NoStop}%
\bibitem [{\citenamefont {Tsang}\ \emph {et~al.}(2012)\citenamefont {Tsang}
  \emph {et~al.}}]{Tsang:2012se}%
  \BibitemOpen
  \bibfield  {author} {\bibinfo {author} {\bibfnamefont {M.~B.}\ \bibnamefont
  {Tsang}} \emph {et~al.},\ }\href {\doibase 10.1103/PhysRevC.86.015803}
  {\bibfield  {journal} {\bibinfo  {journal} {Phys. Rev.}\ }\textbf {\bibinfo
  {volume} {C86}},\ \bibinfo {pages} {015803} (\bibinfo {year} {2012})},\
  \Eprint {http://arxiv.org/abs/1204.0466} {arXiv:1204.0466 [nucl-ex]}
  \BibitemShut {NoStop}%
\bibitem [{\citenamefont {Danielewicz}\ \emph {et~al.}(2002)\citenamefont
  {Danielewicz}, \citenamefont {Lacey},\ and\ \citenamefont
  {Lynch}}]{Danielewicz:2002pu}%
  \BibitemOpen
  \bibfield  {author} {\bibinfo {author} {\bibfnamefont {P.}~\bibnamefont
  {Danielewicz}}, \bibinfo {author} {\bibfnamefont {R.}~\bibnamefont {Lacey}},
  \ and\ \bibinfo {author} {\bibfnamefont {W.~G.}\ \bibnamefont {Lynch}},\
  }\href {\doibase 10.1126/science.1078070} {\bibfield  {journal} {\bibinfo
  {journal} {Science}\ }\textbf {\bibinfo {volume} {298}},\ \bibinfo {pages}
  {1592} (\bibinfo {year} {2002})},\ \Eprint
  {http://arxiv.org/abs/nucl-th/0208016} {arXiv:nucl-th/0208016 [nucl-th]}
  \BibitemShut {NoStop}%
\bibitem [{\citenamefont {Fuchs}\ \emph {et~al.}(2001)\citenamefont {Fuchs},
  \citenamefont {Faessler}, \citenamefont {Zabrodin},\ and\ \citenamefont
  {Zheng}}]{Fuchs:2000kp}%
  \BibitemOpen
  \bibfield  {author} {\bibinfo {author} {\bibfnamefont {C.}~\bibnamefont
  {Fuchs}}, \bibinfo {author} {\bibfnamefont {A.}~\bibnamefont {Faessler}},
  \bibinfo {author} {\bibfnamefont {E.}~\bibnamefont {Zabrodin}}, \ and\
  \bibinfo {author} {\bibfnamefont {Y.-M.}\ \bibnamefont {Zheng}},\ }\href@noop
  {} {\bibfield  {journal} {\bibinfo  {journal} {Phys. Rev. Lett.}\ }\textbf
  {\bibinfo {volume} {86}},\ \bibinfo {pages} {1974} (\bibinfo {year}
  {2001})},\ \Eprint {http://arxiv.org/abs/nucl-th/0011102} {nucl-th/0011102}
  \BibitemShut {NoStop}%
\bibitem [{\citenamefont {Lynch}\ \emph {et~al.}(2009)\citenamefont {Lynch}
  \emph {et~al.}}]{Lynch:2009vc}%
  \BibitemOpen
  \bibfield  {author} {\bibinfo {author} {\bibfnamefont {W.~G.}\ \bibnamefont
  {Lynch}} \emph {et~al.},\ }\href {\doibase 10.1016/j.ppnp.2009.01.001}
  {\bibfield  {journal} {\bibinfo  {journal} {Prog. Part. Nucl. Phys.}\
  }\textbf {\bibinfo {volume} {62}},\ \bibinfo {pages} {427} (\bibinfo {year}
  {2009})},\ \Eprint {http://arxiv.org/abs/0901.0412} {arXiv:0901.0412
  [nucl-ex]} \BibitemShut {NoStop}%
\bibitem [{\citenamefont {Negreiros}\ \emph {et~al.}(2018)\citenamefont
  {Negreiros}, \citenamefont {Tolos}, \citenamefont {Centelles}, \citenamefont
  {Ramos},\ and\ \citenamefont {Dexheimer}}]{Negreiros:2018cho}%
  \BibitemOpen
  \bibfield  {author} {\bibinfo {author} {\bibfnamefont {R.}~\bibnamefont
  {Negreiros}}, \bibinfo {author} {\bibfnamefont {L.}~\bibnamefont {Tolos}},
  \bibinfo {author} {\bibfnamefont {M.}~\bibnamefont {Centelles}}, \bibinfo
  {author} {\bibfnamefont {A.}~\bibnamefont {Ramos}}, \ and\ \bibinfo {author}
  {\bibfnamefont {V.}~\bibnamefont {Dexheimer}},\ }\href {\doibase
  10.3847/1538-4357/aad049} {\bibfield  {journal} {\bibinfo  {journal}
  {Astrophys. J.}\ }\textbf {\bibinfo {volume} {863}},\ \bibinfo {pages} {104}
  (\bibinfo {year} {2018})},\ \Eprint {http://arxiv.org/abs/1804.00334}
  {arXiv:1804.00334 [astro-ph.HE]} \BibitemShut {NoStop}%
\bibitem [{\citenamefont {Zacchi}\ \emph {et~al.}(2016)\citenamefont {Zacchi},
  \citenamefont {Hanauske},\ and\ \citenamefont
  {Schaffner-Bielich}}]{Zacchi:2015oma}%
  \BibitemOpen
  \bibfield  {author} {\bibinfo {author} {\bibfnamefont {A.}~\bibnamefont
  {Zacchi}}, \bibinfo {author} {\bibfnamefont {M.}~\bibnamefont {Hanauske}}, \
  and\ \bibinfo {author} {\bibfnamefont {J.}~\bibnamefont
  {Schaffner-Bielich}},\ }\href {\doibase 10.1103/PhysRevD.93.065011}
  {\bibfield  {journal} {\bibinfo  {journal} {Phys. Rev.}\ }\textbf {\bibinfo
  {volume} {D93}},\ \bibinfo {pages} {065011} (\bibinfo {year} {2016})},\
  \Eprint {http://arxiv.org/abs/1510.00180} {arXiv:1510.00180 [nucl-th]}
  \BibitemShut {NoStop}%
\bibitem [{\citenamefont {Ranea-Sandoval}\ \emph {et~al.}(2016)\citenamefont
  {Ranea-Sandoval}, \citenamefont {Han}, \citenamefont {Orsaria}, \citenamefont
  {Contrera}, \citenamefont {Weber},\ and\ \citenamefont
  {Alford}}]{Ranea-Sandoval:2015ldr}%
  \BibitemOpen
  \bibfield  {author} {\bibinfo {author} {\bibfnamefont {I.~F.}\ \bibnamefont
  {Ranea-Sandoval}}, \bibinfo {author} {\bibfnamefont {S.}~\bibnamefont {Han}},
  \bibinfo {author} {\bibfnamefont {M.~G.}\ \bibnamefont {Orsaria}}, \bibinfo
  {author} {\bibfnamefont {G.~A.}\ \bibnamefont {Contrera}}, \bibinfo {author}
  {\bibfnamefont {F.}~\bibnamefont {Weber}}, \ and\ \bibinfo {author}
  {\bibfnamefont {M.~G.}\ \bibnamefont {Alford}},\ }\href {\doibase
  10.1103/PhysRevC.93.045812} {\bibfield  {journal} {\bibinfo  {journal} {Phys.
  Rev.}\ }\textbf {\bibinfo {volume} {C93}},\ \bibinfo {pages} {045812}
  (\bibinfo {year} {2016})},\ \Eprint {http://arxiv.org/abs/1512.09183}
  {arXiv:1512.09183 [nucl-th]} \BibitemShut {NoStop}%
\bibitem [{\citenamefont {Chamel}\ \emph {et~al.}(2013)\citenamefont {Chamel},
  \citenamefont {Fantina}, \citenamefont {Pearson},\ and\ \citenamefont
  {Goriely}}]{Chamel:2012ea}%
  \BibitemOpen
  \bibfield  {author} {\bibinfo {author} {\bibfnamefont {N.}~\bibnamefont
  {Chamel}}, \bibinfo {author} {\bibfnamefont {A.~F.}\ \bibnamefont {Fantina}},
  \bibinfo {author} {\bibfnamefont {J.~M.}\ \bibnamefont {Pearson}}, \ and\
  \bibinfo {author} {\bibfnamefont {S.}~\bibnamefont {Goriely}},\ }\href
  {\doibase 10.1051/0004-6361/201220986} {\bibfield  {journal} {\bibinfo
  {journal} {Astron. Astrophys.}\ }\textbf {\bibinfo {volume} {553}},\ \bibinfo
  {pages} {A22} (\bibinfo {year} {2013})},\ \Eprint
  {http://arxiv.org/abs/1205.0983} {arXiv:1205.0983 [nucl-th]} \BibitemShut
  {NoStop}%
\bibitem [{\citenamefont {Zdunik}\ and\ \citenamefont
  {Haensel}(2013)}]{Zdunik:2012dj}%
  \BibitemOpen
  \bibfield  {author} {\bibinfo {author} {\bibfnamefont {J.~L.}\ \bibnamefont
  {Zdunik}}\ and\ \bibinfo {author} {\bibfnamefont {P.}~\bibnamefont
  {Haensel}},\ }\href {\doibase 10.1051/0004-6361/201220697} {\bibfield
  {journal} {\bibinfo  {journal} {Astron. Astrophys.}\ }\textbf {\bibinfo
  {volume} {551}},\ \bibinfo {pages} {A61} (\bibinfo {year} {2013})},\ \Eprint
  {http://arxiv.org/abs/1211.1231} {arXiv:1211.1231 [astro-ph.SR]} \BibitemShut
  {NoStop}%
\bibitem [{\citenamefont {Alford}\ \emph {et~al.}(2013)\citenamefont {Alford},
  \citenamefont {Han},\ and\ \citenamefont {Prakash}}]{Alford:2013aca}%
  \BibitemOpen
  \bibfield  {author} {\bibinfo {author} {\bibfnamefont {M.~G.}\ \bibnamefont
  {Alford}}, \bibinfo {author} {\bibfnamefont {S.}~\bibnamefont {Han}}, \ and\
  \bibinfo {author} {\bibfnamefont {M.}~\bibnamefont {Prakash}},\ }\href
  {\doibase 10.1103/PhysRevD.88.083013} {\bibfield  {journal} {\bibinfo
  {journal} {Phys. Rev.}\ }\textbf {\bibinfo {volume} {D88}},\ \bibinfo {pages}
  {083013} (\bibinfo {year} {2013})},\ \Eprint {http://arxiv.org/abs/1302.4732}
  {arXiv:1302.4732 [astro-ph.SR]} \BibitemShut {NoStop}%
\bibitem [{\citenamefont {{Rezzolla}}\ and\ \citenamefont
  {{Zanotti}}(2013)}]{Rezzolla_book:2013}%
  \BibitemOpen
  \bibfield  {author} {\bibinfo {author} {\bibfnamefont {L.}~\bibnamefont
  {{Rezzolla}}}\ and\ \bibinfo {author} {\bibfnamefont {O.}~\bibnamefont
  {{Zanotti}}},\ }\href {\doibase 10.1093/acprof:oso/9780198528906.001.0001}
  {\emph {\bibinfo {title} {Relativistic Hydrodynamics}}}\ (\bibinfo
  {publisher} {Oxford University Press},\ \bibinfo {address} {Oxford, UK},\
  \bibinfo {year} {2013})\BibitemShut {NoStop}%
\bibitem [{\citenamefont {Alford}\ and\ \citenamefont
  {Sedrakian}(2017)}]{AlfordSedrakian2017}%
  \BibitemOpen
  \bibfield  {author} {\bibinfo {author} {\bibfnamefont {M.}~\bibnamefont
  {Alford}}\ and\ \bibinfo {author} {\bibfnamefont {A.}~\bibnamefont
  {Sedrakian}},\ }\href {\doibase 10.1103/PhysRevLett.119.161104} {\bibfield
  {journal} {\bibinfo  {journal} {Phys. Rev. Lett.}\ }\textbf {\bibinfo
  {volume} {119}},\ \bibinfo {pages} {161104} (\bibinfo {year}
  {2017})}\BibitemShut {NoStop}%
\bibitem [{\citenamefont {Christian}\ \emph {et~al.}(2018)\citenamefont
  {Christian}, \citenamefont {Zacchi},\ and\ \citenamefont
  {Schaffner-Bielich}}]{Christian:2017jni}%
  \BibitemOpen
  \bibfield  {author} {\bibinfo {author} {\bibfnamefont {J.-E.}\ \bibnamefont
  {Christian}}, \bibinfo {author} {\bibfnamefont {A.}~\bibnamefont {Zacchi}}, \
  and\ \bibinfo {author} {\bibfnamefont {J.}~\bibnamefont
  {Schaffner-Bielich}},\ }\href {\doibase 10.1140/epja/i2018-12472-y}
  {\bibfield  {journal} {\bibinfo  {journal} {Eur. Phys. J.}\ }\textbf
  {\bibinfo {volume} {A54}},\ \bibinfo {pages} {28} (\bibinfo {year} {2018})},\
  \Eprint {http://arxiv.org/abs/1707.07524} {arXiv:1707.07524 [astro-ph.HE]}
  \BibitemShut {NoStop}%
\bibitem [{\citenamefont {Kurkela}\ \emph {et~al.}(2010)\citenamefont
  {Kurkela}, \citenamefont {Romatschke},\ and\ \citenamefont
  {Vuorinen}}]{Kurkela:2009gj}%
  \BibitemOpen
  \bibfield  {author} {\bibinfo {author} {\bibfnamefont {A.}~\bibnamefont
  {Kurkela}}, \bibinfo {author} {\bibfnamefont {P.}~\bibnamefont {Romatschke}},
  \ and\ \bibinfo {author} {\bibfnamefont {A.}~\bibnamefont {Vuorinen}},\
  }\href@noop {} {\bibfield  {journal} {\bibinfo  {journal} {Phys. Rev.}\
  }\textbf {\bibinfo {volume} {D81}},\ \bibinfo {pages} {105021} (\bibinfo
  {year} {2010})},\ \Eprint {http://arxiv.org/abs/0912.1856} {arXiv:0912.1856
  [hep-ph]} \BibitemShut {NoStop}%
\bibitem [{\citenamefont {Glendenning}(1992)}]{Glendenning:1992vb}%
  \BibitemOpen
  \bibfield  {author} {\bibinfo {author} {\bibfnamefont {N.~K.}\ \bibnamefont
  {Glendenning}},\ }\href {\doibase 10.1103/PhysRevD.46.1274} {\bibfield
  {journal} {\bibinfo  {journal} {Phys.Rev.}\ }\textbf {\bibinfo {volume}
  {D46}},\ \bibinfo {pages} {1274} (\bibinfo {year} {1992})}\BibitemShut
  {NoStop}%
\bibitem [{\citenamefont {Macher}\ and\ \citenamefont
  {Schaffner-Bielich}(2005)}]{Macher:2004vw}%
  \BibitemOpen
  \bibfield  {author} {\bibinfo {author} {\bibfnamefont {J.}~\bibnamefont
  {Macher}}\ and\ \bibinfo {author} {\bibfnamefont {J.}~\bibnamefont
  {Schaffner-Bielich}},\ }\href {\doibase 10.1088/0143-0807/26/3/003}
  {\bibfield  {journal} {\bibinfo  {journal} {Eur.J.Phys.}\ }\textbf {\bibinfo
  {volume} {26}},\ \bibinfo {pages} {341} (\bibinfo {year} {2005})},\ \Eprint
  {http://arxiv.org/abs/astro-ph/0411295} {arXiv:astro-ph/0411295 [astro-ph]}
  \BibitemShut {NoStop}%
\bibitem [{\citenamefont {{Ayriyan}}\ \emph {et~al.}(2018)\citenamefont
  {{Ayriyan}} \emph {et~al.}}]{Ayriyan2018}%
  \BibitemOpen
  \bibfield  {author} {\bibinfo {author} {\bibfnamefont {A.}~\bibnamefont
  {{Ayriyan}}} \emph {et~al.},\ }\href {\doibase 10.1103/PhysRevC.97.045802}
  {\bibfield  {journal} {\bibinfo  {journal} {\prc}\ }\textbf {\bibinfo
  {volume} {97}},\ \bibinfo {eid} {045802} (\bibinfo {year}
  {2018})}\BibitemShut {NoStop}%
\bibitem [{\citenamefont {Bodmer}(1971)}]{Bodmer:1971we}%
  \BibitemOpen
  \bibfield  {author} {\bibinfo {author} {\bibfnamefont {A.~R.}\ \bibnamefont
  {Bodmer}},\ }\href@noop {} {\bibfield  {journal} {\bibinfo  {journal} {Phys.
  Rev. D}\ }\textbf {\bibinfo {volume} {4}},\ \bibinfo {pages} {1601} (\bibinfo
  {year} {1971})}\BibitemShut {NoStop}%
\bibitem [{\citenamefont {Witten}(1984)}]{Witten:1984rs}%
  \BibitemOpen
  \bibfield  {author} {\bibinfo {author} {\bibfnamefont {E.}~\bibnamefont
  {Witten}},\ }\href@noop {} {\bibfield  {journal} {\bibinfo  {journal} {Phys.
  Rev. D}\ }\textbf {\bibinfo {volume} {30}},\ \bibinfo {pages} {272} (\bibinfo
  {year} {1984})}\BibitemShut {NoStop}%
\bibitem [{\citenamefont {{Drago}}\ \emph {et~al.}(2014)\citenamefont
  {{Drago}}, \citenamefont {{Lavagno}},\ and\ \citenamefont
  {{Pagliara}}}]{Drago2014}%
  \BibitemOpen
  \bibfield  {author} {\bibinfo {author} {\bibfnamefont {A.}~\bibnamefont
  {{Drago}}}, \bibinfo {author} {\bibfnamefont {A.}~\bibnamefont {{Lavagno}}},
  \ and\ \bibinfo {author} {\bibfnamefont {G.}~\bibnamefont {{Pagliara}}},\
  }\href {\doibase 10.1103/PhysRevD.89.043014} {\bibfield  {journal} {\bibinfo
  {journal} {Phys. Rev. D}\ }\textbf {\bibinfo {volume} {89}},\ \bibinfo {eid}
  {043014} (\bibinfo {year} {2014})},\ \Eprint {http://arxiv.org/abs/1309.7263}
  {arXiv:1309.7263 [nucl-th]} \BibitemShut {NoStop}%
\bibitem [{\citenamefont {{Bombaci}}\ \emph {et~al.}(2016)\citenamefont
  {{Bombaci}}, \citenamefont {{Logoteta}}, \citenamefont {{Vida{\~n}a}},\ and\
  \citenamefont {{Provid{\^e}ncia}}}]{Bombaci2016}%
  \BibitemOpen
  \bibfield  {author} {\bibinfo {author} {\bibfnamefont {I.}~\bibnamefont
  {{Bombaci}}}, \bibinfo {author} {\bibfnamefont {D.}~\bibnamefont
  {{Logoteta}}}, \bibinfo {author} {\bibfnamefont {I.}~\bibnamefont
  {{Vida{\~n}a}}}, \ and\ \bibinfo {author} {\bibfnamefont {C.}~\bibnamefont
  {{Provid{\^e}ncia}}},\ }\href {\doibase 10.1140/epja/i2016-16058-5}
  {\bibfield  {journal} {\bibinfo  {journal} {European Physical Journal A}\
  }\textbf {\bibinfo {volume} {52}},\ \bibinfo {eid} {58} (\bibinfo {year}
  {2016})},\ \Eprint {http://arxiv.org/abs/1601.04559} {arXiv:1601.04559
  [astro-ph.HE]} \BibitemShut {NoStop}%
\bibitem [{\citenamefont {{Lugones}}(2016)}]{Lugones2016}%
  \BibitemOpen
  \bibfield  {author} {\bibinfo {author} {\bibfnamefont {G.}~\bibnamefont
  {{Lugones}}},\ }\href {\doibase 10.1140/epja/i2016-16053-x} {\bibfield
  {journal} {\bibinfo  {journal} {European Physical Journal A}\ }\textbf
  {\bibinfo {volume} {52}},\ \bibinfo {eid} {53} (\bibinfo {year} {2016})},\
  \Eprint {http://arxiv.org/abs/1508.05548} {arXiv:1508.05548 [astro-ph.HE]}
  \BibitemShut {NoStop}%
\bibitem [{\citenamefont {{Burgio}}\ \emph {et~al.}(2018)\citenamefont
  {{Burgio}}, \citenamefont {{Drago}}, \citenamefont {{Pagliara}},
  \citenamefont {{Schulze}},\ and\ \citenamefont {{Wei}}}]{Drago2018}%
  \BibitemOpen
  \bibfield  {author} {\bibinfo {author} {\bibfnamefont {G.~F.}\ \bibnamefont
  {{Burgio}}}, \bibinfo {author} {\bibfnamefont {A.}~\bibnamefont {{Drago}}},
  \bibinfo {author} {\bibfnamefont {G.}~\bibnamefont {{Pagliara}}}, \bibinfo
  {author} {\bibfnamefont {H.-J.}\ \bibnamefont {{Schulze}}}, \ and\ \bibinfo
  {author} {\bibfnamefont {J.-B.}\ \bibnamefont {{Wei}}},\ }\href@noop {}
  {\bibfield  {journal} {\bibinfo  {journal} {arXiv:1803.09696}\ } (\bibinfo
  {year} {2018})},\ \Eprint {http://arxiv.org/abs/1803.09696} {arXiv:1803.09696
  [astro-ph.HE]} \BibitemShut {NoStop}%
\bibitem [{\citenamefont {Tews}\ \emph
  {et~al.}(2018{\natexlab{b}})\citenamefont {Tews}, \citenamefont {Carlson},
  \citenamefont {Gandolfi},\ and\ \citenamefont {Reddy}}]{Tews:2018kmu}%
  \BibitemOpen
  \bibfield  {author} {\bibinfo {author} {\bibfnamefont {I.}~\bibnamefont
  {Tews}}, \bibinfo {author} {\bibfnamefont {J.}~\bibnamefont {Carlson}},
  \bibinfo {author} {\bibfnamefont {S.}~\bibnamefont {Gandolfi}}, \ and\
  \bibinfo {author} {\bibfnamefont {S.}~\bibnamefont {Reddy}},\ }\href
  {\doibase 10.3847/1538-4357/aac267} {\bibfield  {journal} {\bibinfo
  {journal} {Astrophys. J.}\ }\textbf {\bibinfo {volume} {860}},\ \bibinfo
  {pages} {149} (\bibinfo {year} {2018}{\natexlab{b}})},\ \Eprint
  {http://arxiv.org/abs/1801.01923} {arXiv:1801.01923 [nucl-th]} \BibitemShut
  {NoStop}%
\bibitem [{\citenamefont {Alford}\ \emph {et~al.}(2015)\citenamefont {Alford},
  \citenamefont {Burgio}, \citenamefont {Han}, \citenamefont {Taranto},\ and\
  \citenamefont {Zappalà}}]{Alford:2015dpa}%
  \BibitemOpen
  \bibfield  {author} {\bibinfo {author} {\bibfnamefont {M.~G.}\ \bibnamefont
  {Alford}}, \bibinfo {author} {\bibfnamefont {G.~F.}\ \bibnamefont {Burgio}},
  \bibinfo {author} {\bibfnamefont {S.}~\bibnamefont {Han}}, \bibinfo {author}
  {\bibfnamefont {G.}~\bibnamefont {Taranto}}, \ and\ \bibinfo {author}
  {\bibfnamefont {D.}~\bibnamefont {Zappalà}},\ }\href {\doibase
  10.1103/PhysRevD.92.083002} {\bibfield  {journal} {\bibinfo  {journal} {Phys.
  Rev.}\ }\textbf {\bibinfo {volume} {D92}},\ \bibinfo {pages} {083002}
  (\bibinfo {year} {2015})},\ \Eprint {http://arxiv.org/abs/1501.07902}
  {arXiv:1501.07902 [nucl-th]} \BibitemShut {NoStop}%
\bibitem [{\citenamefont {{Seidov}}(1971)}]{1971SvA....15..347S}%
  \BibitemOpen
  \bibfield  {author} {\bibinfo {author} {\bibfnamefont {Z.~F.}\ \bibnamefont
  {{Seidov}}},\ }\href@noop {} {\bibfield  {journal} {\bibinfo  {journal}
  {Soviet Astronomy}\ }\textbf {\bibinfo {volume} {15}},\ \bibinfo {pages}
  {347} (\bibinfo {year} {1971})}\BibitemShut {NoStop}%
\bibitem [{\citenamefont {{Schaeffer}}\ \emph {et~al.}(1983)\citenamefont
  {{Schaeffer}}, \citenamefont {{Zdunik}},\ and\ \citenamefont
  {{Haensel}}}]{1983A&A...126..121S}%
  \BibitemOpen
  \bibfield  {author} {\bibinfo {author} {\bibfnamefont {R.}~\bibnamefont
  {{Schaeffer}}}, \bibinfo {author} {\bibfnamefont {L.}~\bibnamefont
  {{Zdunik}}}, \ and\ \bibinfo {author} {\bibfnamefont {P.}~\bibnamefont
  {{Haensel}}},\ }\href@noop {} {\bibfield  {journal} {\bibinfo  {journal}
  {Astronomy and Astrophysics}\ }\textbf {\bibinfo {volume} {126}},\ \bibinfo
  {pages} {121} (\bibinfo {year} {1983})}\BibitemShut {NoStop}%
\bibitem [{\citenamefont {Lindblom}(1998)}]{Lindblom:1998dp}%
  \BibitemOpen
  \bibfield  {author} {\bibinfo {author} {\bibfnamefont {L.}~\bibnamefont
  {Lindblom}},\ }\href {\doibase 10.1103/PhysRevD.58.024008} {\bibfield
  {journal} {\bibinfo  {journal} {Phys.Rev.}\ }\textbf {\bibinfo {volume}
  {D58}},\ \bibinfo {pages} {024008} (\bibinfo {year} {1998})},\ \Eprint
  {http://arxiv.org/abs/gr-qc/9802072} {arXiv:gr-qc/9802072 [gr-qc]}
  \BibitemShut {NoStop}%
\bibitem [{\citenamefont {Watts}\ \emph {et~al.}(2016)\citenamefont {Watts}
  \emph {et~al.}}]{Watts:2016uzu}%
  \BibitemOpen
  \bibfield  {author} {\bibinfo {author} {\bibfnamefont {A.~L.}\ \bibnamefont
  {Watts}} \emph {et~al.},\ }\href {\doibase 10.1103/RevModPhys.88.021001}
  {\bibfield  {journal} {\bibinfo  {journal} {Rev. Mod. Phys.}\ }\textbf
  {\bibinfo {volume} {88}},\ \bibinfo {pages} {021001} (\bibinfo {year}
  {2016})},\ \Eprint {http://arxiv.org/abs/1602.01081} {arXiv:1602.01081
  [astro-ph.HE]} \BibitemShut {NoStop}%
\bibitem [{\citenamefont {Bose}\ \emph {et~al.}(2018)\citenamefont {Bose},
  \citenamefont {Chakravarti}, \citenamefont {Rezzolla}, \citenamefont
  {Sathyaprakash},\ and\ \citenamefont {Takami}}]{Bose2017}%
  \BibitemOpen
  \bibfield  {author} {\bibinfo {author} {\bibfnamefont {S.}~\bibnamefont
  {Bose}}, \bibinfo {author} {\bibfnamefont {K.}~\bibnamefont {Chakravarti}},
  \bibinfo {author} {\bibfnamefont {L.}~\bibnamefont {Rezzolla}}, \bibinfo
  {author} {\bibfnamefont {B.~S.}\ \bibnamefont {Sathyaprakash}}, \ and\
  \bibinfo {author} {\bibfnamefont {K.}~\bibnamefont {Takami}},\ }\href
  {\doibase 10.1103/PhysRevLett.120.031102} {\bibfield  {journal} {\bibinfo
  {journal} {Phys. Rev. Lett.}\ }\textbf {\bibinfo {volume} {120}},\ \bibinfo
  {pages} {031102} (\bibinfo {year} {2018})},\ \Eprint
  {http://arxiv.org/abs/1705.10850} {arXiv:1705.10850 [gr-qc]} \BibitemShut
  {NoStop}%
\bibitem [{\citenamefont {Zhao}\ and\ \citenamefont
  {Lattimer}(2018)}]{Zhao:2018nyf}%
  \BibitemOpen
  \bibfield  {author} {\bibinfo {author} {\bibfnamefont {T.}~\bibnamefont
  {Zhao}}\ and\ \bibinfo {author} {\bibfnamefont {J.~M.}\ \bibnamefont
  {Lattimer}},\ }\href {\doibase 10.1103/PhysRevD.98.063020} {\bibfield
  {journal} {\bibinfo  {journal} {Phys. Rev.}\ }\textbf {\bibinfo {volume}
  {D98}},\ \bibinfo {pages} {063020} (\bibinfo {year} {2018})},\ \Eprint
  {http://arxiv.org/abs/1808.02858} {arXiv:1808.02858 [astro-ph.HE]}
  \BibitemShut {NoStop}%
\bibitem [{\citenamefont {Postnikov}\ \emph {et~al.}(2010)\citenamefont
  {Postnikov}, \citenamefont {Prakash},\ and\ \citenamefont
  {Lattimer}}]{Postnikov:2010yn}%
  \BibitemOpen
  \bibfield  {author} {\bibinfo {author} {\bibfnamefont {S.}~\bibnamefont
  {Postnikov}}, \bibinfo {author} {\bibfnamefont {M.}~\bibnamefont {Prakash}},
  \ and\ \bibinfo {author} {\bibfnamefont {J.~M.}\ \bibnamefont {Lattimer}},\
  }\href {\doibase 10.1103/PhysRevD.82.024016} {\bibfield  {journal} {\bibinfo
  {journal} {Phys. Rev.}\ }\textbf {\bibinfo {volume} {D82}},\ \bibinfo {pages}
  {024016} (\bibinfo {year} {2010})},\ \Eprint {http://arxiv.org/abs/1004.5098}
  {arXiv:1004.5098 [astro-ph.SR]} \BibitemShut {NoStop}%
\bibitem [{\citenamefont {Schertler}\ \emph {et~al.}(2000)\citenamefont
  {Schertler}, \citenamefont {Greiner}, \citenamefont {Schaffner-Bielich},\
  and\ \citenamefont {Thoma}}]{Schertler:2000xq}%
  \BibitemOpen
  \bibfield  {author} {\bibinfo {author} {\bibfnamefont {K.}~\bibnamefont
  {Schertler}}, \bibinfo {author} {\bibfnamefont {C.}~\bibnamefont {Greiner}},
  \bibinfo {author} {\bibfnamefont {J.}~\bibnamefont {Schaffner-Bielich}}, \
  and\ \bibinfo {author} {\bibfnamefont {M.~H.}\ \bibnamefont {Thoma}},\
  }\href@noop {} {\bibfield  {journal} {\bibinfo  {journal} {Nucl. Phys.}\
  }\textbf {\bibinfo {volume} {A677}},\ \bibinfo {pages} {463} (\bibinfo {year}
  {2000})},\ \Eprint {http://arxiv.org/abs/astro-ph/0001467} {astro-ph/0001467}
  \BibitemShut {NoStop}%
\bibitem [{\citenamefont {Flanagan}\ and\ \citenamefont
  {Hinderer}(2008)}]{Flanagan:2007ix}%
  \BibitemOpen
  \bibfield  {author} {\bibinfo {author} {\bibfnamefont {E.~E.}\ \bibnamefont
  {Flanagan}}\ and\ \bibinfo {author} {\bibfnamefont {T.}~\bibnamefont
  {Hinderer}},\ }\href {\doibase 10.1103/PhysRevD.77.021502} {\bibfield
  {journal} {\bibinfo  {journal} {Phys. Rev.}\ }\textbf {\bibinfo {volume}
  {D77}},\ \bibinfo {pages} {021502} (\bibinfo {year} {2008})},\ \Eprint
  {http://arxiv.org/abs/0709.1915} {arXiv:0709.1915 [astro-ph]} \BibitemShut
  {NoStop}%
\bibitem [{\citenamefont {Favata}(2014)}]{Favata:2013rwa}%
  \BibitemOpen
  \bibfield  {author} {\bibinfo {author} {\bibfnamefont {M.}~\bibnamefont
  {Favata}},\ }\href {\doibase 10.1103/PhysRevLett.112.101101} {\bibfield
  {journal} {\bibinfo  {journal} {Phys. Rev. Lett.}\ }\textbf {\bibinfo
  {volume} {112}},\ \bibinfo {pages} {101101} (\bibinfo {year} {2014})},\
  \Eprint {http://arxiv.org/abs/1310.8288} {arXiv:1310.8288 [gr-qc]}
  \BibitemShut {NoStop}%
\bibitem [{\citenamefont {{K{\"o}ppel}}\ \emph {et~al.}(2019)\citenamefont
  {{K{\"o}ppel}}, \citenamefont {{Bovard}},\ and\ \citenamefont
  {{Rezzolla}}}]{Koeppel2019}%
  \BibitemOpen
  \bibfield  {author} {\bibinfo {author} {\bibfnamefont {S.}~\bibnamefont
  {{K{\"o}ppel}}}, \bibinfo {author} {\bibfnamefont {L.}~\bibnamefont
  {{Bovard}}}, \ and\ \bibinfo {author} {\bibfnamefont {L.}~\bibnamefont
  {{Rezzolla}}},\ }\href {\doibase 10.3847/2041-8213/ab0210} {\bibfield
  {journal} {\bibinfo  {journal} {Astrophys. J. Lett.}\ }\textbf {\bibinfo
  {volume} {872}},\ \bibinfo {eid} {L16} (\bibinfo {year} {2019})},\ \Eprint
  {http://arxiv.org/abs/1901.09977} {arXiv:1901.09977 [gr-qc]} \BibitemShut
  {NoStop}%
\bibitem [{\citenamefont {{Takami}}\ \emph {et~al.}(2014)\citenamefont
  {{Takami}}, \citenamefont {{Rezzolla}},\ and\ \citenamefont
  {{Baiotti}}}]{Takami:2014}%
  \BibitemOpen
  \bibfield  {author} {\bibinfo {author} {\bibfnamefont {K.}~\bibnamefont
  {{Takami}}}, \bibinfo {author} {\bibfnamefont {L.}~\bibnamefont
  {{Rezzolla}}}, \ and\ \bibinfo {author} {\bibfnamefont {L.}~\bibnamefont
  {{Baiotti}}},\ }\href {\doibase 10.1103/PhysRevLett.113.091104} {\bibfield
  {journal} {\bibinfo  {journal} {Phys. Rev. Lett.}\ }\textbf {\bibinfo
  {volume} {113}},\ \bibinfo {eid} {091104} (\bibinfo {year} {2014})},\ \Eprint
  {http://arxiv.org/abs/1403.5672} {arXiv:1403.5672 [gr-qc]} \BibitemShut
  {NoStop}%
\bibitem [{\citenamefont {Takami}\ \emph {et~al.}(2015)\citenamefont {Takami},
  \citenamefont {Rezzolla},\ and\ \citenamefont
  {Baiotti}}]{takami2015spectral}%
  \BibitemOpen
  \bibfield  {author} {\bibinfo {author} {\bibfnamefont {K.}~\bibnamefont
  {Takami}}, \bibinfo {author} {\bibfnamefont {L.}~\bibnamefont {Rezzolla}}, \
  and\ \bibinfo {author} {\bibfnamefont {L.}~\bibnamefont {Baiotti}},\
  }\href@noop {} {\bibfield  {journal} {\bibinfo  {journal} {Physical Review
  D}\ }\textbf {\bibinfo {volume} {91}},\ \bibinfo {pages} {064001} (\bibinfo
  {year} {2015})}\BibitemShut {NoStop}%
\bibitem [{\citenamefont {{Rezzolla}}\ and\ \citenamefont
  {{Takami}}(2016)}]{Rezzolla2016}%
  \BibitemOpen
  \bibfield  {author} {\bibinfo {author} {\bibfnamefont {L.}~\bibnamefont
  {{Rezzolla}}}\ and\ \bibinfo {author} {\bibfnamefont {K.}~\bibnamefont
  {{Takami}}},\ }\href {\doibase 10.1103/PhysRevD.93.124051} {\bibfield
  {journal} {\bibinfo  {journal} {Phys. Rev. D}\ }\textbf {\bibinfo {volume}
  {93}},\ \bibinfo {eid} {124051} (\bibinfo {year} {2016})},\ \Eprint
  {http://arxiv.org/abs/1604.00246} {arXiv:1604.00246 [gr-qc]} \BibitemShut
  {NoStop}%
\bibitem [{\citenamefont {{Hanauske}}\ \emph {et~al.}(2018)\citenamefont
  {{Hanauske}}, \citenamefont {{Yilmaz}}, \citenamefont {{Mitropoulos}},
  \citenamefont {{Rezzolla}},\ and\ \citenamefont
  {{St{\"o}cker}}}]{Hanauske2018EPJWC}%
  \BibitemOpen
  \bibfield  {author} {\bibinfo {author} {\bibfnamefont {M.}~\bibnamefont
  {{Hanauske}}}, \bibinfo {author} {\bibfnamefont {Z.~S.}\ \bibnamefont
  {{Yilmaz}}}, \bibinfo {author} {\bibfnamefont {C.}~\bibnamefont
  {{Mitropoulos}}}, \bibinfo {author} {\bibfnamefont {L.}~\bibnamefont
  {{Rezzolla}}}, \ and\ \bibinfo {author} {\bibfnamefont {H.}~\bibnamefont
  {{St{\"o}cker}}},\ }in\ \href {\doibase 10.1051/epjconf/201817120004} {\emph
  {\bibinfo {booktitle} {European Physical Journal Web of Conferences}}},\
  Vol.\ \bibinfo {volume} {171}\ (\bibinfo {year} {2018})\ p.\ \bibinfo {pages}
  {20004}\BibitemShut {NoStop}%
\bibitem [{\citenamefont {{Hanauske}}\ and\ \citenamefont
  {{Bovard}}(2018)}]{Hanauske2018JApA}%
  \BibitemOpen
  \bibfield  {author} {\bibinfo {author} {\bibfnamefont {M.}~\bibnamefont
  {{Hanauske}}}\ and\ \bibinfo {author} {\bibfnamefont {L.}~\bibnamefont
  {{Bovard}}},\ }\href {\doibase 10.1007/s12036-018-9536-3} {\bibfield
  {journal} {\bibinfo  {journal} {Journal of Astrophysics and Astronomy}\
  }\textbf {\bibinfo {volume} {39}},\ \bibinfo {eid} {45} (\bibinfo {year}
  {2018})}\BibitemShut {NoStop}%
\bibitem [{\citenamefont {Most}\ \emph
  {et~al.}(2018{\natexlab{b}})\citenamefont {Most} \emph
  {et~al.}}]{Most:2018eaw}%
  \BibitemOpen
  \bibfield  {author} {\bibinfo {author} {\bibfnamefont {E.~R.}\ \bibnamefont
  {Most}} \emph {et~al.},\ }\href@noop {} {\  (\bibinfo {year}
  {2018}{\natexlab{b}})},\ \Eprint {http://arxiv.org/abs/1807.03684}
  {arXiv:1807.03684 [astro-ph.HE]} \BibitemShut {NoStop}%
\bibitem [{\citenamefont {Bauswein}\ \emph {et~al.}(2018)\citenamefont
  {Bauswein} \emph {et~al.}}]{Bauswein:2018bma}%
  \BibitemOpen
  \bibfield  {author} {\bibinfo {author} {\bibfnamefont {A.}~\bibnamefont
  {Bauswein}} \emph {et~al.},\ }\href@noop {} {\  (\bibinfo {year} {2018})},\
  \Eprint {http://arxiv.org/abs/1809.01116} {arXiv:1809.01116 [astro-ph.HE]}
  \BibitemShut {NoStop}%
\bibitem [{\citenamefont {{Amati}}\ \emph {et~al.}(2018)\citenamefont {{Amati}}
  \emph {et~al.}}]{Amati2018}%
  \BibitemOpen
  \bibfield  {author} {\bibinfo {author} {\bibfnamefont {L.}~\bibnamefont
  {{Amati}}} \emph {et~al.},\ }\href {\doibase 10.1016/j.asr.2018.03.010}
  {\bibfield  {journal} {\bibinfo  {journal} {Advances in Space Research}\
  }\textbf {\bibinfo {volume} {62}},\ \bibinfo {pages} {191} (\bibinfo {year}
  {2018})},\ \Eprint {http://arxiv.org/abs/1710.04638} {arXiv:1710.04638
  [astro-ph.IM]} \BibitemShut {NoStop}%
\end{thebibliography}%
\bibliographystyle{apsrev4-1}
  
\end{document}